\documentclass[apj]{emulateapj}

\usepackage{graphicx}
\usepackage{txfonts}
\usepackage{lscape}
\usepackage{aas_macros}
\usepackage{amssymb}
\usepackage{natbib}
\bibpunct{(}{)}{;}{a}{,}{,}
\usepackage{longtable}
\usepackage[hang,bf,footnotesize]{subfigure}
\usepackage[figuresright]{rotating}
\begin{document}

\title{The protoplanetary disks in the nearby massive star forming region Cygnus~OB2}

\author{M. G. Guarcello\altaffilmark{1}, J. J. Drake\altaffilmark{1}, N. J. Wright\altaffilmark{1,2}, J. E. Drew\altaffilmark{2}, R. A. Gutermuth\altaffilmark{3,4}, J. L. Hora\altaffilmark{1}, T. Naylor\altaffilmark{5}, T. Aldcroft\altaffilmark{1}, A. Fruscione\altaffilmark{1}, D. Garc\'{i}a-Alvarez\altaffilmark{6,7,8}, V. L. Kashyap\altaffilmark{1}, R. King\altaffilmark{5}}

\altaffiltext{1}{Smithsonian Astrophysical Observatory, MS-67, 60 Garden Street, Cambridge, MA 02138, USA}
\altaffiltext{2}{School of Physics, Astronomy \& Mathematics, University of Hertfordshire, College Lane, Hatfield, AL10 9AB, UK}
\altaffiltext{3}{Five College Astronomy Department, Smith College, Northampton, MA 01063, USA} 
\altaffiltext{4}{Department of Astronomy, University of Massachusetts, Amherst, MA, USA} 
\altaffiltext{5}{School of Physics, University of Exeter, Stocker Road, Exeter EX4 4QL, UK}
\altaffiltext{6}{Dpto. de Astrof\'{\i}sica, Universidad de La Laguna, 38206 - La Laguna, Tenerife, Spain}
\altaffiltext{7}{Grantecan CALP, 38712 Bre{\~na} Baja, La Palma, Spain}
\altaffiltext{8}{Instituto de Astrof\'{i}sica de Canarias, E-38205 La Laguna, Tenerife, Spain}

\begin{abstract}

The formation of stars in massive clusters is one of the main modes of the star formation process. However, the study of massive star forming regions is hampered by their typically large distances to the Sun. One exception to this is the massive star forming region Cygnus~OB2 in the Cygnus~X region, at the distance of $\sim1400\,pc$. Cygnus~OB2 hosts very rich populations of massive and low-mass stars, being the best target in our Galaxy to study the formation of stars, circumstellar disks, and planets in presence of massive stars. In this paper we combine a wide and deep set of photometric data, from the $r$ band to $24\,\mu m$, in order to select the disk bearing population of stars in Cygnus~OB2 and identify the class~I, class~II, and stars with transition and pre-transition disks. We selected 1843 sources with infrared excesses in an area of $1^{\circ} \times 1^{\circ}$ centered on Cyg~OB2 in several evolutionary stages: 8.4\% class~I, 13.1\% flat-spectrum sources, 72.9\% class~II, 2.3\% pre-transition disks, and 3.3\% transition disks. The spatial distribution of these sources shows a central cluster surrounded by a annular overdensity and some clumps of recent star formation in the outer region. Several candidate subclusters are identified, both along the overdensity and in the rest of the association. 
\end{abstract}



\section{Introduction}
\label{intro}

The vast majority of stars form in clusters \citep{LadaLada2003} of different size and stellar population, and a significant fraction of star formation occurs in proximity of OB stars. The energetic radiation from even just a few of massive stars in a cluster have dramatic effects on the star formation process, and the evolution of circumstellar disks, the entire parental cluster, and the molecular cloud. Circumstellar disks close to massive stars can be quickly dissipated by photoevaporation induced by the intense UV radiation (i. e. \citealp{Johnstone1998}). The loss of gas from the photoevaporating disk is evident in the images of the proplyds (i.e. protoplanetary disks) observed by the Hubble Space Telescope in Orion \citep{ODell1994}, and the effects of photoevaporation have been observed in other massive clusters (i.e. \citealp{Balog2007, Guarcello2007, Wright2012}). It has been shown (i.e. \citealp{AdamsPFM2006}) that during the early evolution of young clusters hosting massive stars, even the low-mass stars formed at large distance from the OB members may experience intense UV radiation during their orbital motion around the cluster center.There is also strong evidence that the massive stars can trigger the formation of new generations of low-mass stars in the parental clouds \citep{ElmegreenLada1977,Elmegreen2011}. \par
  In order to properly understand these processes, and also thanks to the growing evidence that our Solar System formed in presence of nearby O stars \citep{Adams2010}, in the last decades the astronomical community put a great effort in the study of how star formation proceeds in presence of massive stars. However, the study of star formation in the most massive star-forming environments, containing thousands massive and low-mass stars, is hampered by the large distances of these associations, which usually lie nearby the Galactic center (such as the Arches cluster, at only $\sim 30\,pc$ from the Galactic Center, \citealp{Nagata1995}), in distant regions of our Galaxy (such as Westerlund~1, with a distance to the Sun ranging between 4 and $5\,kpc$, \citealt{Negueruela2010,Gennaro2011}), or in other Galaxies, such as 30~Doradus in the Large Magellanic Cloud. \par
The massive association Cygnus~OB2, in the Cygnus-X molecular cloud, is the best target to study star formation in presence of massive stars. In fact, it hosts thousands OB stars and low-mass young sources, with a distance of $\sim 1400\, pc$ to the Sun \citep{Rygl2012}, which is $\sim0.6$ times the distance to the second closest massive star forming region (the Carina Nebula, \citealt{Smith2002}). The massive population of Cyg~OB2 was first estimated by \citet{Reddish67} to about 300 OB stars. Later, \citet{Knodlseder2000} estimated a population of about 2600 B stars and 120 O stars, but this result was downsized by following studies \citep{Comeron2002,Hanson2003,Drew2008}. Cyg~OB2 also contains some of the most massive stars known in our Galaxy \citep{WalbornHLM2002}. \par
  The population of the central area of the association has an age ranging mainly between $\sim3\,Myrs$ and $\sim5\,Myrs$ \citep{Wright2009,Wright2010}. The cluster is affected by a large extinction, which is highly variable across the field. The main extinction range is $3^m\leq A_V\leq8^m$ (i.e. \citealp{Drew2008,Sale09,Wright2010}, and \citealp{Guarcello2012B}). Since Cyg~OB2 has cleared most of the intracluster medium creating a central cavity \citep{Schneider2006}, this extinction is mainly due to the foreground dust associated with the Cygnus Rift. \par
In their study based on an archival Chandra/ACIS-I observation and 2MASS data, \citet{AlbaceteColombo2007} found a low fraction of members still bearing a disk ($\sim4.4\%$), much lower than other coeval clusters, suggesting that disks and star formation have been halted prematurely by the extremely intense ionizing radiation produced by the OB stars of Cyg~OB2. The disk fraction in the association has been recalculated by \citet{Wright2010} by using deeper NIR data from the UKIDSS/GPS catalog, finding a larger fraction (5.9\% - 7.9\%) more compatible with the age of the association. \par
  In this paper, we aim to finally identify the disk-bearing population of the association, in an area of $\sim1$ square degree centered on Cyg~OB2. We will select stars with disks combining several criteria and using deep optical and infrared data. We will also study their evolutionary status and obtain information about the structure of the association and the presence of recent star formation sites. 
Fig. \ref{rgb_im} shows a RGB image of Cyg~OB2, where green is the $H\alpha$ emission from hot gas, blue the stellar emission in the $r^{\prime}$ band \citep{Drew2005}, red marks the $8.0\mu m$ emission from dust (from Spitzer/IRAC images of Cyg~OB2, \citealp{Beerer2010}). This image maps the regions where the dust emission prevails over the gas emission and vice versa. Gas emission is prominent in the area at north-west cleared from dust, and particularly prominent in front of the northern trunk, suggesting the presence of a photodissociation front along the trunk where the gas is heated by the incident UV radiation from nearby massive stars. Dust emission is more prominent in the south-east direction and in the regions classified as DR18, and Globules \#1 and \#2 \citep{Schneider2006}. The dashed box encompasses the one square degree field of the Chandra Cygnus~OB2 Legacy Survey \citep{Drake2009}. This area contains the central part of Cyg~OB2 and most of the important nebular structures in this area. The white box encompasses the area observed with OSIRIS \citep{Guarcello2012B}. \par
  The paper is organized as follow: in the Sect. \ref{cat_sec} and \ref{match_sec} we present the data and discuss the technique adopted to merge them in an unique multiwavelength catalog; in Sect. \ref{diskselection} we select the preliminary list of candidate stars with disks; in Sect. \ref{contselection} we select the candidate contaminants (foreground stars, background giants, extragalactic sources and stars with unreliable excesses) and produce the final catalog, which is analyzed in Sect. \ref{merg_list}. The evolutionary status of the disk-bearing stars is studied in Sect. \ref{evol}, while the subclustering, the morphology of the associations and the sites of recent star formation are studied in Sect. \ref{spadis} and \ref{activeSF}.

\section{Data and multi-wavelength catalog}
\label{cat_sec}

Young Stellar Objects (YSOs) are stars with a complex morphology, which depends on their evolutionary status. In class~I YSOs,  the central star is surrounded by an accreting circumstellar disk and envelope; the envelope is completely or partially dissipated in the class~II phase; in the more evolved class~III objects the circumstellar disk is also dissipated, and usually what remains is a young planetary system and/or a debris disk orbiting around a contracting pre-Main Sequence star. The transition between the class~II and class~III phase is thought to involve a disk with an evacuated inner region (``transition disks'' or ``disks with large inner holes'') and/or a low mass disk populated by processed dust. \par
  The best way to study and select a sample of YSOs in a star forming region is by adopting a multiwavelength approach. In fact, the study of their stellar properties (e.g. mass, age, etc...) requires the use of optical data, the emission from the inner part of the disk is usually prominent in the NIR, while the emission from the outer disk and the envelope become important at longer wavelengths. Besides, different disk morphologies and geometric properties may result in different SEDs (Spectral Energy Distributions), making some approach more effective than others for their detection and study. For these reasons, to obtain a reliable and as more complete as possible list of disk-bearing members of Cyg~OB2, we compiled a multi-wavelength catalog based on optical and infrared data. The catalogs employed are described in following list. \par
  
$ $\par
{\it GTC/OSIRIS catalog} (65349 sources). \par

Our main optical catalog in $r,\,i,\,z$ bands is obtained from observations with OSIRIS, mounted in the $10.4\,m$ Gran Telescopio CANARIAS (GTC) telescope of the Spanish Observatorio del Roque de los Muchachos in La Palma \citep{Cepa2000} and compiled by \citet{Guarcello2012B}. If a source meets the criteria for good photometry defined in \citet{Guarcello2012B}, then its optical photometry in $r,\,i,\,z$ bands is taken from this catalog independently from the quality of other available optical data. This catalog reaches $r=27^m$, while the limit for good photometry is approximately at $r=25^m$ (i.e. the limit reached in the color-magnitude diagrams involving these data, where the limits of $\sigma_{mag}<0.1^m$ and $\sigma_{col}<0.1^m$ have been adopted). At the distance of $1400\,pc$, and using a $3.5\,Myrs$ isochrone \citep{Wright2010} with the average extinction $A_V=4.3^m$ \citep{Guarcello2012B}, this limit corresponds to a $0.16\,M_{\odot}$ star. \par
$ $\par

{\it IPHAS catalog} (24072 sources). \par
The catalog in the $r^{\prime},\,i^{\prime},\,H_{\alpha}$ bands is obtained from observations with the Wide Field Camera (WFC) on the $2.5\,m$ Isaac Newton Telescope (INT) for the INT Photometric H$\alpha$ Survey (IPHAS, \citealt{Drew2005}). The limit for good photometry is approximately $r^{\prime}=21.5^m$. When necessary, the IPHAS data are converted from the Vega system to the AB system (the photometric system of both the OSIRIS and SDSS data) by using the transformations defined in \citet{Guarcello2012B}. \par
$ $\par

{\it SDSS/DR8 catalog} (27531 sources). \par
The Eighth Data Release (DR8) of the SDSS catalog \citep{Aihara2011} in $u,\,g,\,r,\,i,\,z$ bands. This catalog reaches $r=22^m$ with good photometry ($\sim 0.4\,M_{\odot}$ at the distance of Cyg~OB2), but it is three magnitudes brighter than the saturation limit adopted for the OSIRIS catalog ($r=16^m$, corresponding to a $2.5\,M_{\odot}$ star), allowing us to study stars brighter than those in the OSIRIS catalog. \par
$ $\par

{\it UKIDSS/GPS catalog} (273478 sources). \par
The main NIR catalog in the JHK bands is the Galactic Plane Survey (GPS, \citealp{Lucas2008}) of the United Kingdom Infrared Deep Sky Survey (UKIDSS; \citealp{LawrenceWAE2007}), based on observations with the Wide Field Camera (WFCAM, \citealp{Casali2007}) on the United Kingdom InfraRed Telescope (UKIRT).  We undertook our own photometric extraction from the processed WFCAM images (\citealp{DyeWHC2012}, Irwin et al in preparation) obtained from the WFCAM Science Archive \citep{HamblyCCM2008}. This optimal extraction \citep{Naylor1998} is described in detail in \citet{KingNBG2013}, and results in photometry on the system of \citet{HodgkinIHW2009}. It improves the quality of the final data compared with the UKIDSS extraction pipeline, allowing us to reach deeper magnitudes with acceptable uncertainties. The UKIDSS data reach $J=21^m$, corresponding to stars with $M< 0.1 M_{\odot}$ at the distance and extinction of Cyg~OB2. \par
$ $\par

{\it 2MASS/PSC catalog} (43485 sources). \par
Complementary NIR data in $JHK$ bands are obtained from the 2 Micron All Sky Survey (2MASS, \citealp{Cutri2003}) Point Source Catalog (PSC),  and are mainly used to retrieve the JHK photometry of the NIR sources brighter than the saturation limits in UKIDSS. \par
$ $\par

{\it Spitzer Legacy Survey of the Cygnus~X region} (149381 sources). \par
Spitzer data in the IRAC $3.6,\,4.5,\,5.8,\,8.0\,\mu m$ and MIPS $24\mu m$ bands are retrieved from the first data delivery of the Spitzer Legacy Survey of the Cygnus-X region \citep{Beerer2010}, with sources detected down to $0.5\, M_{\odot}$. \par
$ $\par

All the data used here, except those from OSIRIS, are available in the entire field studied. The OSIRIS data are only available in a central $\sim40^{\prime}\times 40^{\prime}$ field. 

\section{Cross-correlation procedure}
\label{match_sec}

The multiwavelength catalog is built by matching  all the catalogs listed in Sect. \ref{cat_sec}. Table \ref{match_tb} shows the intermediate matches, each involving two of these catalogs, with the number of matched sources, the multiple matches, and the used matching radii. In the fourth and fifth rows, ``NIR'' means the joined UKIDSS+2MASS catalog, ``OPTICAL'' the OSIRIS+SDSS+IPHAS catalog. We first merged the UKIDSS and 2MASS catalogs in a master $JHK$ catalog (first match in table \ref{match_tb}); then we created an optical catalog by matching the OSIRIS and IPHAS catalogs (second match), and then the OSIRIS+IPHAS with the SDSS one (third match). In the fourth match the optical and $JHK$ catalogs are merged together, and finally matched with the Spitzer catalog in the last step. \par
	In each intermediate match shown in Table \ref{match_tb}, we first converted the astrometry of catalog ``B'' to that of the catalog ``A'' by matching the two catalogs using $r_{match}=2^{\prime \prime}$ and then correcting the coordinates of catalog ``B'' by the median of the positional offsets. The following step consisted in the calculation of the appropriate matching radius between the two catalogs ``A'' and ``B''. We performed 30 test-matches, increasing the test value of the matching radius by $0.1^{\prime \prime}$ at each step. In each test match, we calculated the expected number of spurious matches using the method developed by \citet{Damiani2006}, which is appropriate for correlated catalogs whose sources are almost uniformly distributed in the field. With this approach, it was possible to obtain the differential distributions of the expected real and spurious matches as a function of the increasing matching radius. We simply adopted as matching radius the highest value at which the differential distribution of the spurious matches is small (few percentiles) compared to that of the real matches. We then found the sources in common between the two catalogs merging the pairs separated by a distance smaller than the matching radius. The results of each match have then been visually reviewed. \par
Table \ref{cntrp_tb} shows the distribution of the optical, $JHK$, and Spitzer counterparts in our catalog. Given the depth of the UKIDSS catalog, the sample of the sources with only UKIDSS counterparts forms the largest group.

\section{Disks selection}
\label{diskselection}

The procedure we adopted to select the disk-bearing population of Cyg~OB2 involves several criteria, whose results are merged in a unique list of stars with disks. This list is then pruned from all the stars meeting the criteria for being one of the following contaminants: foreground stars, background giants, extragalactic sources, and stars with unreliable excesses. In the following, we first describe each of these criteria and illustrate them with appropriate diagrams. Each selection is applied only to the stars with errors in the involved colors smaller than $0.15^m$. This corresponds to the limit of ``good photometry'' defined in Sect. \ref{cat_sec}. Some of the loci shown in the diagrams will be defined later in the paper (i.e. those of giants and other contaminants). In this section we will describe all the selection criteria we adopted. In Section \ref{contselection} we will show how we identified candidate contaminants among the selected stars and how we obtained the final list of stars with disk. \par

  \subsection{The Gutermuth et al. 2009 scheme}
  \label{gut_scheme}

This scheme is an updated version of that introduced by \citet{Gutermuth2008}. It involves several IRAC colors, which are used for the selection both of contaminants and disk-bearing stars in different evolutionary stages. In \citet{Gutermuth2009} (hereafter GMM09) this method is first applied to select candidate background galaxies and AGN, and then stars with disks are selected among the remaining sources. Since we need to compare the efficiency of each selection criterion, we applied in this step only the definitions of stars with disks presented in GMM09, while the contaminants have been removed in later step (Sect. \ref{contselection}). {\bf Since an estimate of the individual extinction is available only for $\sim 21\%$ of the Spitzer sources, we have not attempted an extinction correction of the IRAC colors. In Sect. \ref{contselection} false positives that have been selected also because of this approximation will be identified and removed from the list.} The GMM09 method is our main selection criterion, resulting in a selection which is larger and less affected by contamination than the other methods, as we will show later. The initial selection obtained with this scheme consists in a total of 1448 sources with excesses. Fig. \ref{1324_im} shows the $[3.6]-[5.8]$ vs. $[4.5]-[8.0]$ diagram, which is one of those used in this scheme, overplotted with the stars with disks selected using this method. \par

  \subsection{Other methods}
  \label{oth_scheme}

In order to take full advantage of the large set of photometric data we have available, we will merge the selection obtained with the GMM09 method with those obtained from several other criteria defined by several authors. Since different selection methods may be sensitive to different morphology, geometry, and evolutionary status of the disks, and also to different contaminants, this procedure will allow us to obtain a highly reliable and complete selection. For instance, small excesses due to low-mass disks can be more easily detected by combining optical and infrared colors \citep{GuarcelloDMP2010}; background giants can be more easily identified using the optical and $JHK$ colors, that are more sensitive to the extinction, while Spitzer colors are more sensitive to the presence of infrared excesses; or, finally, the nature of a source with IRAC colors typical of a background galaxy can be determined by its position in the optical or $JHK$ color-magnitude diagrams. \par

$ $\par
{\bf The [4.5] vs. [4.5]-[8.0] diagram}. A useful method to select disk-bearing objects is provided by the $[4.5]$ vs. $[4.5]-[8.0]$ diagram \citep{Harvey2007}. In this diagram the direction of the reddening vector is almost vertical, resulting in a small overlap between the loci of reddened and intrinsically red sources. Unfortunately, the faint-red part of the diagram can be contaminated by background galaxies, so this method is less useful for selecting faint stars. Most of the stars in the gap between the disks and AGN loci, however, are selected by other methods. A total of 667 candidate stars with disks have been selected with this method. Fig. \ref{224_im} shows the $[4.5]$ vs. $[4.5]-[8.0]$ diagram with the defined loci and the selected candidate stars with disks. \par
$ $\par
 
{\bf $K$-excess stars}. Stars with excess in the $K$ band have been selected with three criteria. The $J-H$ vs. $H-K$ diagram is a good method to select stars whose $H-K$ is more red than expected photospheric values \citep{Meyer1997}. To select stars with excesses in $K$ band, we defined a locus in the $J-H$ vs. $H-K$ diagram taking into account the direction of the reddening vector, and selected only those stars having $J-K>2^m$. Fig. \ref{jhhk_im} shows the $J-H$ vs. $H-K$ diagram with the loci of candidate stars with disks and background giants, together with the selected stars (385 stars).  
  An additional selection of candidate $K$-excess stars is provided by the use of two reddening-free color indices $Q_{JHHK}$ \citep{Damiani2006,Guarcello2007} and $Q_{riHK}$. The general definition of the index $Q_{ABCD}$ is:

\begin{equation}
Q_{ABCD}=\left( A-B \right) - \left( C-D \right) \times E_{A-B}/E_{C-D}
\label{ind_eq}
\end{equation}

where $A,\,B,\,C,\,D$ are four generic bands, and $E_{A-B}$ and $E_{C-D}$ are the color excesses in the two colors $A-B$ and $C-D$. In this case, $C-D$ is always $H-K$, while $A-B$ is equal to $J-H$ in the former index and $r-i$ in the latter. The index is reddening-free, so its numerical value is not affected by interstellar absorption. If compared to photospheric values, the index become more negative if $C-D$ is more red or $A-B$ is more blue. These two properties make the index a very effective method to select stars with excesses in the $D$ band ($K$ in this case), and it can distinguish less intense excesses than the color-color diagram \citep{GuarcelloDMP2010}. Besides, using an optical color such as $A-B$ the index can select stars with intense accretion or an optical SED dominated by scattered light, since both of these effects may result in blue excesses (\citealt{GuarcelloDMP2010}, and Bonito et al., in preparation). \par
  The main caveat in using these indices is that the ratio between the color excesses must be exactly known in order for the indices to be really reddening-free. This is hard to achieve, considering the high interstellar extinction in the direction of Cyg~OB2. Thus, this selection can suffer significant contamination from normal stars and background giants. In Sect. \ref{contselection} we will describe how we discarded these contaminants. \par
Fig. \ref{qk_im} shows $Q_{JHHK}$ vs. $H-K$ and $Q_{riHK}$ vs. $H-K$ diagrams with the defined loci and the selected stars with excess in $K$: 247 and 124 for each index, respectively. In both diagrams, an excess was defined to be present for those stars whose index is smaller than the chosen limit of photospheric emission by more than $3\sigma_Q$. The photospheric limits were chosen based on the colors of normal stars predicted by the \citet{Siess2000} and PADOVA isochrones. The $J-K>2^m$ filter was also applied to these two selections. Combining the three diagnostics, a total of 518 candidate $K$-excess stars have been selected. \par

$ $\par
{\bf IRAC $Q$ indices}. To select stars with excess in each IRAC band taking advantage of the properties of the $Q$ indices, we defined 4 indices $Q_{riJ[sp]}$, which are similar to those defined in \citet{Guarcello2009}. Using Eq. \ref{ind_eq}, in these indices $A-B$ is the optical color $r-i$ while $C-D$ is equal to $J-[sp]$ with $[sp]$ being one of the IRAC bands. Fig. \ref{qir_im} shows the four diagrams used to select stars with excesses based on the $Q_{riJ[sp]}$ indices. In these diagrams, the PADOVA isochrones can not be used to define a lower limit for the photospheric indices in the $Q_{rij[sp]}$ diagrams, since some of the models of giants with circumstellar dust are characterized by very negative $Q$ values. However, these background sources hardly contaminate our selection since they are found at very high extinctions, as shown in the diagram, not accessible to our optical data. The contamination by background giants is, then, a less important issue here than for the $Q_{JHHK}$ and $Q_{riHK}$ indices. To define the limit for photospheric colors, we selected all the optical+IRAC sources lying around the origin in the IRAC $[3.6]-[4.5]$ vs. $[5.8]-[8.0]$ color-color diagrams. This population is known to be dominated by stars with photospheric colors at moderate extinction \citep{Allen2004}. We then set as limit for the photospheric colors the lower limit of the $Q$ indices obtained from these stars.\par
  The main advantage of using the $Q_{riJ[sp]}$ indices is that the selection is not affected by the decreasing sensitivity in the IRAC bands at increasing wavelength as using the IRAC color-color diagrams. With the $Q_{riJ[sp]}$ indices this problem is not an important issue being the selection of stars with excesses in [3.6], for instance, independent from that in [8.0]. The main drawback of this method is the uncertainty on the color excesses ratio at large extinction. In principle, this problem may result in the selection of several background contaminants. This problem will be tackled by our selection of contaminants and a detailed analysis of the SED of the stars selected only with these indices. A total of 800 sources have been selected with this method.

 $ $\par
{\bf MIPS [24] excesses}. Photospheric emission at $24\, \mu m$ is generally very small, so even weak disk emission in this band may result in large excesses. For this reason, criteria based on the [24] MIPS band are very powerful for the selection of stars with disks. It is also possible to select very embedded objects that can not be observed at the shorter infrared wavelengths. The main caveat in this approach is that the distribution of the MIPS sources with good photometry is not uniform in our field, with a larger sensitivity in the central-west area where the dust emission is fainter. \par
  We used four criteria based on the MIPS [24] band, as shown in Fig. \ref{mipsec_im}. In the $[3.6]$ vs. $[3.6]-[24]$ diagram 401 sources fall within the disk locus, defined with the conditions $[3.6]< 14^m$ and $[3.6]-[24]>2^m$ \citep{Rebull2011}. The lower limit of this locus was defined by those authors by studying the distribution in this diagram of the sources in the 6.1 square degrees SWIRE survey \citep{Lonsdale2003}. In the other three diagrams used to select stars with excess in [24], the reddening vector is almost vertical, resulting powerful method to select stars with disks typically falling in the region of the diagrams to the right of the normal-colors stars. In the $[24]$ vs. $[8.0]-[24]$ diagram 234 sources fall in the disks region defined with $[8.0]-[24]>1^m$ (except for the very faint sources). The disks region in the $[24]$ vs. $[4.5]-[8.0]$ contains 249 candidate sources with disks. In the $[4.5]-[5.8]$ vs. $[5.8]-[24]$ diagram we adopted a slightly more restrictive selection criterion ($[5.8]-[24]>2^m$) than that adopted in \citet{Gutermuth2008} ($[5.8]-[24]>1.5^m$) in order to reduce the chances of contamination from normal-colors stars, selecting 374 sources.

\subsection{Robustness of the adopted criteria}
\label{robu_sect}
  
    All the criteria described in the previous subsections use infrared and optical+infrared diagrams where regions typically occupied by disk-bearing stars are defined. In some cases, the limits of these regions have been defined by other authors, and applied to clusters which have stellar population and extinction different than Cyg~OB2; in some other cases they have been defined trying to exclude normal color stars as judged against models of Main-sequence and giant stars, such as in the $Q_{JHHK}$ diagram. Some of these choices are partly arbitrary, so a better understanding of the effects induced in the selection by slightly changing these limits is required. We verify the criteria by repeating the selection adopting less strict limits, and verifying the nature of the new sources that are selected. We do not need to repeat the process applying more strict limits, since the level of contamination of the criteria we defined will be analyzed later in detail. 
Table \ref{relax_tb} shows the number of new selections adopting the ``relaxed'' limits in each criterion (second column), together with the number of the stars which are not selected by some of the other criteria defined in Sect. \ref{diskselection} (``unique new selections''). In all cases, with the exception of the $J-H$ vs. $H-K$ diagram, most of the new selections are candidate stars with disks already selected  with one or more of the adopted criteria. A significant fraction, however, are definitively new selections. By studying their nature in the various color-color and color-magnitude diagrams, we have indication that these stars are mainly highly absorbed sources with intrinsic normal colors. Most of the unique new selections with the $J-H$ vs. $H-K$ diagram and $Q_{JHHK}$ index likely are very low-mass objects. Most of them have $J-K<2^m$ which is hard to achieve for stars which are supposed to have excess in the $K$ band at the distance and extinction of Cyg~OB2. In conclusion, adopting several criteria for the selection of stars with disks reduces the impact on the final selection of the uncertainty due to the adopted disk regions in the used diagrams. However, the relaxed conditions are less reliable than those we used, and they result in selections which are strongly contaminated. \par    

\section{Contaminants}
\label{contselection}

Merging all the outputs of the criteria defined in the previous section, we obtained a list of 2703 candidate stars with disks. The next steps consists in removing the contaminants and pruning the list of stars with disks. This process has been done following a five-steps selection procedure, shown in the block diagram in Fig. \ref{algo_im}. Each star has to pass all 5 checks in order to be accepted in our final list of stars with disks. These steps are explained in the following sections.

\subsection{Stars with unreliable excesses}
\label{firststep}

In the first step we selected and discarded the sources whose infrared excesses are not reliable, and whose selection may have been induced by inaccurate photometry. \par
  In sect. \ref{bwe} we will define a class of stars with disks showing optical colors bluer than the other stars associated with Cyg~OB2. As we will discuss, these colors can be explained by the presence of the disk, and the $Q$ indices are very sensitive to them. However, the presence of an IR excess is demanded since the presence of a circumstellar disk must produce one at NIR or longer wavelengths. Contaminated optical photometry (i.e. due to blends, or proximity to either CCD edges or bright stars) may result in selections via $Q_{riJ[sp]}$ indices that are erroneous. In these cases, no excesses are expected to be found with purely NIR methods. For this reason, we removed from the list all the $Q-excess$ sources having blue optical colors and that are not selected by other NIR methods. \par
We performed a further test for the stars selected only with the $Q$ indices or showing excesses only in $K$ band {\bf in order to exclude stars with SEDs that can be matched to reddened photospheres}. The fit with photospheric models has been done using the SED fitting tool introduced by \citet{Robitaille2007}. With this tool it is possible to compare the observed SEDs with those of YSO models covering an extensive parameter space: 20000 models, each at 10 different inclination angles, for a total of 200000 distinct SEDs of YSOs at different evolutionary phases. This tool also allows to fit the observed SED with photospheric models from \citet{Kurucz1993} and \citet{Brott2005} for a large range of distance and extinction, which is especially useful for this test. Using this tool, we selected and discarded those sources whose SED is compatible with any reddened photospheric model with distances between $10\,pc$ and $10\,kpc$ and extinction ranging from $A_V=0.1^m$ to $A_V=1000^m$. This test was necessary in order to verify the excesses in $K$ band of those faint UKIDSS sources which lack other counterparts. Background giants may also have been selected by the $Q$ indices because the used color ratio $E_{A-B}/E_{C-D}$ is not appropriate for background highly extinguished objects. {\bf The color ratio we adopted is surely reliable for stars at low reddening, for which the $Q$ indices can be safely assumed to be reddening free. Taking into account that the reddening law changes at high extinction and distances, as well as that the reddening vector deviates from linearity, the color ratio may not be appropriate and the $Q$ indices are not reddening free anymore. In these cases, they may give erroneous selections of background stars at high extinction as disk-bearing objects.} \par

\subsection{Stars selected with one criterion}
\label{secondstep}

The second step selects good candidate stars with disks among those showing infrared excess only in one band ($\lambda_{ex}$). Among these sources, we retained those {\bf faint sources not detected at $\lambda > \lambda_{ex}$ for the decrease of sensitivity at longer wavelength in our data}. We also retained the stars with excess in one band but having colors typical of YSOs in two diagrams which have not been used for the selection: the $[3.6]-[4.5]$ vs. $[5.8]-[8.0]$ diagram (see Sect. \ref{thirdstep}), and the $r-H_{\alpha}$ vs. $r-i$ diagram (see Sect. \ref{ha}), the latter being compatible with accreting sources. As further test, among the remaining sources we removed those whose SED is compatible with a reddened photosphere as verified with the \citet{Robitaille2007} SED fitting tool. Finally, we also removed the sources showing excesses only in $Q_{riJ[5.8]}$, which very likely have the [5.8] magnitude contaminated by nebular PAH emission along the line of sight. 

\subsection{Candidate giants}
\label{thirdstep}

Within the third step we selected and discarded candidate background giants that, especially when they are surrounded by circumstellar dust, may occupy the same region of disk-bearing objects in several diagrams. To take into account this source of contamination, we used the PADOVA isochrones \citep{Girardi2002} to obtain the expected optical and infrared colors and magnitudes of giants, both with and without circumstellar dust, with ages ranging from $0.5\,Gyrs$ to $13\,Gyrs$ and masses ranging from $0.15\,M_{\odot}$ to $2.9\,M_{\odot}$. PADOVA isochrones are suitable for our scope since they are constantly updated in order to provide data in several photometric systems and they use models for the emission from evolved stars with circumstellar dust in the Spitzer bands \citep{Groenewegen2006}. We defined the typical loci of giants in 6 diagrams:

\begin{itemize}
\item in the $[3.6]-[4.5]$ vs. $[5.8]-[8.0]$ diagram (shown in the left panel of Fig. \ref{giant_im}) the giant locus overlaps both the region typically populated by normal photospheres and reddened class~II/I objects;
\item in the $[3.6]-[5.8]$ vs. $[4.5]-[8.0]$ diagram (Fig. \ref{1324_im}) it contains only the blue end of the sample of candidate stars with disks;
\item in the $[4.5]$ vs $[4.5]-[8.0]$ diagram (Fig. \ref{224_im}) it mostly corresponds to the region populated by normal photospheres, with $[4.5]-[8.0]<0.8^m$, overlapping the disk region only in a bright section scarcely populated by candidate stars with disks;
\item also in the $[8.0]$ vs $[4.5]-[8.0]$ diagram (shown in the right panel of Fig. \ref{giant_im}) it corresponds to the normal colors locus;
\item in the $[24]$ vs. $[8.0]-[24]$ diagram (Fig. \ref{mipsec_im}), the models of giants with circumstellar dust are red and bright, so that the giant locus ($[8.0]-[24]<2^m$) significantly overlaps with the disks locus.
\item in the $[4.5]-[5.8]$ vs. $[5.8]-[24]$ diagram (Fig. \ref{mipsec_im}), models of giants with dust can be very red, so that the giant locus roughly corresponds to the entire diagram, with the exception of the box delimited by $[5.8]-[24] > 3^m$ and $[4.5]-[5.8]<1^m$
\end{itemize}

Candidate background contaminants are first selected as those sources falling in the region defined as ``giant locus'' in each of these diagrams where they can be plotted. Adopting this criterion, a total of 235 candidate giant contaminants were identified. Later, we estimated their extinction from the optical and $JHK$ color-color diagrams, and we retained in the disks list those candidate background giants showing low to intermediate extinction in the optical and infrared color-color diagrams.

\subsection{Extragalactic contaminants}
\label{fourthstep}

With the fourth step we selected and discarded candidate background galaxies. These sources have been selected with four different criteria:

\begin{itemize}
\item \citet{Donley2012} updated the AGN classification made by \citet{Stern2005} based on the $[3.6]-[4.5]$ vs. $[5.8]-[8.0]$ diagram (Fig. \ref{giant_im}). In this diagram the AGNs overlap both with giants and YSOs, but only faint sources ($[3.6]>14.5^m$) are classified as possible AGN.
\item Another criterion for the selection of AGN is provided by GMM09, using the $[4.5]$ vs $[4.5]-[8.0]$ diagram and a combination of six different conditions involving these two IRAC bands. \par
\item GMM09 selected PAH dominated galaxies from the $[4.5]-[5.8]$ vs. $[5.8]-[8.0]$ diagram and $[3.6]-[5.8]$ vs. $[4.5]-[8.0]$ (shown in Fig. \ref{1324_im}). In the latter the PAH galaxies partially overlap the distributions of candidate stars with disks;
\item \citet{Harvey2007} defined the typical locus of background galaxies in the $[24]$ vs. $[8.0]-[24]$, and $[24]$ vs. $[4.5]-[8.0]$ diagrams (Figures \ref{224_im} and \ref{mipsec_im}) by using data from the SWIRE survey.
\end{itemize}

In order to classify a source as a candidate background galaxy we require that it falls in the galaxies loci in each of these diagram where it can be plotted. We selected in this way a total of 37 candidate AGN, 6 PAH-galaxies, and 7 background galaxies. We then removed all those selected sources without further information from optical or UKIDSS/2MASS data. The nature of the remaining sources having optical or $JHK$ counterpart can be assessed using these data. We discarded those showing optical colors typical of quasars \citep{Richards2004}, JHK colors typical of AGN (derived from the \citealp{PethRS2011} catalog) or having an extended shape in the optical and NIR images. 

\subsection{Foreground contaminants}
\label{fifthstep}

Foreground optical sources may contaminate our selection as a  consequence of a mismatch with the infrared catalog. This contamination mainly affects the selection made with the $Q$ indices involving optical and infrared colors. The most likely mismatch, in fact, occurs between a foreground optical source and a background infrared source. The combined SED of such a mismatch can mimic an infrared excess detectable with the $Q$ indices \citep{GuarcelloDMP2010}. The best way to select these sources is by using the $r-i$ vs. $i-z$ diagram \citep{Guarcello2012B} to compare their extinction with that of the members of the association. Fig. \ref{riiz_im} shows the $r-i$ vs. $i-z$ diagram with all the sources with good photometry and all the candidate stars with infrared excesses. The two solid lines are the $3.5\,Myrs$ isochrones drawn with the 95\% and 5\% quantiles of the visual extinction distribution of the candidate members optically and X-ray selected in \citet{GuarcelloDMP2010}. A small group of candidate disk-bearing stars lie leftward of the isochrone drawn using the 5\% quantile extinction, showing extinction smaller than the cluster range. If these stars are selected only with the $Q$ indices, then they are good candidate mismatches between the optical and infrared catalogs. \par

\section{Final list of stars with disks}
\label{merg_list}

All the sources accepted by each of the 5 steps in the block diagram shown in Fig. \ref{algo_im} form the final list of candidate stars with disks in Cyg~OB2, which contains 1843 stars. A total of 337 of these sources fall in the central area observed with Chandra/ACIS-I ($99.7\,ksec$, P.I. Flaccomio, centered at $\alpha=20:33:11.0$ and $\delta= +41:15:10.0$), where small populations of 23 and 63 disk-bearing stars have been previously reported by \citet{AlbaceteColombo2007} and \citet{Wright2010}, respectively. The large improvement brought by our study in the knowledge of the disk population in the central area of Cyg~OB2 is evident, and it is due to our extensive use of optical and infrared data, while in these previous studies stars with disks have been selected only among the X-ray sources using 2MASS and UKIDSS $J-H$ vs. $H-K$ diagram. \par
  Table \ref{criteria_tb} shows the initial number of sources selected with each of the adopted criteria, and the number of those that are in the final list of stars with disks. The most efficient method we adopted is the GMM09 scheme, with a total of 1461 stars selected, only 56 discarded and 478 detected only with this criterion. This is not surprising, given that this method is the only one which combines several IRAC colors. The sources with excesses in [24] typically show excesses also in the IRAC bands, at least at [8.0]. Only 16 sources show excesses only in [24]. The diagnostics that selected the largest fraction of discarded stars are those based on the $K$ band. This is mainly due to the depth of the UKIDSS catalog. Most of the discarded sources are, in fact, very low mass stars detected only with UKIDSS, for which it was not possible to verify the reliability of the selection, or candidate background giants observed with very high extinction, wrongly selected by the $Q$ indices. In the final list, there are 59 sources showing excesses only in $K$. Fig \ref{dk_spadis_fig} compares the spatial distributions of the stars with disks selected with the GMM09 and those selected with other criteria. With few exceptions (such as the tip of DR18), the two samples of stars show the same spatial distribution, indicating that the inhomogeneity of the optical and MIPS data will not affect any inference on the morphology of the association. 

Fig. \ref{vis_crit} compares the sample of stars selected by each criterion. Each circle represents a criterion, and all the stars present in the final list of disk-bearing objects lie along the circumferences, always in the same order.  In each circle, a line from the center to the position corresponding to a given source is drawn if this star is selected by the particular criterion. All the stars not selected by the GMM09 scheme are collected in the upper right quarter.  This quarter is filled by selections made with the criteria based on MIPS and the $Q_{riJ[sp]}$ indices. In fact, stars having a strong excess in [24] combined with weak IRAC excesses are, then, successfully selected with the $Q_{riJ[sp]}$ indices, confirming their sensitivity to  small excesses \citep{GuarcelloDMP2010}. \par
  The number of disks identified with the $Q_{riJ[sp]}$ indices increases at longer increasing wavelength of the $[sp]$ band, and there are not many ``unique'' selections among the first three $Q_{riJ[sp]}$ indices. We then conclude that these three indices are not essential to obtain a reliable selection of disk bearing stars.  Analogously, the selections made with the $J-H$ vs. $H-K$ diagram and the $Q_{JHHK}$ indices are similar {\bf (with 21 stars selected with the $Q_{JHHK}$ index but not in the $J-H$ vs. $H-K$ diagram and 25 stars in the opposite way)}, and disks selected with $[4.5]-[5.8]$ vs. $[5.8]-[24]$ are almost completely contained in the set obtained using the other MIPS criteria. Fig. \ref{vis_crit} is a powerful diagram, then, to study how different selection criteria compare with each other and what contributions of the different criteria are to the to the final list of selected stars. \par
The catalog containing all the photometric data of the disk-bearing objects in Cyg~OB2, their classification and the selection criteria is explained in Appendix \ref{cata_sec}.

\subsection{Cluster locus in the optical diagram and BWE stars}
\label{bwe}

Fig. \ref{rri_im} shows the $r$ vs. $r-i$ diagram of all the sources with good optical photometry falling in the studied field. The stars have been plotted using the OSIRIS photometry; the SDSS photometry has been used for those stars with bad or saturated OSIRIS photometry, while the IPHAS photometry, properly converted into the SDSS photometric system \citep{Guarcello2012B} has been used for those source with bad OSIRIS and SDSS data. The distance of $1.4\,kpc$ \citep{Rygl2012} and the extinction estimated in \citet{Guarcello2012B} for the cluster members and the most distant foreground stars have been used to draw the isochrones and ZAMS from \citet{Siess2000}. The extinction vector has been calculated from the reddening law defined in \citet{Donnell94}. The large dots in this diagram mark the stars with disk with good optical photometry. They share the same region of the diagram with the optical sources with X-ray detection (Fig. 10 in \citealt{Guarcello2012B}), limited in the blue end by the $5\,Myrs$ isochrone. The individual extinction of the disk-bearing sources can be derived from their position in the $r-i$ vs. $i-z$ diagram (Fig. \ref{riiz_im}) with the same procedure adopted for the optical+X-ray sources in \citet{Guarcello2012B}. The extinction map derived from the disk population (not shown here) shares similar properties with that derived using the X-ray selected candidate cluster members, indicating an increase of the optical extinction in the west and south-west direction. The median extinction of the candidate stars with disks is $A_V=4.82^m$, only $0.5^m$ larger than that of the X-ray population. This does not mean that the presence of a disk does not have a strong impact on the extinction of the central star, since this result is biased toward low and intermediate extinctions for the use of the sources with good optical photometry, which are usually not heavily embedded in circumstellar material. \par
	Fig. \ref{rri_im} also shows a sample of 39 candidate stars with disks with optical colors bluer than the cluster locus, these being more compatible with the foreground population than with the association. It must be pointed out that these stars are not AGN, or foreground contaminants or mismatches, which have been removed during the review phase. Similar stars (called Blue stars With Excesses, or $BWE$, stars) have been observed and studied in NGC~6611 by \citet{GuarcelloDMP2010} and Bonito et al. (in preparation), and observed in other clusters (i.e. in the Orion Nebula Cluster, \citealt{Hillenbrand1997}). Blue optical colors in stars with disks can be produced by intense accretion from the disk to the central star \citep{HartmannKenyon1990} and scattering of stellar optical light by small dust in the circumstellar material \citep{GuarcelloDMP2010}. \citet{DeMarchiPoster2012} also suggests that the BWE stars in NGC~6611 can be part of a disk-bearing population older than $10\,Myrs$. To support this hypothesis, stars with disks older than $10\,Myrs$ have been observed and well studied \citep{Palla2005,Argiroffi2007}. \par
The nature of these stars can be studied by comparing their observed SEDs with those of the YSO models presented in \citet{Robitaille2007}. In the fit, we fixed the distance equal to $1.4\,kpc$ and selected for each star the models for which the reduced chi-square satisfies the condition $\chi^2-\chi_{best}^2\leq 3$, where $\chi_{best}^2$ is the reduced chi-square of the best-fit model ( as suggested by \citealp{Robitaille2007}). Among these BWE stars, 31 are compatible with SED models, both of ``phaseI'' and ``phaseII'' YSOs. Following the definition in \citet{Robitaille2007}, the former models still have a dense protostellar envelope (i.e. $\dot{M}_{envelope}>10^{-6}\,M_{\odot}\, yr^{-1}$), which may scatter a significant amount of stellar light into the line of sight \citep{Damiani2006}. For those fitting ``phaseII'' models (i.e. $M_{disk}>10^{-6}\,M_{\odot}$), the scattering of optical radiation by an almost edge on disk is the main explanation for their optical colors, as found in \citet{GuarcelloDMP2010}. This interpretation is valid for 21 of these sources whose disk is observed with a large inclination angle (the SED models have only ten possible values for the inclination angle, and the SEDs compatible with the observed SED of these 21 stars have inclination angles larger than $81^{\circ}$). Only in 6 cases the SED is compatible with YSO models with the right combination of high accretion rate ($\dot{M}=10^{-7} - 10^{-8}\,M_{\odot}\,yr^{-1}$) and low stellar mass. 

\subsection{Sources with $H\alpha$ emission}
\label{ha}

\citet{Vink2008} identified about 50 emission line stars in Cyg~OB2 and its periphery. Among those falling in the field that we are studying (18 stars), 17 are present in our catalog of disk bearing sources, which are then candidate to bear an actively accreting disk. We can select other stars with both a disk and emission in $H\alpha$, which is a signature of intense gas accretion, using the IPHAS  $r^{\prime}-H_{\alpha}$ vs. $r^{\prime}-i^{\prime}$ diagram, shown in Fig. \ref{rhari_im}. In this diagram the small dots mark all the IPHAS sources in our catalog with good photometry. The solid lines are the {\bf colors of normal stars derived by \citet{Drew2005} using the solar metallicity stellar SEDs of \citet{Pickles1998}}, drawn with $E_{B-V}=1^m$, $E_{B-V}=2^m$, and $E_{B-V}=3^m$, corresponding to $A_V=3.23^m$, $A_V=6.45^m$, and $A_V=9.68^m$ \citep{MunariCarraro1996} which is roughly the extinction range of the optically+X-ray selected cluster members \citep{Guarcello2012B}. \par
	We selected the disk-bearing objects with $H\alpha$ emission as those whose $r^{\prime}-H_{\alpha}$ is larger by more than $3\sigma_{r^{\prime}-H_{\alpha}}$ than the value expected from the $E_{B-V}=1^m$ ZAMS. With this approach, we found 52 sources with emission in $H\alpha$. Among these sources, 13 have been previously classified as ``emission-line stars" by \citet{Vink2008}. {\bf We do not expect large contamination of our sample by chromospheric active stars, given that the H$_{\alpha}$ luminosity of these stars is expected to be $\sim 2$ order of magnitudes fainter than in actively accreting stars \citep{DeMarchi2010}. Besides, weak lines T-Tauri stars are defined as having H$_{\alpha}$ equivalent width} $<10\,A$, {\bf which in the IPHAS color-color space corresponds to about a 0.1 increase of the $r^{\prime}-H_{\alpha}$ color \citep{Drew2005}}. Note that 7 out of 39 BWE stars have intense emission in $H\alpha$, {\bf suggesting the blue optical colors observed in these stars can be a consequence of intense accretion from the disk (see Sect. \ref{bwe}).} \par 
	
\section{The evolutionary status of the stars with disks}
\label{evol}

\subsection{The alpha index}
\label{alph}

The evolutionary status of disk-bearing objects can be studied using the infrared spectral index $\alpha=d\,log\left(\lambda \,F_{\lambda} \right) / d\,log \left(\lambda \right)$. Following the classification scheme of \citet{Wilking2001}, a class~I YSO still embedded in an accreting envelope is characterized by a rising infrared SED with $\alpha > 0.3$. During YSO evolution, the circumstellar envelope is dissipated by the accretion onto the forming circumstellar disk and the radiation from the central star. In the intermediate phase, it may be possible to observe the light from the central star and the disk together with the envelope emission, resulting in a flat or double peaked SED. These ``flat-spectrum'' sources have $-0.3\leq \alpha \leq 0.3$. Once the envelope is completely dissipated and the infrared SED is dominated by the emission of the optically thick circumstellar disk (class~II YSO), the spectral index ranges between -0.3 and -1.6. Alternative classification schemes based on the spectral index have been proposed (i.e. \citealt{Lada1987}), but applied to our sample of stars with disks, they give unreliable selection of optically bright embedded objects. \par
  It has been shown, however, that flat-spectrum sources usually do not consist of a well-defined evolutionary phase in YSO evolution, and normal class~II sources (mainly at high extinction) can be confused as ``flat-spectrum'' sources using $\alpha$ index defined fluxes between $2\,\mu m$ and $24\, \mu m$. Besides, we have to bear in mind that we have not calculated the $\alpha$ indices using reddening-corrected SEDs. We will not consider these ``flat-spectrum'' sources as a distinct evolutionary phase. \par
We calculated the alpha index of disk-bearing objects after converting the $K$, IRAC and [24] magnitudes into flux units and applying a linear fit to the infrared SED. It was possible to calculate the spectral index for 1830 stars. Fig. \ref{alpha_im} shows the distributions of the spectral indices of stars with disks, together with the limits between different evolutionary stages as defined in \citet{Wilking2001}. As expected, the large majority of stars (72.9\%) are classified as class~II objects, but a significant population of embedded objects are found (188 class~I sources). \par
  Following this classification, 92 stars have a spectral index typical of class~III objects. We verified their nature using the \citet{Robitaille2007} SED fitting tool in two steps. In the first step, we verified whether they are normal stars lying in the direction of Cyg~OB2, wrongly selected as stars with excesses. This test has been performed fitting their observed SEDs with the photospheric models provided by the SED fitting tool with distances ranging from $10\,pc$ to $10\,kpc$ the extinctions from $A_V=0.1^m$ to $A_V=1000^m$. In almost all cases, it was not possible to find a photospheric model fitting the observed SEDs of these stars in all the photometric bands, showing excesses in the longest infrared bands. Only 8 stars could be fitted properly by a photospheric model, and they have been discarded from the list of candidate stars with disks. \par
	In order to shed some light on the nature of the remaining 84 stars, we fitted the YSO models from \citet{Robitaille2007} to their observed SEDs. In the 52 cases with a good fit, the presence of small excesses are justified (following the results of the SED fitting) by the presence of very low-mass disks, i.e. $M_{disk}<10^{-5}\,M_{\odot}$, lower than what is usually observed in class~II objects ($10^{-2}\, ,10^{-3}\,M_{\odot}$), but larger than the lower limit of disk masses for the ``phaseII'' YSOs in \citet{Robitaille2007}; and/or disks with large inner hole, i.e. $R_{inner}>5\,AU$. It is not surprising that disks with such properties produce only small excesses in the NIR bands, resulting in low spectral indices. In the rest of the paper, we will consider these stars as normal disk-bearing objects. Fig. \ref{SED_im} shows the SED of one of these stars with a $\alpha$ index typical of class~III objects (-1.80), but showing infrared excesses at [8.0] and being classified as a transition disk (see Sect. \ref{trans}). Following the SED fit, this star has a low-mass disk ($M_{disk}=1.54\times10^{-5}\,M_{\odot}$) and large inner radius ($R_{in}=33\,AU$), compatible with the lack of excesses for $\lambda<8.0\,\mu m$. The lack of detection at [24] is justified by the intense background around the position of this star. The upper limit is the faintest magnitude of the only two MIPS sources within one arcminute from the position of the source.

\subsection{The embedded population}
\label{embed}

{\bf Class~I objects have been selected by the analysis of their SEDs and two photometric criteria}. One is the method defined by \citet{Wilking2001} ($\alpha > 0.3$), resulting in 188 candidate class~I objects. The second method is based on the $[4.5]-[5.8]$ and $[3.6]-[4.5]$ colors, as defined by GMM09, resulting in 146 candidate class~I YSOs. The intersection between these two samples consists of 95 sources, while 93 stars are selected only by the former method and 48 only by the latter.\par
	{\bf By looking at their SEDs and position in the various IRAC diagrams}, the sample of stars selected by both methods form a reliable sample of class~I objects. To properly classify the stars selected only with one photometric method, we used the \citet{Robitaille2007} SED fitting tool, i.e. we classify as class~I objects those stars for which more than 50\% of the models fitting the observed SEDs is classified as a ``phaseI'' YSO, following \citet{Robitaille2007} classification. With this approach, we classified as class~I objects 60 sources among the 93+48 selected only with one method, resulting in a total of 155 class~I stars. Among these stars, 15 have good optical counterparts, which is unusual for class~I YSOs. Their excesses are genuine, as verified during the review phase. \par
Almost all the 48 candidate class~I sources selected with the GMM09 method but not with the \citet{Wilking2001} criterion have an $\alpha$ index typical of flat spectrum sources. Some of them have been classified as embedded objects as explained. Hence, the remaining flat spectrum sources will be considered as normal disk-bearing stars, likely suffering extinctions larger than the stars selected as class~II sources with the $\alpha$ index. \par
  The $24\mu m$ image of Cyg~OB2 from the {\emph Spitzer Cygnus~OB2 Legacy Survey} (not shown here) was visually inspected to identify very embedded objects not classified as stars with disks because of poor photometry in the other bands. We identified 24 sources detected in both [24] and at least one other IRAC band (mainly $5.8\mu m$ and $8.0\mu m$). Their colors are typical of very red members of Cyg~OB2 with disk and they lie in regions with dense nebulosity. These sources have been retained and classified as ``highly embedded'' (Appendix \ref{cata_sec}) in the final list of disk-bearing members of Cyg~OB2.  \par
Our deep photometry allows us to detect the central star inside some of the proplyd-like objects in Cyg~OB2 discovered by \citet{Wright2012}. Given their dimensions (two or three orders of magnitude larger than the proplyds observed in Orion) they have been classified as ``proplyd-like'' objects which are evaporating under the action of the intense UV flux emitted by the massive stars of Cyg~OB2. These objects will be the subject of forthcoming publications, presenting detailed SED analysis and spectroscopic studies (Robinson et al. and Guarcello et al., both in preparation). \par
  Also a proplyd-like object not classified by \citet{Wright2012} has a faint YSO in its center, as shown in Fig. \ref{prop_im}, with excesses in [5.8] and [8.0] and whose SED fits a class~I model younger than $3\times 10^4\,yrs$. The evaporating proplyd is $21^{\prime \prime}$ long, corresponding to $\sim3\times 10^4\,AU$ at the distance of Cyg~OB2. This object lies at the projected distance of $2.3\,pc$ and $7.8\,pc$ from the closest B and O stars, respectively, which may be responsible for the ionization and evaporation of the globule. \par

\subsection{Stars with transition disks}
\label{trans}

Transition disks are considered to be an intermediate evolutionary status between the class~II phase, with intense NIR excesses due to the emission from an optically thick inner disk, and class~III phase, with no NIR excesses as a consequence of the dissipation of the inner disk and/or the depletion of small dust grains in the disk. The general definition of a transition disk is a YSO with photospheric NIR and optically thick MIR emission, due to a disk with a large inner hole and thick outer region \citep{Muzerolle2010}. It is still not clear whether this evolutionary phase represents a necessary step in disk evolution or is a consequence of giant planets formation occurring in the inner disk (i.e. TW Hya, \citealp{Calvet2002}). In the former case, the main mechanism for the creation of the inner hole is claimed to be the photoevaporation driven by the central star \citep{Alexander2006}. Photoevaporation can drive an intense wind from the circumstellar disks, and as soon as the mass-loss rate is higher than the accretion rate, the outer disk is not able to replenish the evaporating material. The first consequence of this is the creation of an intermediate gap between the inner and the outer disk, followed by a rapid dissipation of the inner disk and the creation of an ``inner hole'' phase. Disks with intermediate gap between the optically thick inner and outer disk have been named ``Pre-Transitional Disks'' \citep{Espaillat2007}. If the creation of the inner hole is a consequence of planet formation, it is not surprising that disks which are not forming planets may evolve differently. In this case the disk is expected to evolve from a primordial massive disk, with a mass of some $0.01\,M_{\odot}$ and a large population of small (a few $\mu m$ size) dust grains producing intense NIR excesses, to a low-mass disk, depleted of small dust grains and producing low infrared excesses even if the inner disk is still present. \par
	Both these disk configurations (disks with a large inner hole and with a low-mass inner disk depleted of small grains) produce small infrared excesses, if any. As a consequence, purely photometric selection criteria based on IRAC colors may lead to an incorrect classification of evolved disks still having an inner region as transition disks. The analysis of the SEDs is the most accurate method to select properly a population of transition disks \citep{Merin2010}. We therefore produced a reliable list of candidate transition and pre-transition disks in Cyg~OB2 with the following method. We first selected a sample of stars having the photometric properties expected from transition, pre-transition and low-mass disks: i.e. those showing excesses only at [8.0] and [24] bands, or whose alpha index is compatible with class~III objects following the \citet{Wilking2001} classification, with good photometry in [8.0] and [24], or stars with disks with $1^m<[8.0]-[24]<2^m$ (Fig. \ref{mipsec_im}). The final classification is obtained by fitting the observed SEDs with the YSOs models of \citet{Robitaille2007}.

\begin{itemize}
\item A transition disk must show a declining NIR SED with IRAC emission at or very close to the photospheric values, followed by an intense MIR emission from the outer disk, and its SED has to fit models with large inner holes ($R_{inner}\geq5\,AU$). A total of 62 stars meet these criteria and are classified as transition disks. Their inner radii lie in the range $5\,AU<R_{inner}<91.2\,AU$, with a median value of $17.4\,AU$, and with masses are $M_{disk}<0.0941\,M_{\odot}$ (median value $1.2\times 10^{-4}\,M_{\odot}$).
\item A low-mass disk must show a declining NIR SED with very low excesses which extend to the MIR bands, and its SED has to fit models with $R_{inner} < 5\,AU$, $M_{disk} < 10^{-4}\,M_{\odot}$. The stars selected as YSOs with low-mass disks number 23. Their masses are on average two orders of magnitude smaller than those predicted for transition disks (median value $M_{disk}=8.59\times 10^{-6}\,M_{\odot}$) and they have $R_{inner}<2.3\,AU$. 
\item a pre-transition disk still has an optically thick inner disk, producing significant NIR emission, which is followed by a MIR emission dip; its observed SED must be characterized by NIR excesses, with a [24] flux at the same or lower level than that in [8.0]. Since the \citet{Robitaille2007} models do not include disk models with an intermediate gap, these stars hardly fit the theoretical models. A total of 42 stars are candidates for having a pre-transition disk.
\item We also consider the possibility that low NIR excesses can be produced by a highly inclined disk that masks the emission from the inner disk. {\bf Stars fitting models with $R_{inner}<5\,AU$, $M_{disk}>10^{-4}\,M_{\odot}$, highly inclined disks, and with significant [24] emission are classified in this group. Sixteen stars, fitting models with these properties and inclination angle $\theta>75^{\circ}$, have been classified as having highly inclined disks}. Their SEDs fit models with small inner radii (median value $R_{inner}=0.3\,AU$) and significant mass (median value of $M_{disk}=10^{-4}\,M_{\odot}$, but with 6 cases more massive than $0.01\,M_{\odot}$). These characteristics are required in order to achieve a significant obscuration of the central star and distortion of the optical/NIR SED. 
\end{itemize}

A remaining group of 4 stars shows small excesses and declining SED in the NIR, but very intense MIR emission, well above the emission level at $8.0\mu m$. These stars fit a model where the intense emission at [24] is due to the presence of a dense circumstellar envelope, with accretion rates in the range $4.96\times 10^{-7}\,M_{\odot} yr^{-1}<\dot{M}_{envelope}<1.80\times 10^{-6}\,M_{\odot} yr^{-1}$. Fig. \ref{SED_td} shows five examples of SEDs of stars classified with this method. \par
	It is interesting to verify whether the stars in these five different groups also share different photometric properties. It is not surprising that the stars with high-inclination disks are in general optically very faint, with one exception whose optical SED is dominated by scattered light. Fig. \ref{tdlm_magr} compares the distributions of the magnitudes in the $r$ band of stars with transition and pre-transition disks, of low-mass disks and of the whole disk population. The $r$ magnitude distribution of the entire disk population peaks between $20^m<r<24^m$. Even restricting the sample to the stars with good photometry in [8.0] or [24] (793 stars), which is crucial for our classification, the peak is between $20^m<r<22^m$. Using the colors of a $3.5\,Myrs$ isochrone from \citet{Siess2000} at a distance of $1400\,pc$ \citep{Rygl2012} and $A_V=4.3^m$ \citep{Guarcello2012B}, this range corresponds to stars with $M<1\,M_{\odot}$. Roughly the same distribution is shared by the stars with low-mass disks, though none of them is brighter than $r=18^m$ ($M\sim1.8\,M_{\odot}$), while the distribution for the stars with transition and pre-transition disks peaks at $r<18^m$. This difference cannot be due only to the differential extinction affecting this region, so it should reflect a difference in stellar mass. This suggests that in $3-5\,Myrs$ (which is the general age range of Cyg~OB2 members, \citealp{Wright2010}), the stars more massive than $1.8\,M_{\odot}$ have more chance to create a transition or a pre-transition disk, likely inducing more intense photoevaporation than the less massive stars, whose disks have more chance to evolve and still keep their inner region. \par
The number of transition and pre-transition disks we classified in Cyg~OB2 is low compared to the entire disk population with detections in [8.0] and [24] and good photometry in one of these two bands. Using this sample, the fraction of transition disks is 7.8\%, and that of pre-transition disks 5.3\%. For a $3-5\,Myrs$ old cluster, these fractions are indicative of a lifetime for the transitional phase $\leq0.5\,Myrs$ \citep{Muzerolle2010,Luhman2010,Currie2011}. It must be noted, however, that our sample does not have a well defined age. In this estimate, we are mixing together information from the oldest regions in Cyg~OB2, with age presumably in the range $3-5\,Myrs$, and from the youngest regions with ongoing star formation. A better age restriction of our sample will only be possible after a more detailed study of the star-formation history in this region.\par

\section{Sub-clustering and morphology of the association}
\label{spadis}

\begin{figure*}[]
        \centering
        \includegraphics[width=12cm]{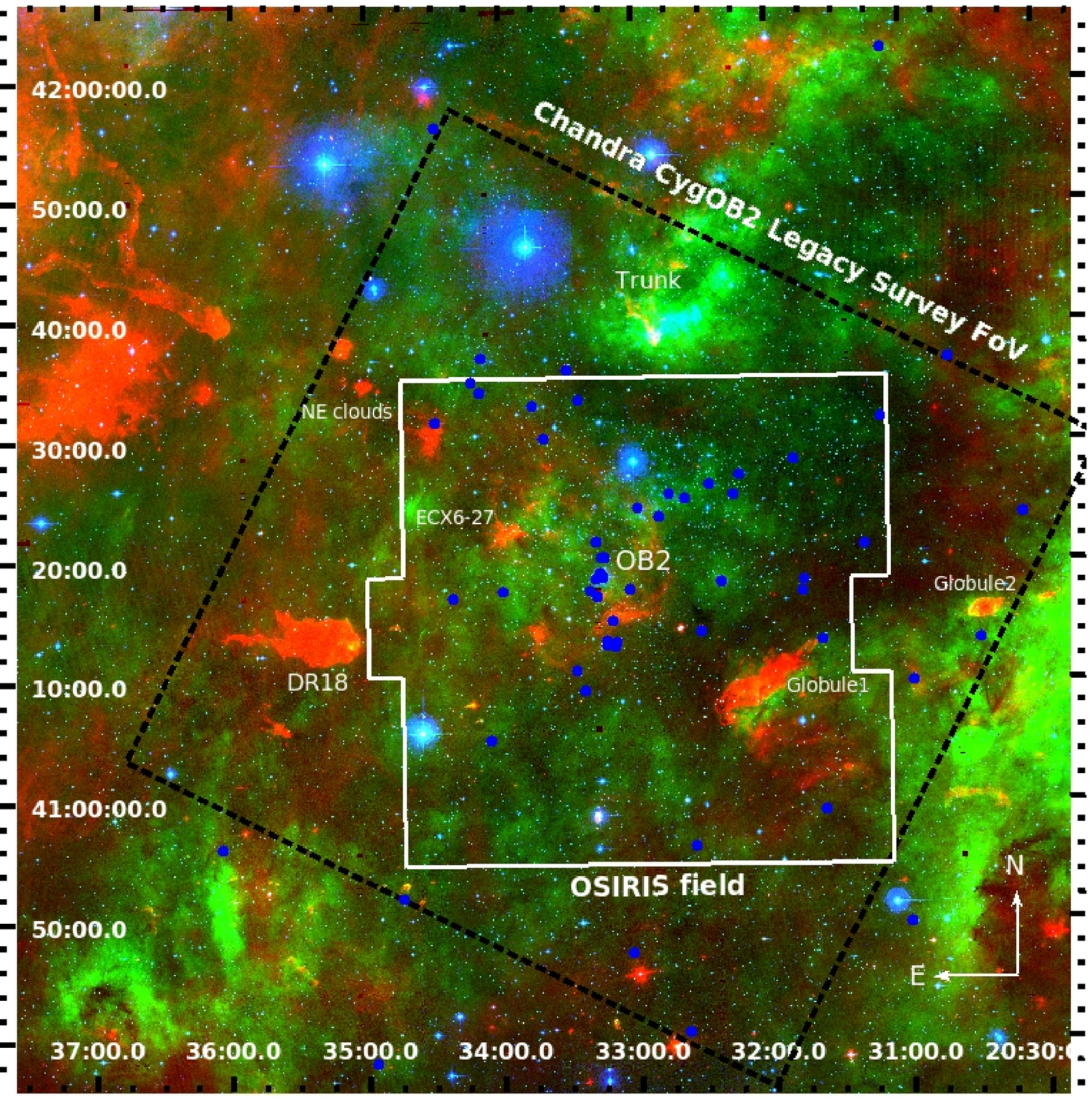}
        \includegraphics[width=6cm]{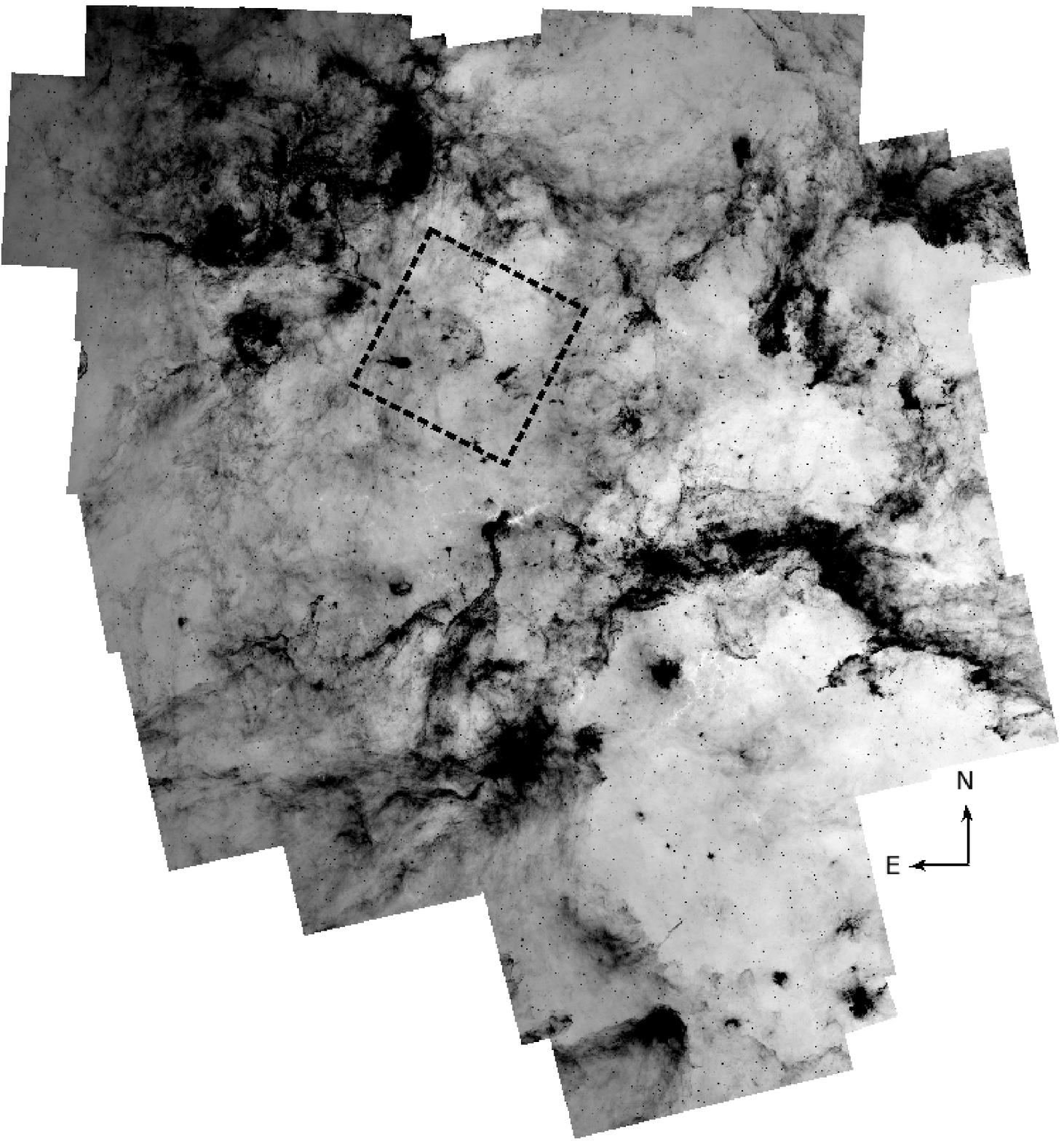}
	\caption{RGB image of the Cygnus~OB2 association. The emission in $H\alpha$ is shown in green, the emission in $8.0\mu m$ in red, and stellar emission in $r^{\prime}$ in blue. The field of view of the Chandra Cygnus~OB2 Legacy Survey, {\bf centered on $\alpha=20:33:14$ and $\delta=+41:18:54$} is encompassed by the dashed box, while the field observed with OSIRIS by the white box. {\bf Spitzer data are available in the entire field}. The positions of the O stars are marked with blue circles (Wright et al. 2013, in preparation). The positions of the relevant structures that are discussed in this paper are also indicated. {\bf The small panel in the right shows the entire field observed for the Cygnus-X Legacy Survey with the portion that it is studied in this paper encompassed by the dashed box.}}
        \label{rgb_im}
        \end{figure*}

\begin{figure}[]
        \centering
        \includegraphics[width=8.5cm]{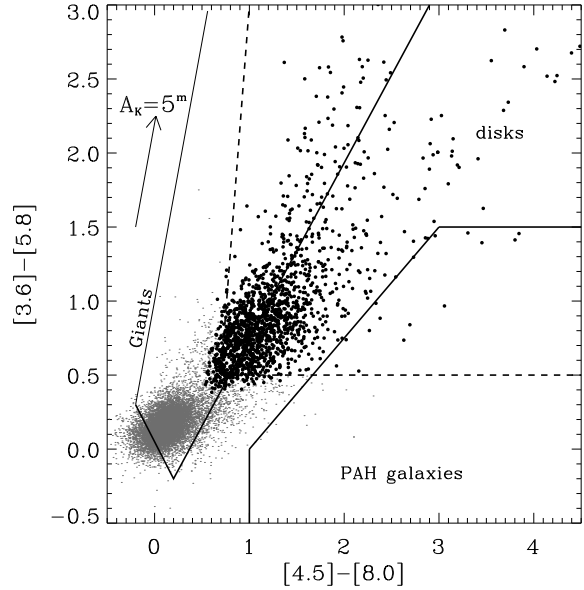}
        \caption{$[3.6]-[5.8]$ vs. $[4.5]-[8.0]$ diagram with the loci of disks and PAH galaxies from GMM09 and the giants locus defined with the PADOVA \citep{Girardi05} isochrones (dotted lines). See Sect. \ref{contselection} for detail on the definition of the giants and PAH galaxies loci. Small gray dots correspond to sources with good photometry, black dots are those selected as disk-bearing objects with the GMM09 scheme. The reddening vector was obtained from \citet{Flaherty2007}.}
        \label{1324_im}
        \end{figure}

\begin{figure}[]
        \centering
        \includegraphics[width=8cm]{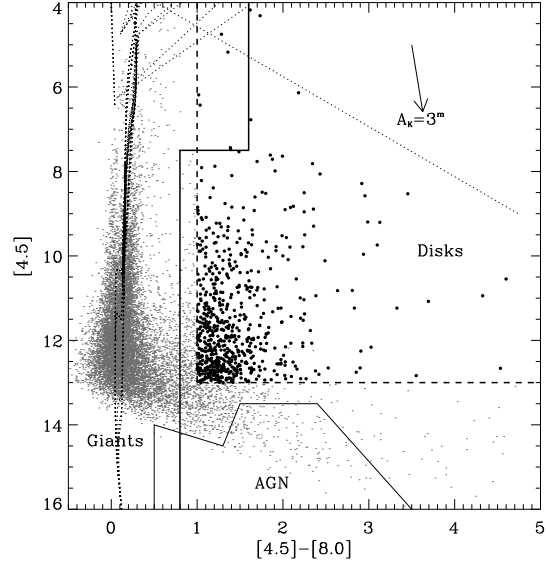}
        \caption{$[4.5]$ vs. $[4.5]-[8.0]$ diagram with the locus of stars with disks from \citet{Harvey2007}, the giants locus defined with the PADOVA isochrones (dotted lines), the locus populated by AGN defined in GMM09. Small gray dots are sources with good photometry, black dots are candidate disk-bearing objects selected with this method. The reddening vector is obtained from \citet{Flaherty2007}. }
        \label{224_im}
        \end{figure}

\begin{figure}[]
        \centering
        \includegraphics[width=8cm]{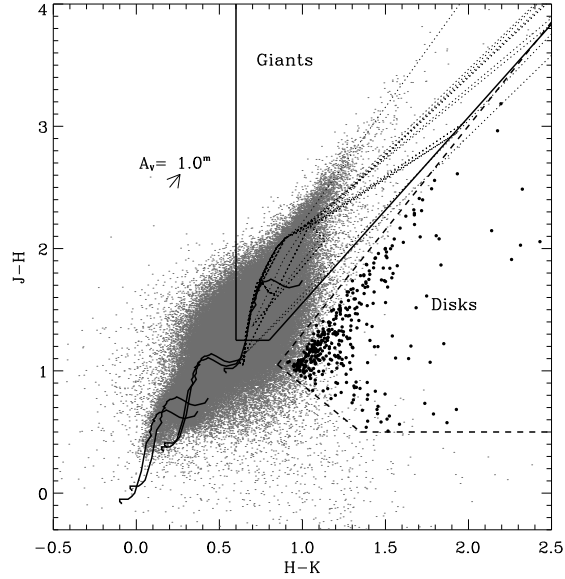}
        \caption{$J-H$ vs. $H-K$ diagram with the loci of disks and giants defined with the PADOVA isochrones (dotted lines). There is only a partial overlap between the two loci at very high extinction. Small gray dots are the sources with good photometry, black dots are candidate disk-bearing objects selected with this method. The reddening vector is obtained from \citet{RiekeLebofsky1985}. The \citet{Siess2000} ZAMS with $A_V$ from $0^m$ to $4.3^m$ are shown (solid lines).}
        \label{jhhk_im}
        \end{figure}

\begin{figure*}[]
        \centering
        \includegraphics[width=8cm]{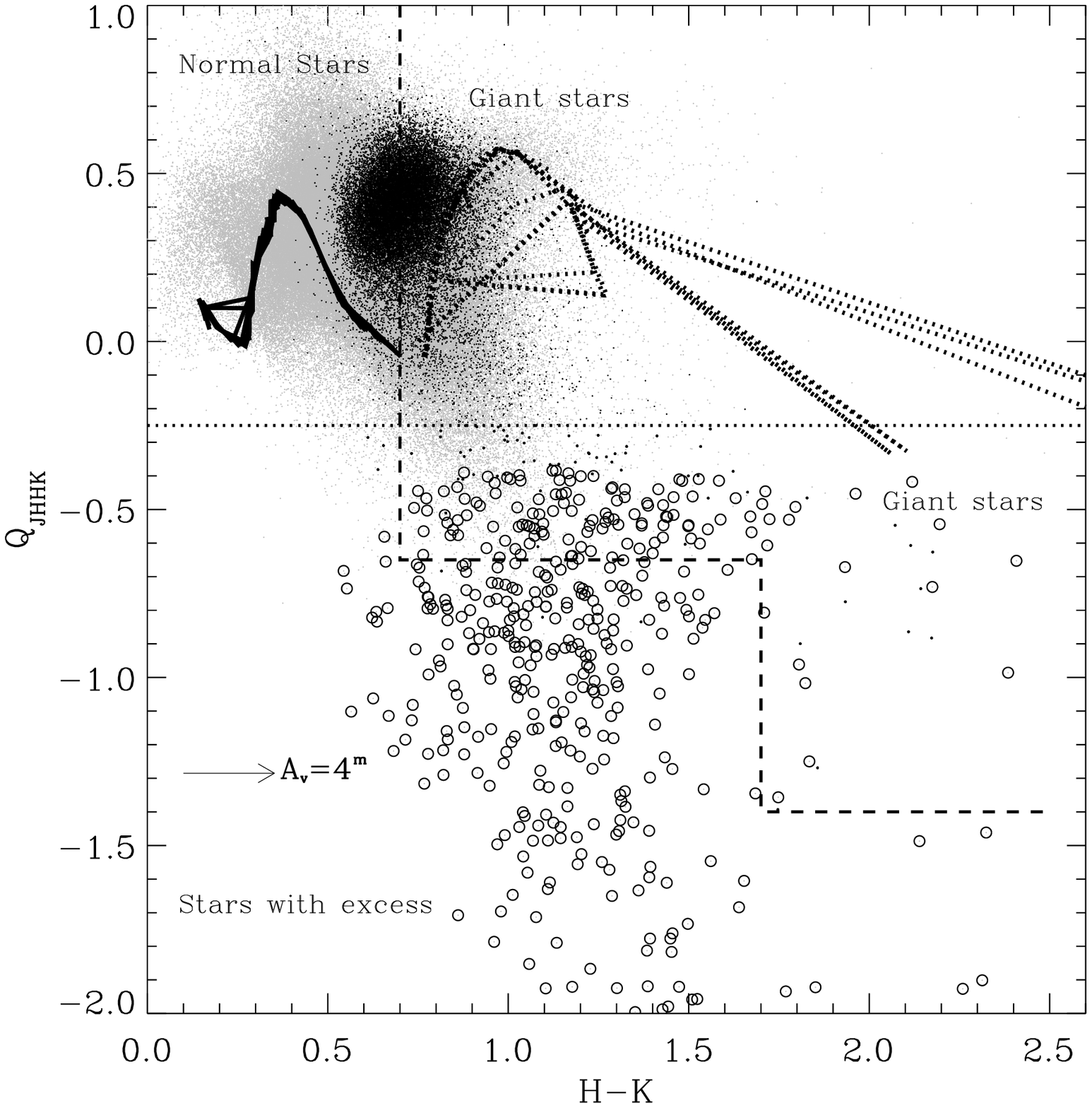}
        \includegraphics[width=8cm]{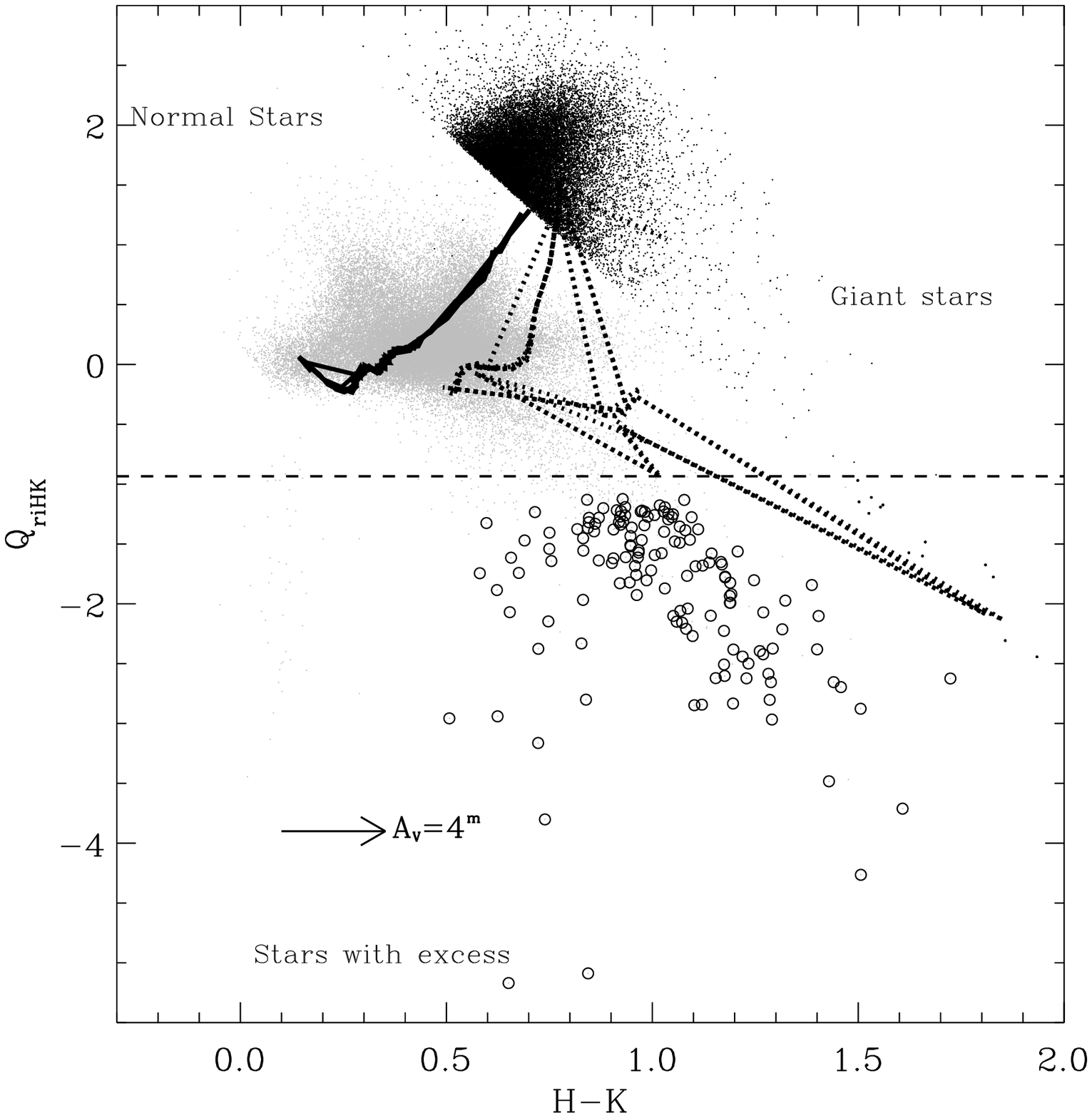}
        \caption{$Q_{JHHK}$ vs. $H-K$ (left panel) and $Q_{riHK}$ vs. $H-K$ (left panel) diagrams with all the stars with good photometry (gray dots) and those candidate to have excess in $K$ (empty circles). The horizontal lines separate the adopted loci of reddened photospheres and sources with excess. The \citet{Siess2000} isochrones with age ranging from $1\,Myr$ to $1\,Gyr$ (solid lines) and the PADOVA models from $500\,Myrs$ to $10\,Gyrs$ (dotted lines) are shown, with different values of extinction. The black small dots are giants of the northern Galactic plane \citep{Wright2008}, some of which fall in the locus of stars with $K$ excess. The dashed line in the left panel delimits the locus that can be contaminated by background giants.}
        \label{qk_im}
        \end{figure*}

\begin{figure*}[]
        \centering
        \includegraphics[width=14cm]{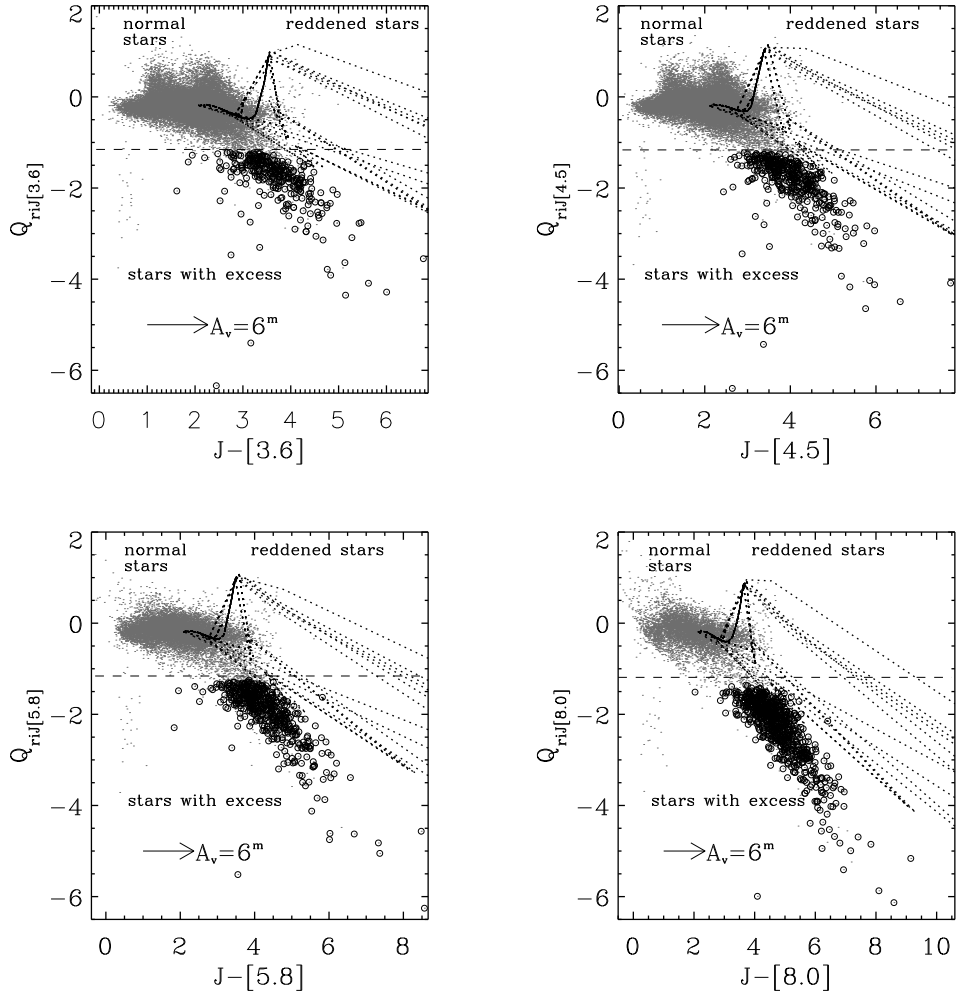}
        \caption{The four $Q_{riJ[sp]}$ vs. $J-[sp]$ diagrams used to select stars with IRAC excesses (open circles). The stars with good photometry are marked with gray dots. The dotted lines are PADOVA isochrones.}
        \label{qir_im}
        \end{figure*}

\begin{figure}[]
        \centering
        \includegraphics[width=7cm]{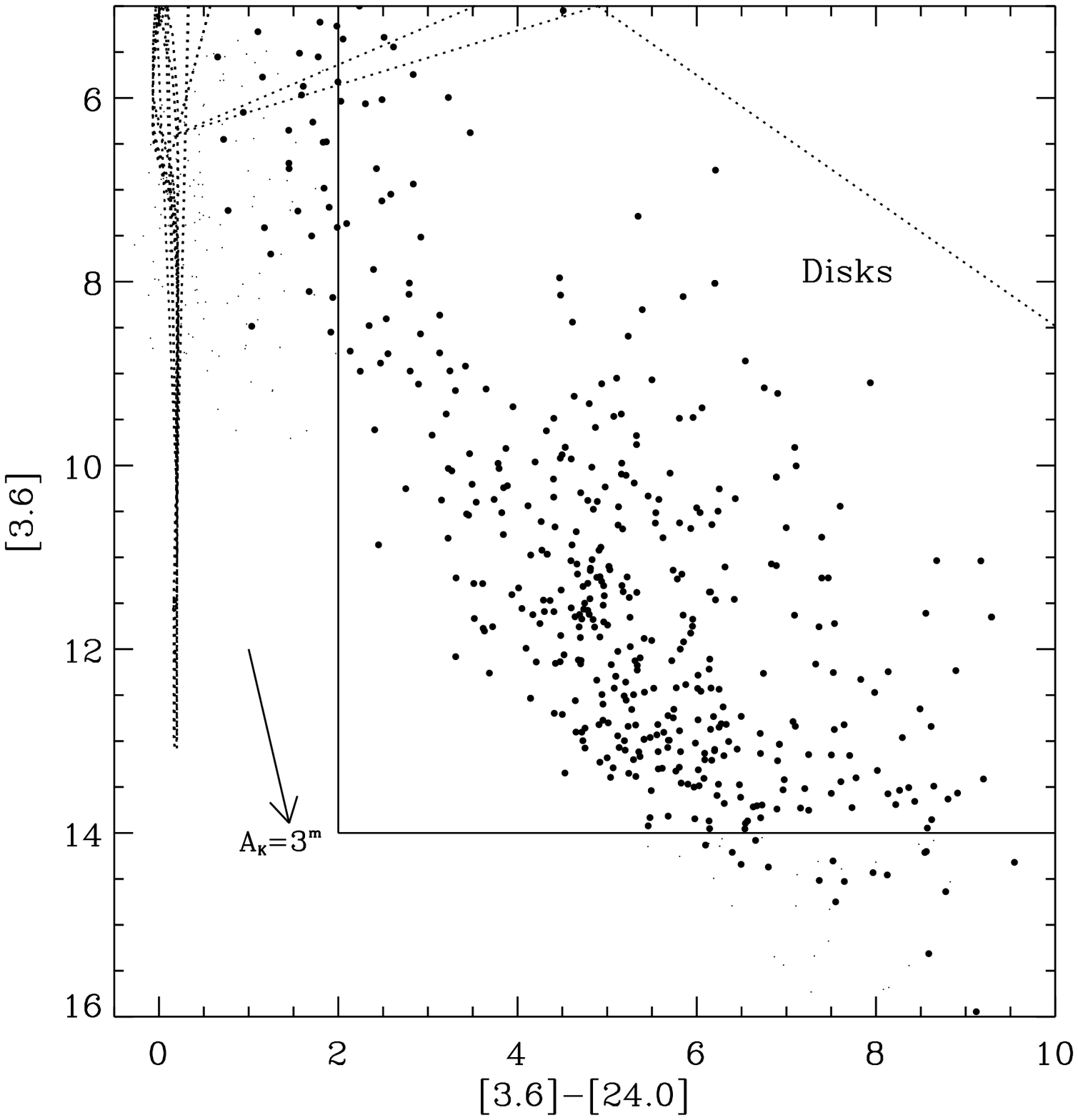}
        \includegraphics[width=7cm]{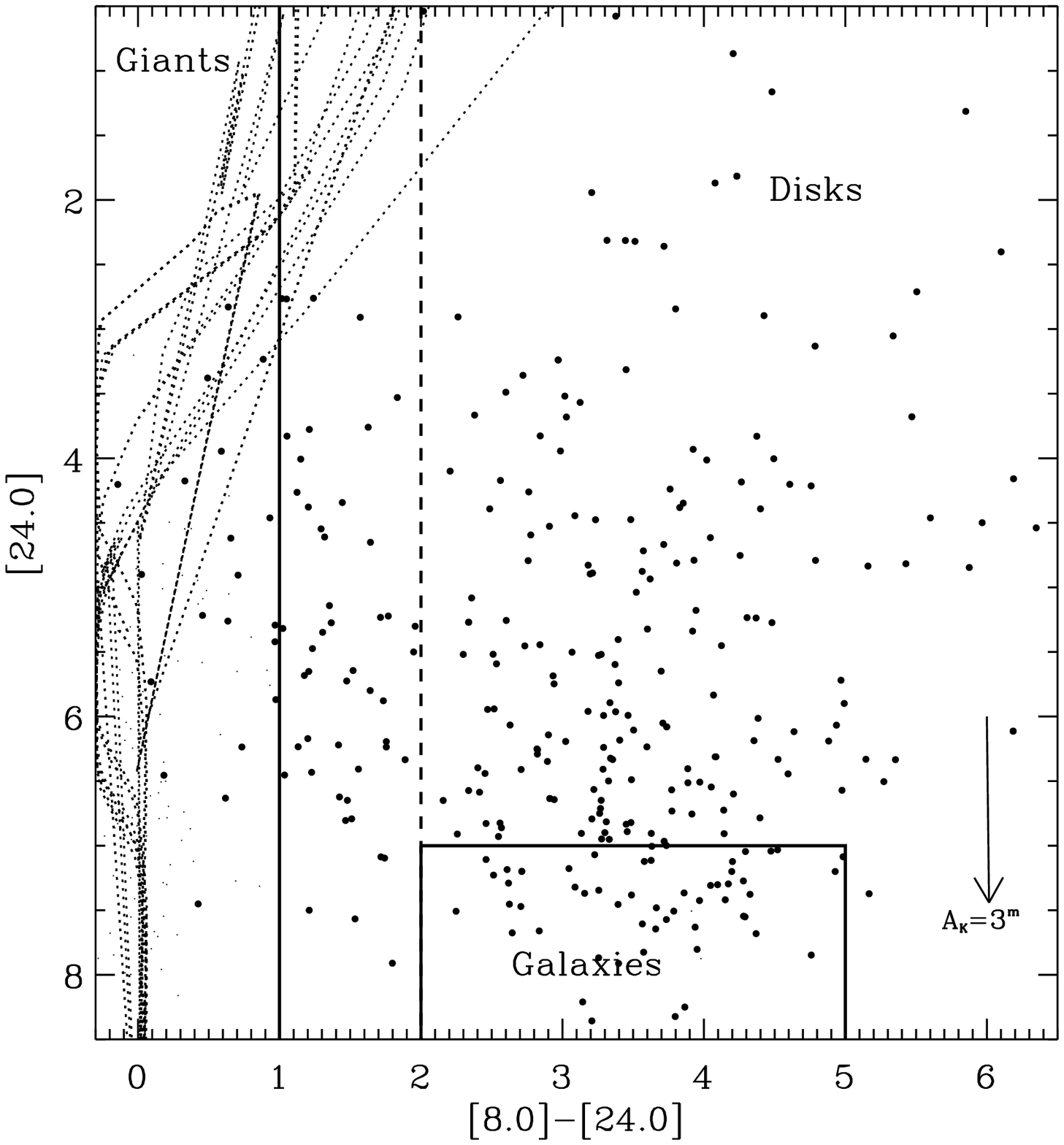}
        \par
        \includegraphics[width=7cm]{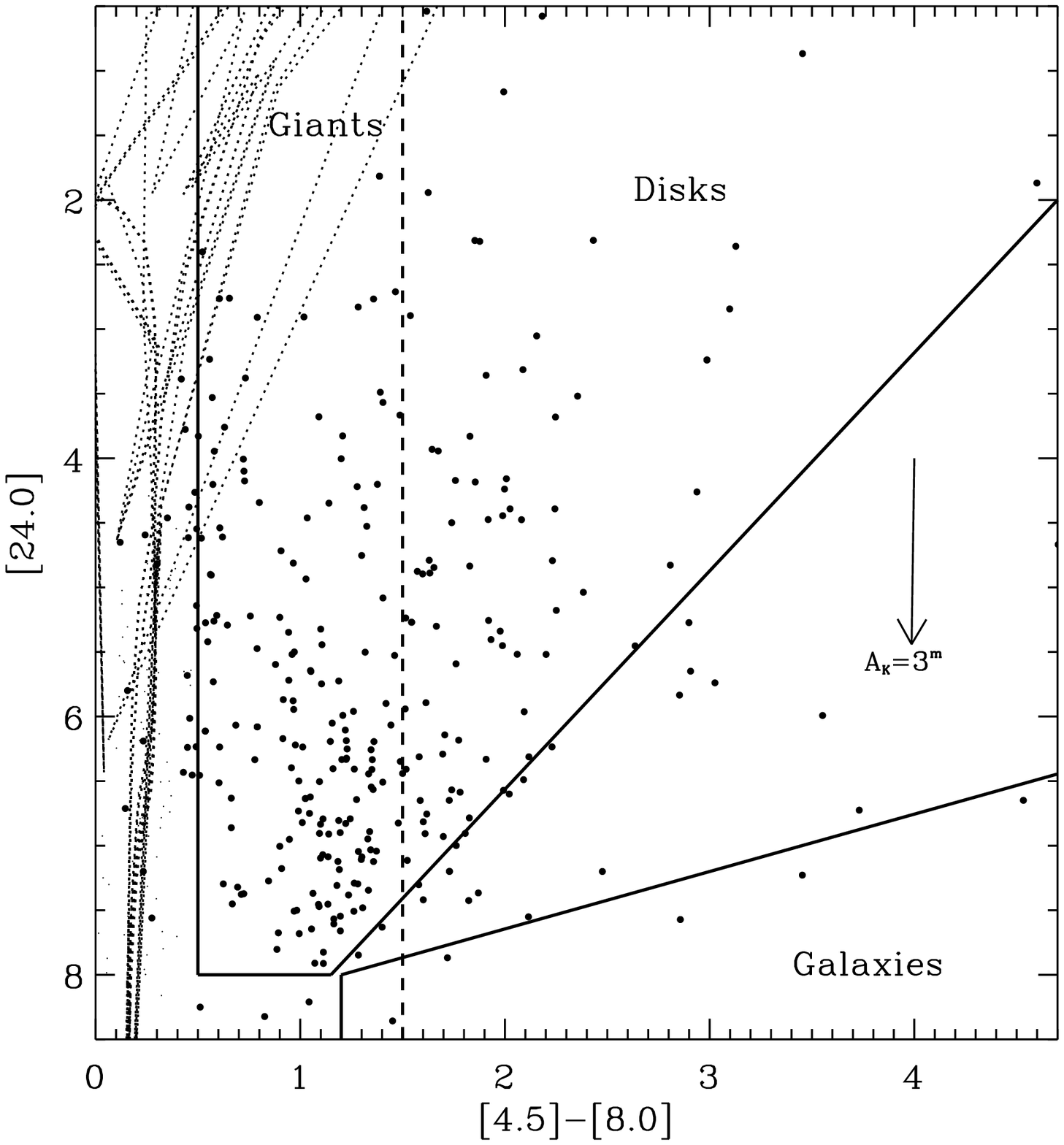}
        \includegraphics[width=7cm]{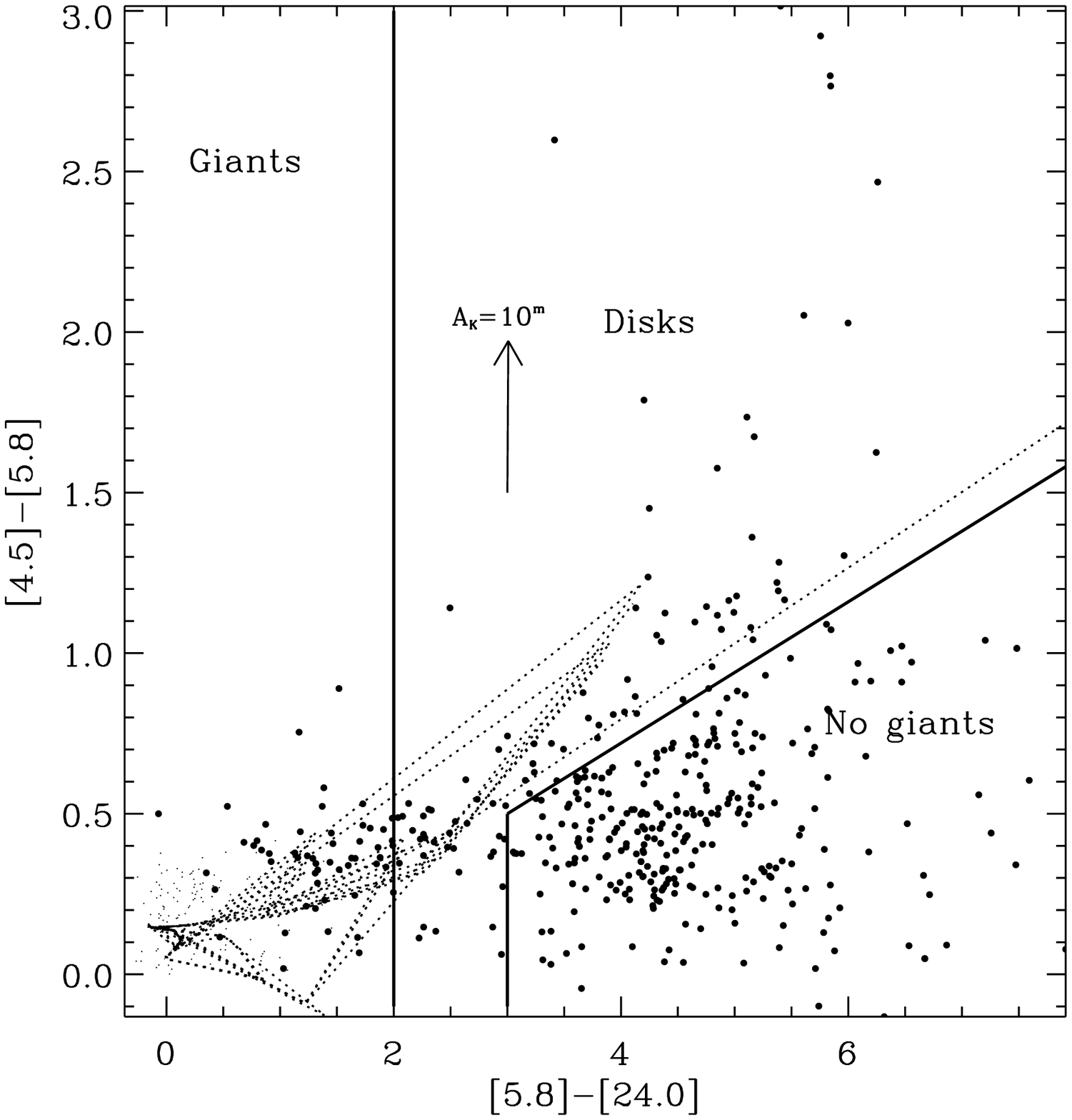}
        \caption{The four diagrams used to select stars with excesses in [24] (black circles). The gray small dots mark all the sources with good photometry. The dotted lines are the PADOVA isochrones. The loci of contaminants and  stars with excesses in the [24] band are indicated.}
        \label{mipsec_im}
        \end{figure}

\begin{figure*}[]
        \centering
        \includegraphics[width=16cm]{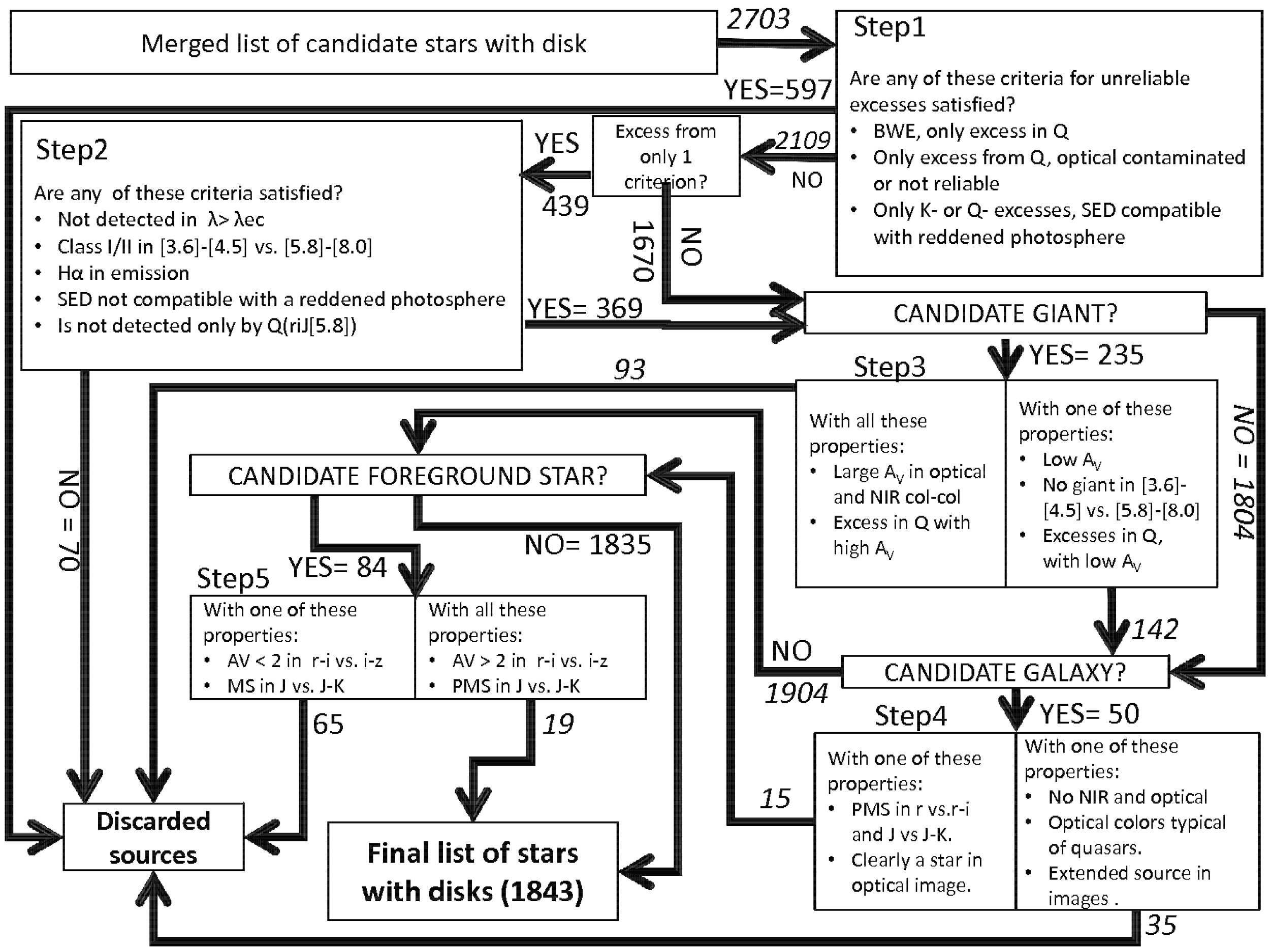}
        \caption{Algorithm used to merge the lists of candidate stars with disks from the various adopted criteria and accounts for possible contaminants.}
        \label{algo_im}
        \end{figure*}

\begin{figure*}[]
        \centering
        \includegraphics[width=14cm]{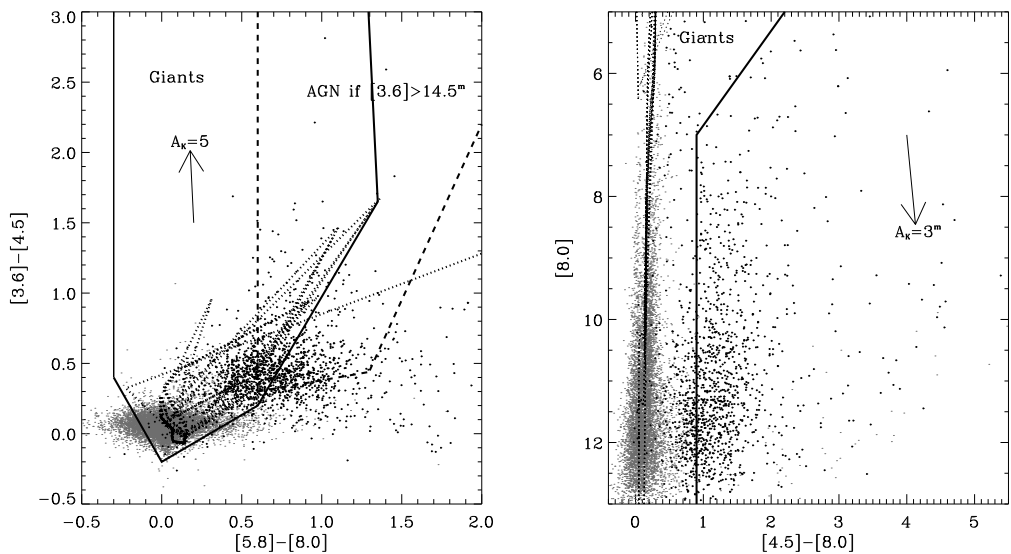}
        \caption{$[3.6]-[4.5]$ vs. $[5.8]-[8.0]$ (left panel) and $[8.0]$ vs $[4.5]-[8.0]$ (right panel) diagrams, with the sources with good photometry (gray dots), all the candidate stars with excesses (black dots) and the giant loci, delimited by solid lines. In the left panel the dashed lines delimit the AGN locus.}
        \label{giant_im}
        \end{figure*}

\begin{figure}[]
        \centering
        \includegraphics[width=8cm]{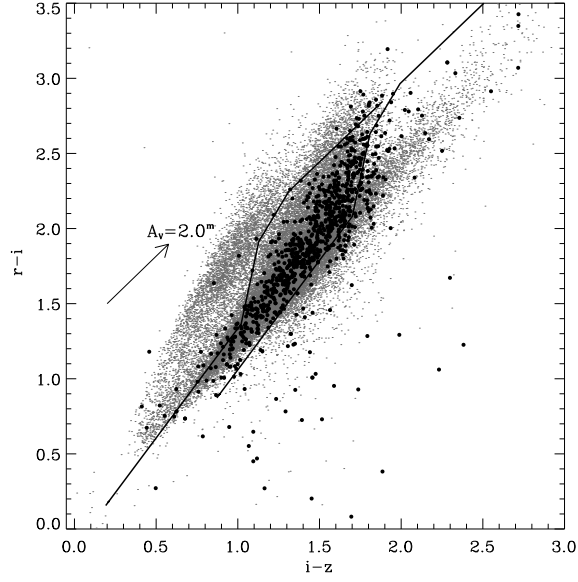}
        \caption{$r-i$ vs. $i-z$ diagram of the optical sources with good photometry (gray dots) and all the candidate stars with excess (black dots). The solid lines are two $3.5\,Myrs$ isochrones of \citet{Siess2000} with $A_V=2.4^m$ and $A_V=6^m$.}
        \label{riiz_im}
        \end{figure}

\begin{figure*}[]
        \centering
        \includegraphics[width=12.5cm]{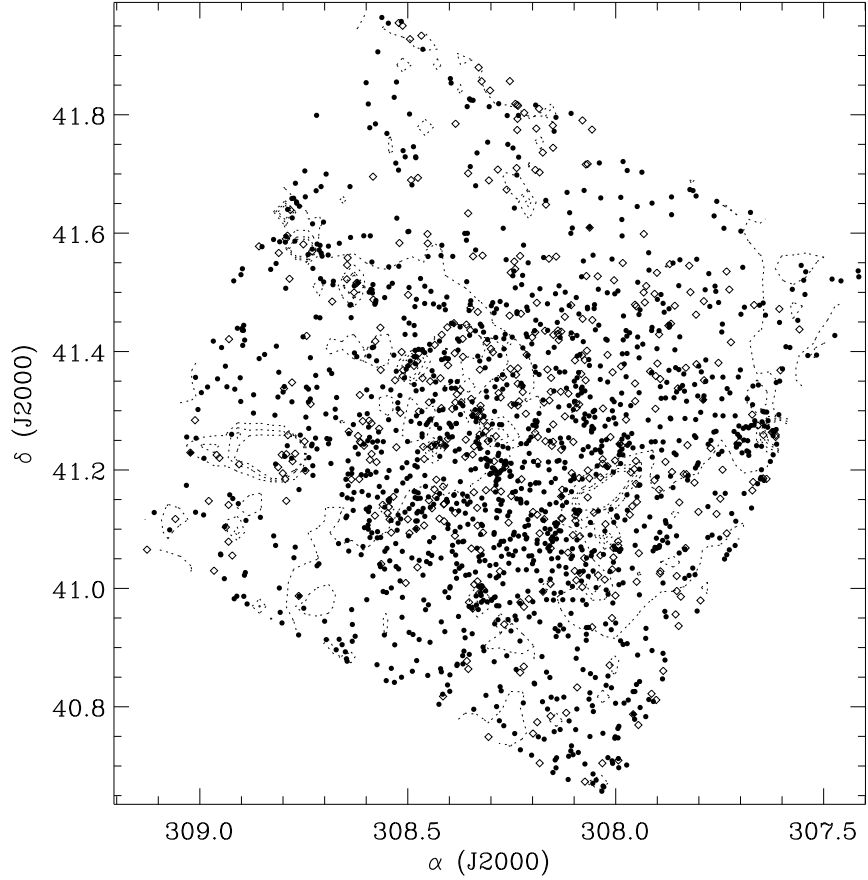}
        \caption{Spatial distribution of the stars with disks selected with the GMM09 method (filled circles) {\bf and only with the other criteria adopted} (empty diamonds). The 16.5\%, 33\%, 49.5\%, 66\%, and 82.5\% emission levels measured at $8.0\mu m$ are marked with dashed lines.}
        \label{dk_spadis_fig}
        \end{figure*}

\begin{figure}[]
        \centering
        \includegraphics[width=10cm]{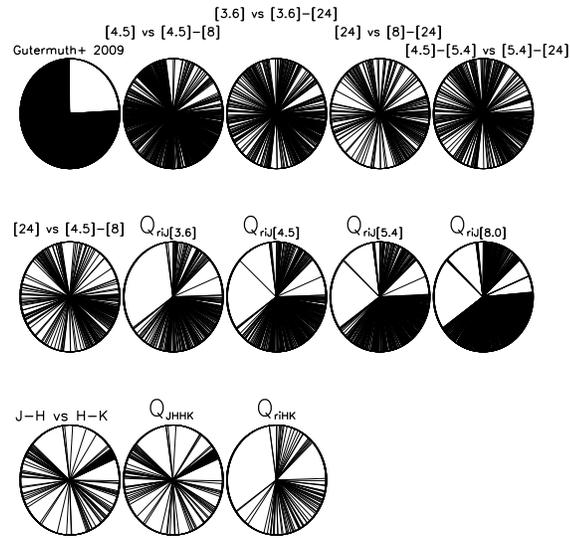}
        \caption{Diagram showing how the samples of the stars selected by each criterion are related each other. Each circle represents one selection criterion, with all the stars with disks along the circumference. A line connects the position of a particular star with the center of the circle if this star is selected by the given criterion. Stars are sorted in the same order along each circle.}
        \label{vis_crit}
        \end{figure}

\begin{figure}[]
        \centering
        \includegraphics[width=9cm]{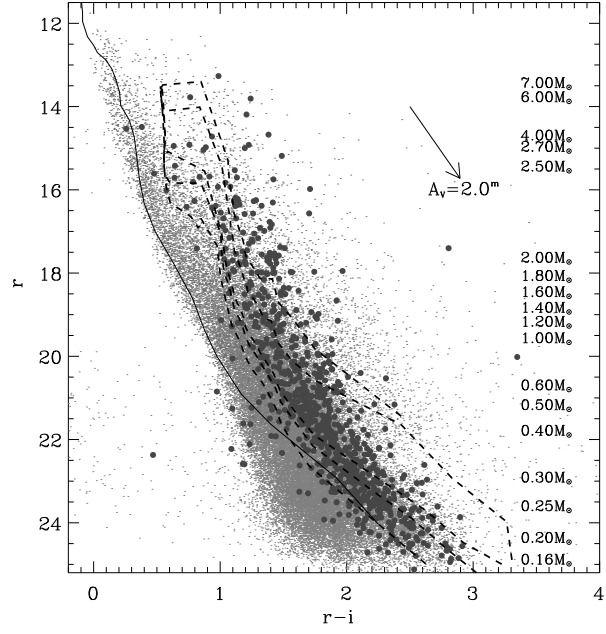}
        \caption{$r$ vs. $r-i$ diagram of the optical sources with good photometry in the studied field (light gray dots) and the candidate stars with disk (dark gray circles). The black line is the ZAMS with a distance $d=850\,pc$ and $A_V=1^m$. The dashed lines are the $0.2,\, 1,\, 3,\, 5,\, 10\,Myrs$ isochrones with $d=1400\,pc$ and $A_V=4.3^m$.}
        \label{rri_im}
        \end{figure}

\begin{figure}[]
        \centering
        \includegraphics[width=8cm]{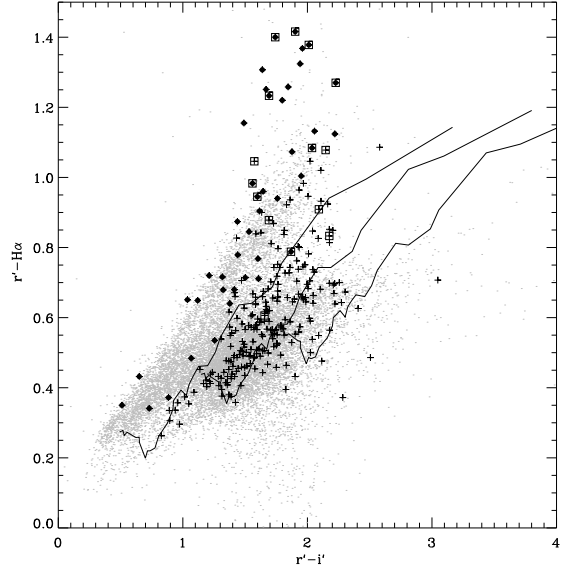}
        \caption{$r^{\prime}-H_{\alpha}$ vs. $r^{\prime}-i^{\prime}$ diagram of the IPHAS sources with good photometry (light gray dots), the candidate stars with disks (black crosses) and stars with disk with emission in $H\alpha$ (black circles) and the emission line stars discovered by \citet{Vink2008} (squares). The black lines are ZAMS with increasing extinction: from $E_{B-V}=1^m$ to $E_{B-V}=3^m$.}
        \label{rhari_im}
        \end{figure}

\begin{figure}[]
        \centering
        \includegraphics[width=9cm]{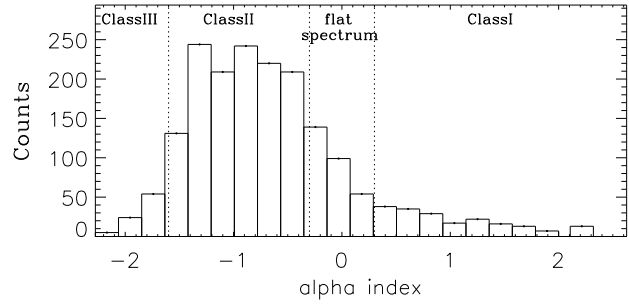}
        \caption{Distribution of the spectral indices of the disk bearing stars.}
        \label{alpha_im}
        \end{figure}

\begin{figure}[]
        \centering
        \includegraphics[width=6cm]{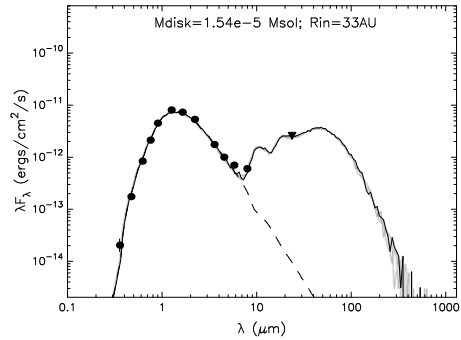}
        \caption{Results of the SED fitting for a disk-bearing star with a low mass disk with large inner radius. The black dots are the observed fluxes, the triangle is the assumed upper limit for the emission at [24]; the dashed line is the SED of the best-fit photosphere, the black line is the SED of the YSO best-fit model.}
        \label{SED_im}
        \end{figure}

\begin{figure}[]
        \centering
        \includegraphics[width=5cm]{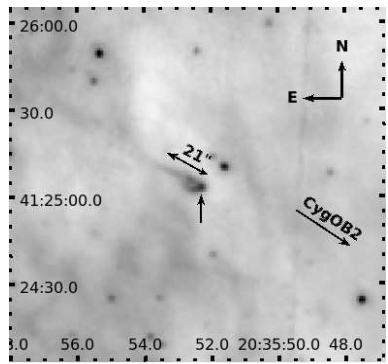}
        \caption{$8.0\mu m$ image of the proplyd not studied by \citet{Wright2012}, with the position of the YSO marked by an arrow. Its size and the approximate direction toward Cyg~OB2 are also indicated.}
        \label{prop_im}
        \end{figure}

\begin{figure*}[]
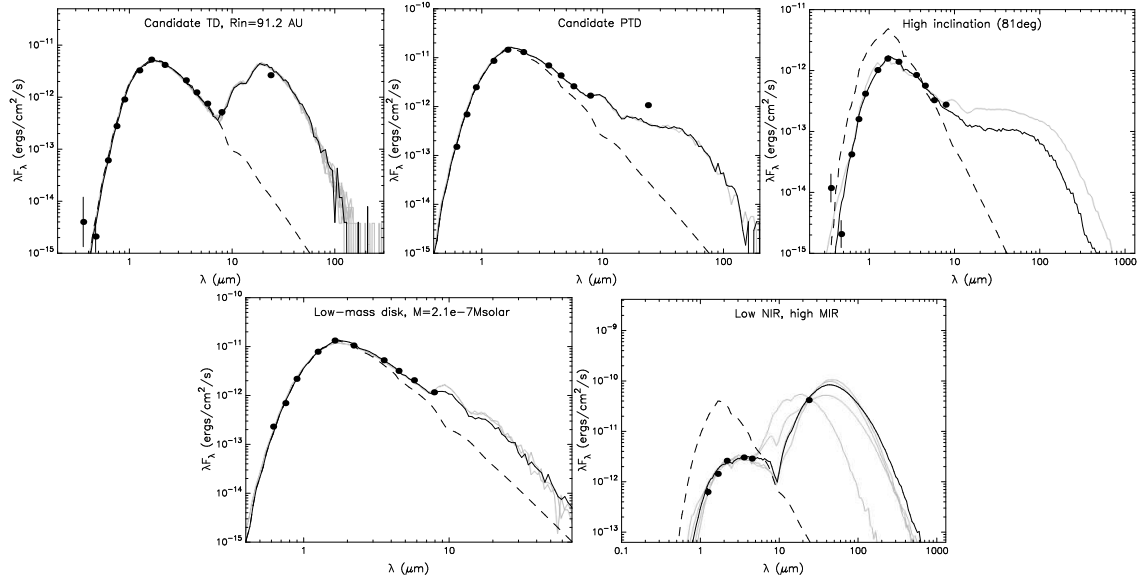

        \centering
        \includegraphics[width=5cm]{fig18a.ps}
        \includegraphics[width=5cm]{fig18b.ps}
        \includegraphics[width=5cm]{fig18c.ps}
        \par
        \includegraphics[width=5cm]{fig18d.ps}
        \includegraphics[width=5cm]{fig18e.ps}
        \caption{Examples of SED of YSOs classified as: transition disks (top left), pre-transition disks (top center), highly inclined disk (top right), low-mass disk (bottom left), and disk with low NIR and high envelope emission (bottom right).}
        \label{SED_td}
        \end{figure*}

\begin{figure}[]
        \centering
        \includegraphics[width=6cm]{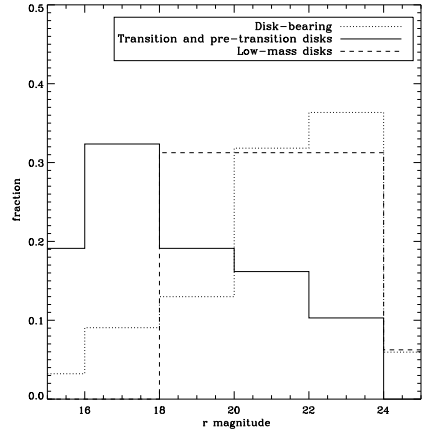}
        \caption{$r$ magnitude distributions of the entire disk population of Cyg~OB2, and of those classified as stars with transition, pre-transition, and low-mass disks.}
        \label{tdlm_magr}
        \end{figure}

\begin{figure*}[]
        \centering
        \includegraphics[width=5.7cm]{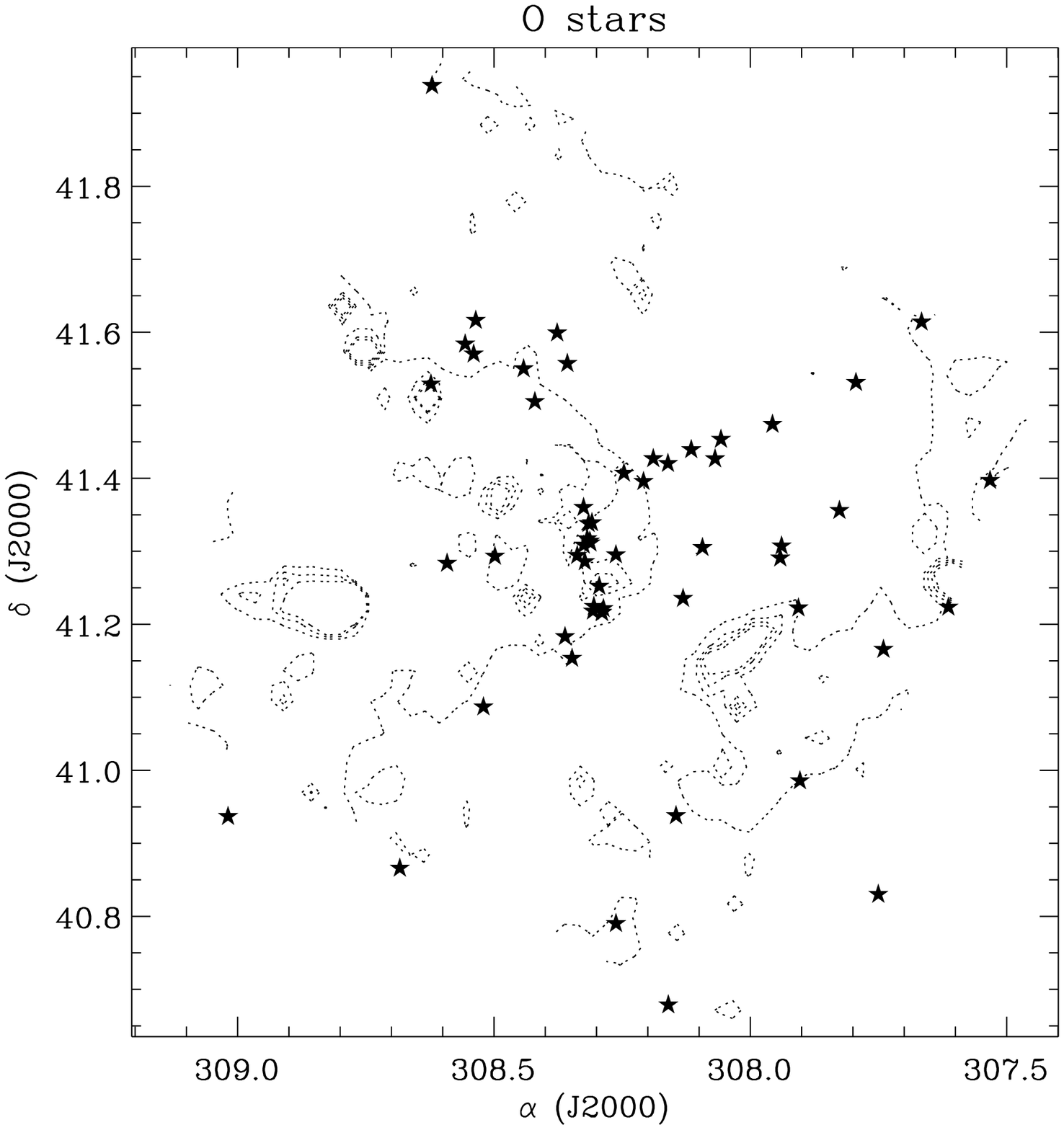}
        \includegraphics[width=5.7cm]{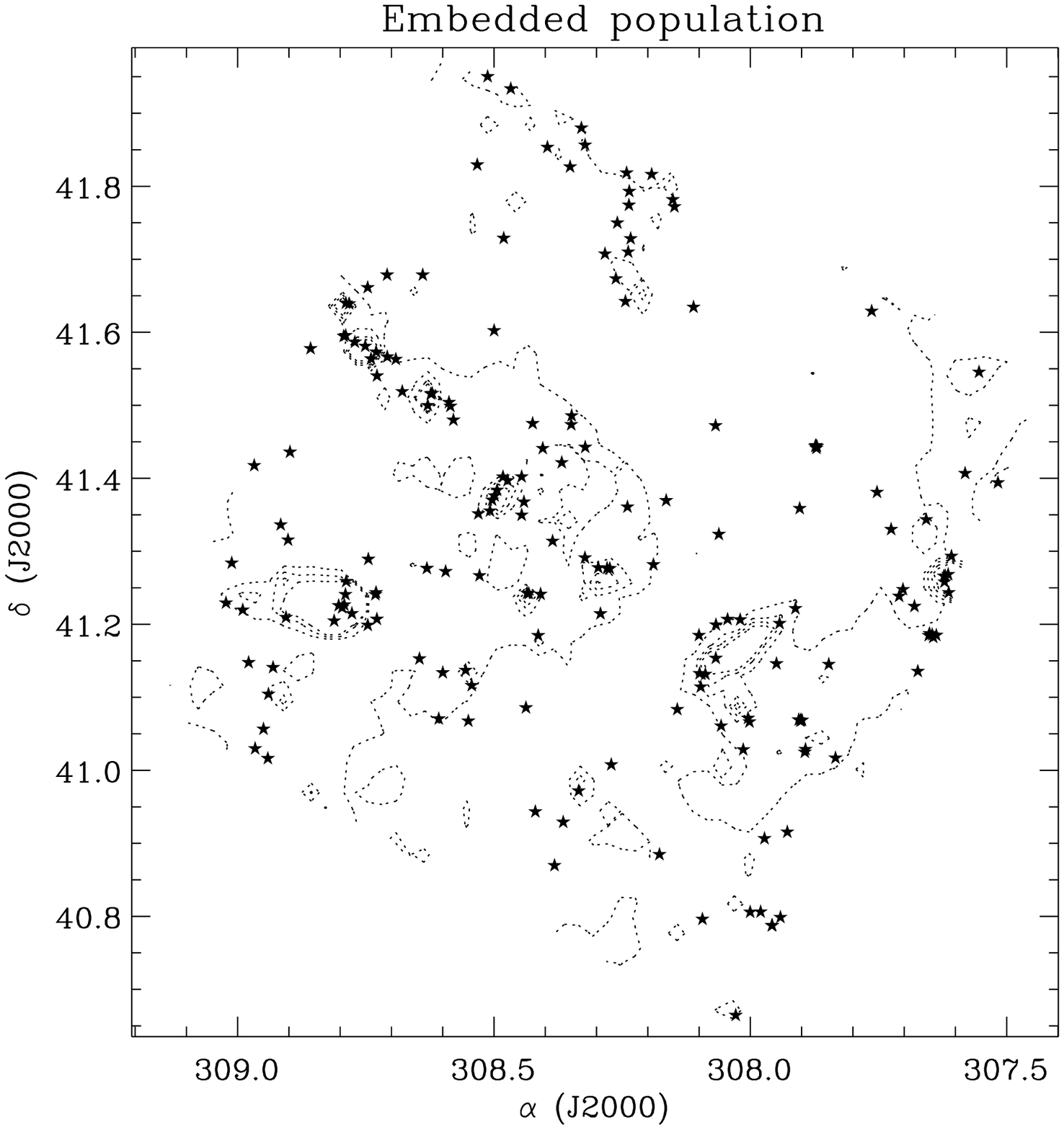}
        \par
        \includegraphics[width=5.7cm]{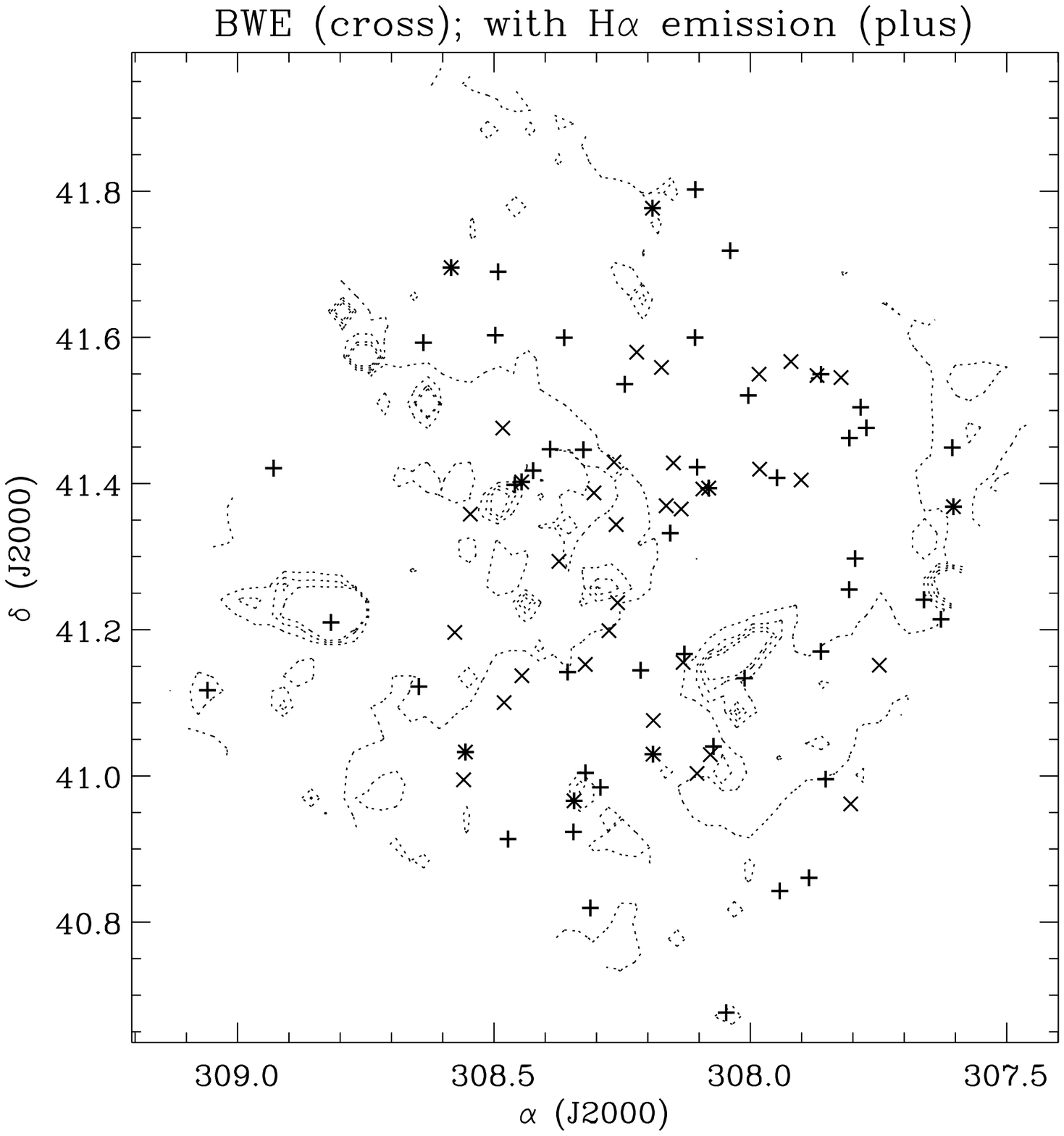}
        \includegraphics[width=5.7cm]{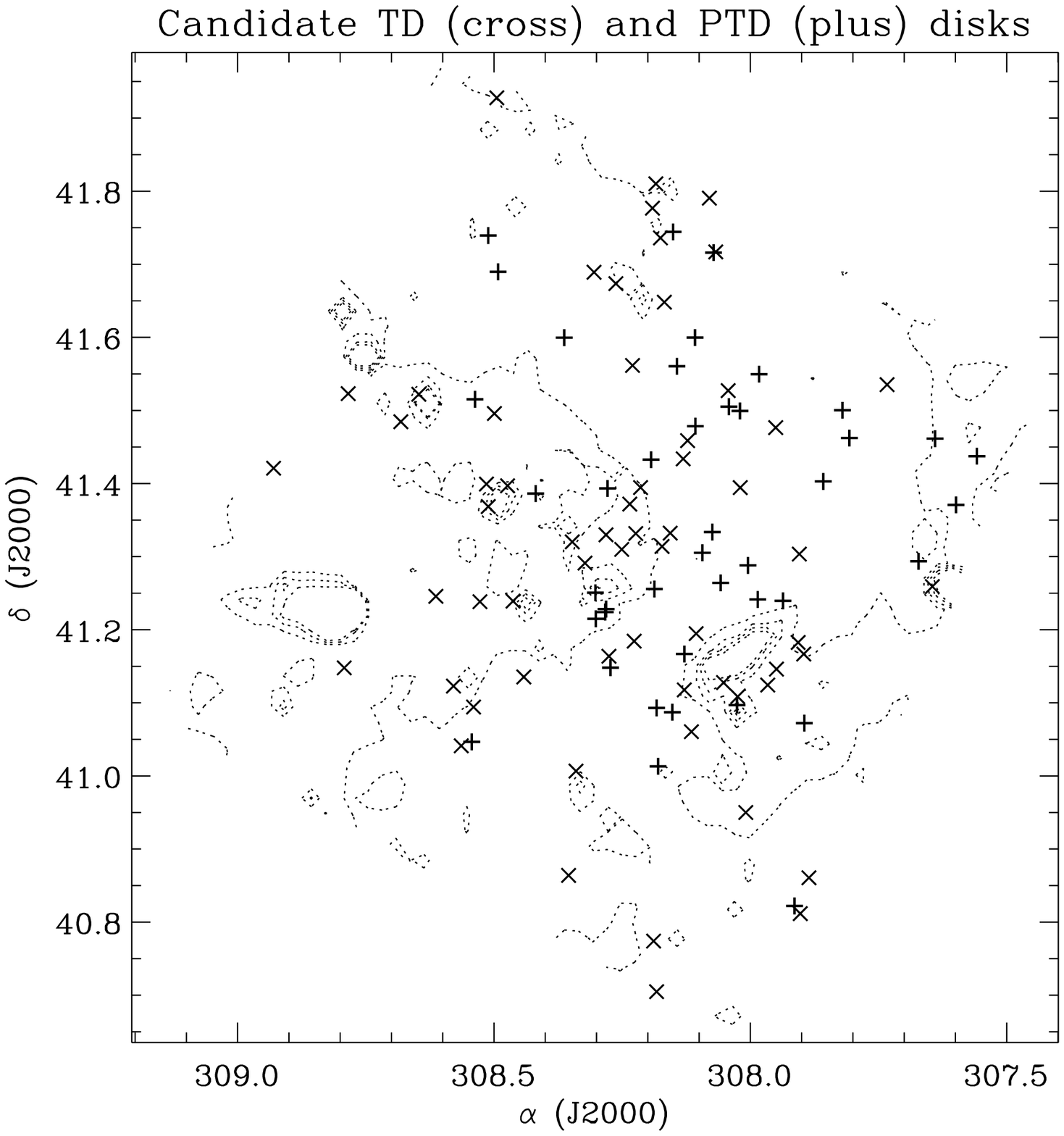}
        \caption{Spatial distribution of different populations of Cyg~OB2: the O stars in the upper left panel, and subsamples of disk-bearing objects in the remaining panels. In order to compare the positions of the stars with the structure of the nebula, the 16.5\%, 33\%, 49.5\%, 66\%, and 82.5\% emission levels measured at $8.0\mu m$ are also indicated.}
        \label{spadis_im}
        \end{figure*}

After the study of \citet{Knodlseder2000}, who concluded that Cyg~OB2 is a spherically symmetric association with a diameter of $\sim2^{\circ}$ and a half-light radius of $13^{\prime}$, more detailed analysis revealed a less uniform structure. For instance, there are strong indications that the population of Cyg~OB2 is the result of different star formation events \citep{Wright2010}; the cloud morphology (Fig. \ref{rgb_im}) reveals a non uniform structure; the spatial distribution of the O stars is far from being spherical; a large population of young A stars has been found southward of the nominal center of Cyg~OB2 \citep{Drew2008}; and the inner part of the association seems to comprise two different optical clusters \citep{Bica2003}. 
  Fig. \ref{spadis_im} shows the spatial distributions of different populations of Cyg~OB2. The overplotted 16.5\%, 33\%, 49.5\%, 66\%, and 82.5\% emission levels at $8.0\mu m$ from the IRAC observations help us to identify the regions with intense dust emission. The left top panel shows the spatial distribution of the O stars identified by different authors \citep{JohnsonMorgan1954,Walborn1973,Massey1991,Comeron2002,Hanson2003,Kiminki2007,Negueruela2008}, that are split into four different groups. Two groups lie in the center of the association, divided by an intense nebular emission. These two groups correspond to the two central clusters identified by \citet{Bica2003}, and it is not clear whether they really are two different clusters or their separation is induced by the peak of extinction  corresponding to the bright nebula between them. A third group is elongated toward the north-west direction, in a region with low dust but high gas emission (Fig. \ref{rgb_im}), while the fourth group lies in the north-east. \par 
The second top panel shows the spatial distribution of the embedded objects (classified as ``class~I'' and ``highly embedded'' objects). They are evidently clustered in dense nebulae, such as the two Globules, the cloud DR18 analyzed in \citet{Schneider2006} and in the northern pillar structure approximately at $\alpha\sim 308.2$ and $\delta\sim 41.7$ (see also Sect. \ref{activeSF}). Both the BWE and the stars with $H\alpha$ emission (bottom left panel) are more numerous in the north-west and south directions with respect to the center of the association. This may be associated with a disk population in these outer regions younger than in the center. A younger disk population, in fact, means on average higher accretion rates and a larger population of small unprocessed grains in the disks, which are both required to have BWE and $H\alpha$ emitting stars. The role of the UV radiation emitted by the O stars can also be invoked \citep{Robberto2004}, but this possibility will be more properly addressed in forthcoming studies. The spatial distributions of transition and pre-transition disks is slightly different, with a larger concentration in the center and westward. 

\subsection{Surface density map}
\label{density_sec}

Cyg~OB2 is a complex association, so a high level of subclustering is expected. However the spatial distribution of the disk-bearing objects shown in Fig. \ref{dk_spadis_fig} does not allow us a clear understanding of the level of subclustering, if any, present in the region. A better way to examine this is through the analysis of stellar density maps and the Minimum Spanning Tree technique \citep{Barrow1985}. \par
\begin{figure}[]
        \centering
        \includegraphics[width=8.5cm]{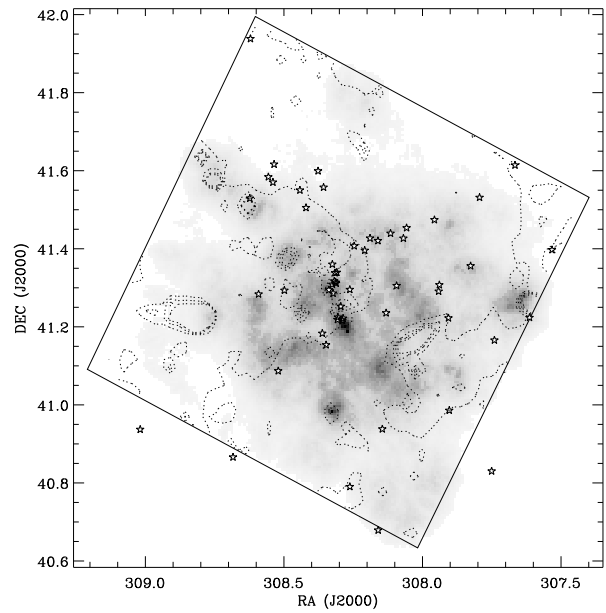}
        \caption{Surface density map of the disk-bearing objects in Cyg~OB2. The contours map the $8.0\mu m$ emission as in Fig. \ref{spadis_im}. The stars symbols mark the positions of the O stars.}
        \label{densitymap}
        \end{figure}

	We calculated the stellar surface density of the candidate disk bearing objects following the method introduced by \citet{CasertanoHut1985}, where the stellar surface density $\sigma(i,j)$ inside a cell of an uniform grid with center at the coordinates $\left( i,j \right)$ is:

\begin{equation}
\sigma(i,j)=\frac{N-1}{\pi r^2_N \left(i,j \right)}
\end{equation}

where $r_N \left(i,j \right)$ is the distance between the center of the cell and the $N^{th}$ most distant source (i.e. $r_3 \left(i,j \right)$ is the distance to the third most distant source). $N=18$ is a good choice to smooth out the smallest scale structures \citep{Gutermuth2008, Winston2012}. The typical uncertainty in density estimation goes as $\sigma/\left(N-2 \right)^{0.5}$ \citep{CasertanoHut1985}, being about 25\% using $N=18$. 
  Fig. \ref{densitymap} shows the surface density map of Cyg~OB2 obtained with a $200\times200$ uniform grid and $N=18$. The central overdensity corresponding to the southern cluster identified by \citet{Bica2003} is clearly evident. Another overdensity corresponds to the central northern cluster. As explained, the two optical clusters may be divided because of the extinction peak located approximately at $\alpha\sim 308.2924,\,\delta\sim 41.2554$, but we found a distribution of disk-bearing objects that connects them, even if with lower density than that of the two clusters. An evident ring-shaped overdensity about 0.4 deg in diameter surrounds the central area. This ring-like structure is divided in three different sections: an arc at south-east $\sim 24^{\prime}$ long and $14^{\prime}$ to $15^{\prime}$ distant from the approximate center of the association; an overdensity in the south-west direction, southward the bright Globule $\#1$ identified by \citet{Schneider2006} and $\sim13.6^{\prime}$ from the southern cluster; and a $\sim14.2^{\prime}$ segment in the west connecting the northern part of the Globule $\#1$ and the elongated distribution of O stars. It is likely that these last two segments form a unique structure together with the Globule $\#1$, whose high density and brightness prevent a deep identification of the embedded low-mass YSOs. \par
	It is not easy to understand the nature of this ring-like overdensity around the two central clusters, without information about the stellar dynamics and the sequence of star formation. These will be the subjects of forthcoming studies. The most likely hypothesis is that star formation has been triggered by the massive stars in the center and it occurred in the expanding shocked nebular front, but in this case these stars along the ring must be young enough to be not dynamically relaxed, and younger than the stars in the central region. An indication in this direction is perhaps provided by the spatial distribution of transition disks and accreting stars shown in Fig. \ref{spadis_im}, where the center of the association shows a deficit of the latter and an overdensity of the former stars that are expected to be older than actively accreting stars. 

\subsection{Subclustering}
\label{subcl_sec}

\begin{figure}[]
        \centering
        \includegraphics[width=6cm]{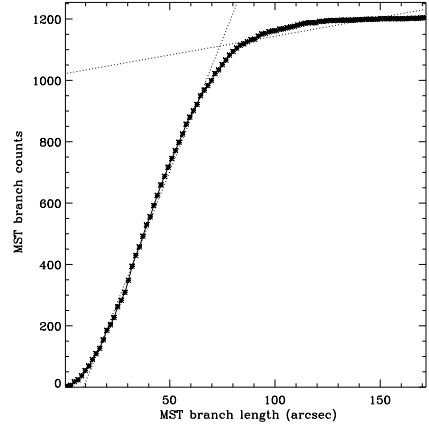}
        \caption{Cumulative distribution of the MST branches length. The dotted lines represent the linear fit of the points smaller and larger than the chosen critical branch length.}
        \label{branch_dis}
        \end{figure}

\begin{figure*}[]
        \centering
        \includegraphics[width=8.5cm]{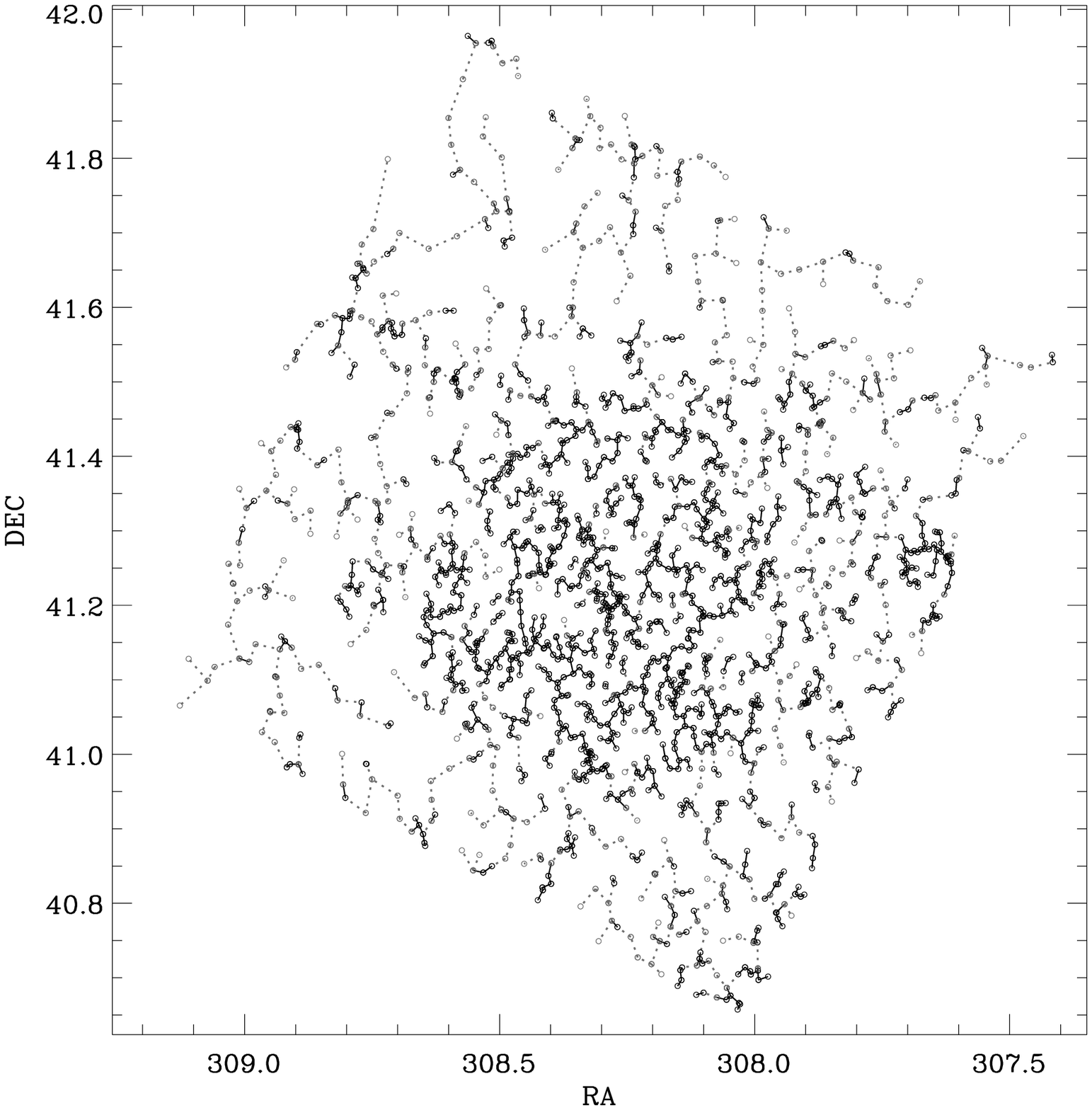}
        \includegraphics[width=8.5cm]{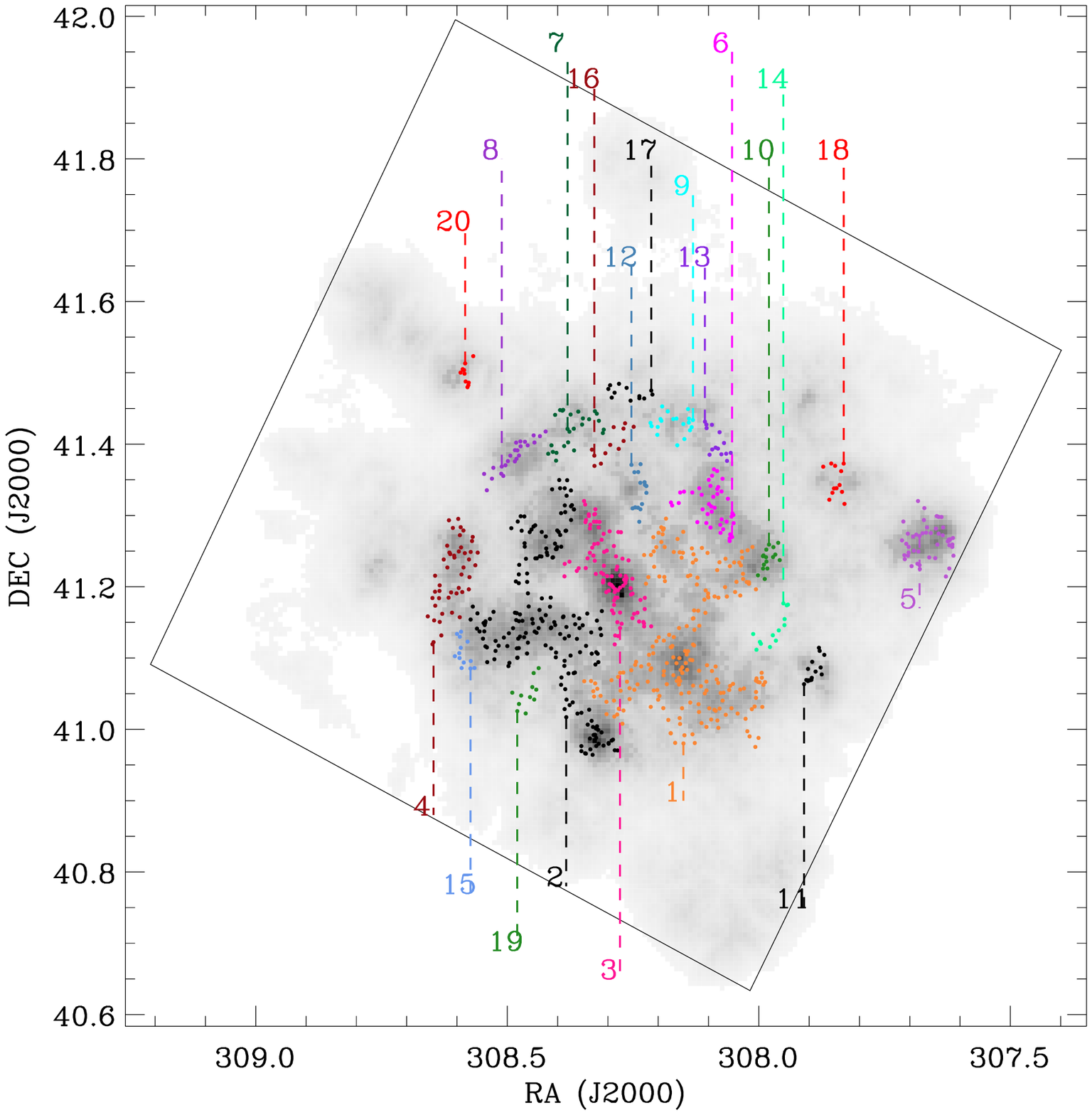}
        \caption{Left panel: the MST obtained with all the disk-bearing YSOs, whose positions are marked by circles. The branches smaller than the critical length and the connected points are marked in black with continuous lines; those larger than the critical length are in gray with dotted lines. Right panel: the surface density map shown in Fig. \ref{densitymap} with the 20 subclusters we found, each marked with a different color and identification number.}
        \label{mst_im}
        \end{figure*}

Subclustering can be studied in a more appropriate way by the use of the {\it Minimum Spanning Tree} (MST) technique \citep{Barrow1985}, which is defined as a set of branches that connect all the points in a given sample (in our case the positions of the disk-bearing objects), minimizing the total length of the branches, and not producing closed loops. The MST has several advantages for studying the spatial distribution of a sample of stars with respect to other methods, such as it does not smooth out the geometry of small groups and allows a direct identification and extraction of subclustered members. For the construction of the MST of the disk-bearing objects and the identification of the subclusters, we used the {\emph R} statistics package {\it nnclust}\footnote{http://cran.r-project.org/web/packages/nnclust/nnclust.pdf}. 
  The identification of the subclusters requires the use of a critical branch length, that we defined adopting the method introduced in GMM09. We first constructed the cumulative distribution of the branches lengths (Fig. \ref{branch_dis}). Then we found the length at which the cumulative distribution changes shape, first by choosing an arbitrary separation length in the center of the distribution, then performing two linear fits of the points smaller and larger than the chosen length, calculating the value at which the two best-fitting lines intersect and using this value as new separation length, repeating the procedure until it converged (after 6 iterations). The final critical length is $71.9^{\prime \prime}$, corresponding to $0.51\,pc$ at the distance to Cyg~OB2.
The result of this procedure is shown in Fig. \ref{mst_im}. The left panel shows the MST connecting the positions of the candidate disk-bearing stars. The branches shorter than the critical length and the connected points are drawn in black, while those larger are shown in gray and dotted lines. The black lines connect the stars that belong to different subclusters. {\bf Those that contains more than 20 stars}, and the positions of their members, are shown overplotted on the surface density map in the right panel. Each subcluster is marked with a different color and an identification number from 1 to 20, in order of number of stars. The spatial distribution of the subclusters mainly follows the annular overdensity, with only one cluster in the center of the association, and the subclusters \#5, \#11, \#18, \#20 spatially separated from the whole central region. \par
   In total, 45.3\% of the entire disk population belongs to the subclusters, while the remainder is spread over the region or constitutes smaller aggregates, as shown in the left panel of Fig. \ref{mst_im}. However, the results obtained with the MST have to be used with caution. Some of the spatial associations shown in Fig. \ref{mst_im} are questionable and slight changes in the critical branch length result in different subclusters. The use of the color-color and color-magnitude diagrams help us to understand whether the physical properties of stars belonging to various subclusters are really different. Table \ref{clust_tbl} summarizes the properties of the stellar population of each subcluster.
The two central clusters found by \citet{Bica2003} are merged into the central subcluster \#3, mainly because of several disk-bearing objects lying in the nebulosity that separates the two optical subclusters. This finding, together with the fact that the disk-bearing stars falling in these two fields populate similar loci in the various diagrams (with no strong evidence for different extinctions or evolutionary status), suggests that they really form a unique cluster, and that the separation between the two optical clusters is mainly due to the bright nebulosity between them. \par
  The subclusters in the northern part of the overdensity ring (i.e. \#6, \#7, \#9, \#13, \#16, and \#17) share similar properties, having a low fraction of embedded members ($\leq 5\%$ in total) and an extinction lower than the rest of the disk-bearing population. Their median $J-H$ color, in fact, is $1.31^m$, while that of the entire population is $1.44^m$ (almost the same of that of the stars in the central subcluster), with a difference corresponding to an extinction difference of $A_V=1.2^m$. These properties suggest that they actually form a unique elongated structure $\sim10.6pc$ long, which runs in the east-west direction from the Globule \#1 northward the central cluster. The lower extinction in this subcluster agrees with the extinction map found in \citet{Guarcello2012B} using optical and X-ray data. \par
The large subcluster \#2 (187 stars) is apparently one example of wrong association made by the MST procedure, since it appears to be divided into three groups: one northward that looks connected to the central cluster; one southward centered on a peak in the surface density map (Fig. \ref{densitymap}), and the central part that looks like being a connection between the subclusters \#1 and \#4. However, the diagrams reveal that the stars associated with the whole subcluster \#2 share the same properties. Also the stars in the subclusters \#4 and \#15 share similar properties, but with differences in the median $J-H$ indicating larger extinction of $A_V\sim0.5^m$ for \#4 and $A_V\sim0.8^m$ for \#15. Analogously, \#19 is formed principally by low mass stars with a slightly higher extinction (by $\sim0.4^m$) than the stars in subclusters \#2, and with a high fraction of H$\alpha$ emitters (9/16). We then conclude than the subclusters \#2, \#4, and \#15 may form an unique elongated structure southward the central cluster, with similar stellar content but extinction slightly increasing toward south-east. \par
  The western branch is mainly formed by the big subcluster \#1, with 213 stars mainly around the Globule\#1. Its northern part has been split from the subcluster \#10, which is slightly less extinguished (with a difference of $A_V\sim0.8^m$) but similar stellar content, while the subcluster \#1 may represent a different small association of stars with a variety of brightness and extinction.  \par
The remaining subclusters mainly form the most embedded and young disk population surrounding the central cluster and the ring. The subcluster \#11 lies westward the Globule\#2, and it is characterized by a large fraction of embedded objects (4/11) but just a slightly larger extinction in the $J-H$ color with respect the entire disk population (a difference of $A_V\sim0.4^m$). The subcluster \#5 is formed by the stars around and within the Globule\#2, and its population can be divided in more evolved class~II stars with a higher extinction than the rest of the disk bearing population (with a difference of $A_V\sim0.5^m$) and an embedded population of class~I YSOs inside the Globule with very red IRAC colors, suggesting extinctions of $A_K>4^m$. The subcluster \#12 has an intermediate position between the central cluster and the northern rim, with the 6.7\% of embedded stars. This affects the estimate of its extinction, inferred from the median $J-H$ color, which is higher than the stars in the northern rim by $A_V\sim1^m$, but lower than the rest of the central cluster by $A_V=0.2^m$. The subcluster \#8 consists of the stars inside and around the bright cloud ECX6-27. Even if the median $J-H$ color of these stars suggests a similar extinction than the rest of the disk bearing population, the stars inside the cloud are in an early evolutionary phase, being more embedded in the cloud. The membership of this population of stars to Cyg~OB2 will be addressed later.  The subcluster \#18, located in the center of the north-west cavity in the molecular cloud, is formed by class~II YSOs with a slightly larger extinction than the rest of the disk-bearing population (by $A_V=0.4^m$) and mainly an early spectral type as suggested by their position in the color-magnitude diagrams. Finally, stars of subcluster \#20 are embedded in one of the bright rimmed clouds in the north-east, with a high fraction of embedded YSOs.\par	
  
\section{Active star formation in Cygnus OB2}
\label{activeSF}

\begin{figure}[]
        \centering
        \includegraphics[width=8cm]{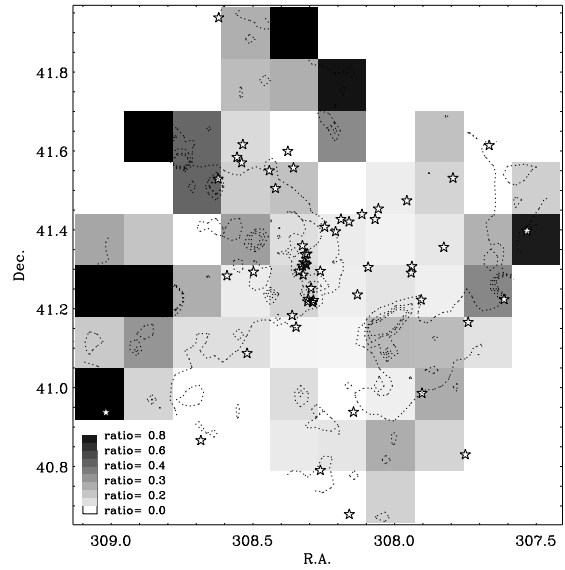}
        \caption{Gray-scale map of the ratio between the number of class~I and class~II YSOs in a uniform grid. The stars symbols mark the positions of the O star, while the dotted lines the [8.0] emission contours.}
        \label{IvsII_im}
        \end{figure}

\begin{figure*}[]
        \centering
        \includegraphics[width=5.5cm]{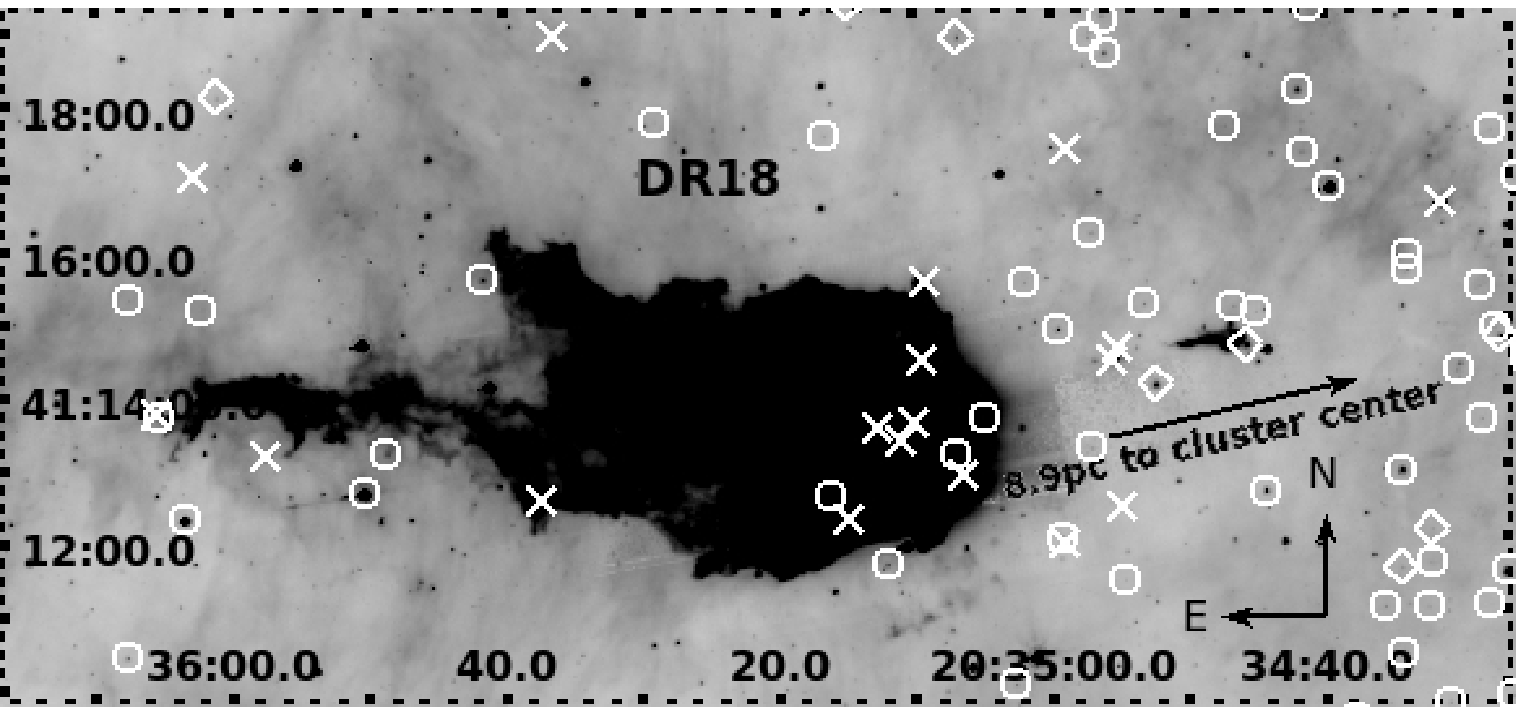}
        \includegraphics[width=5.5cm]{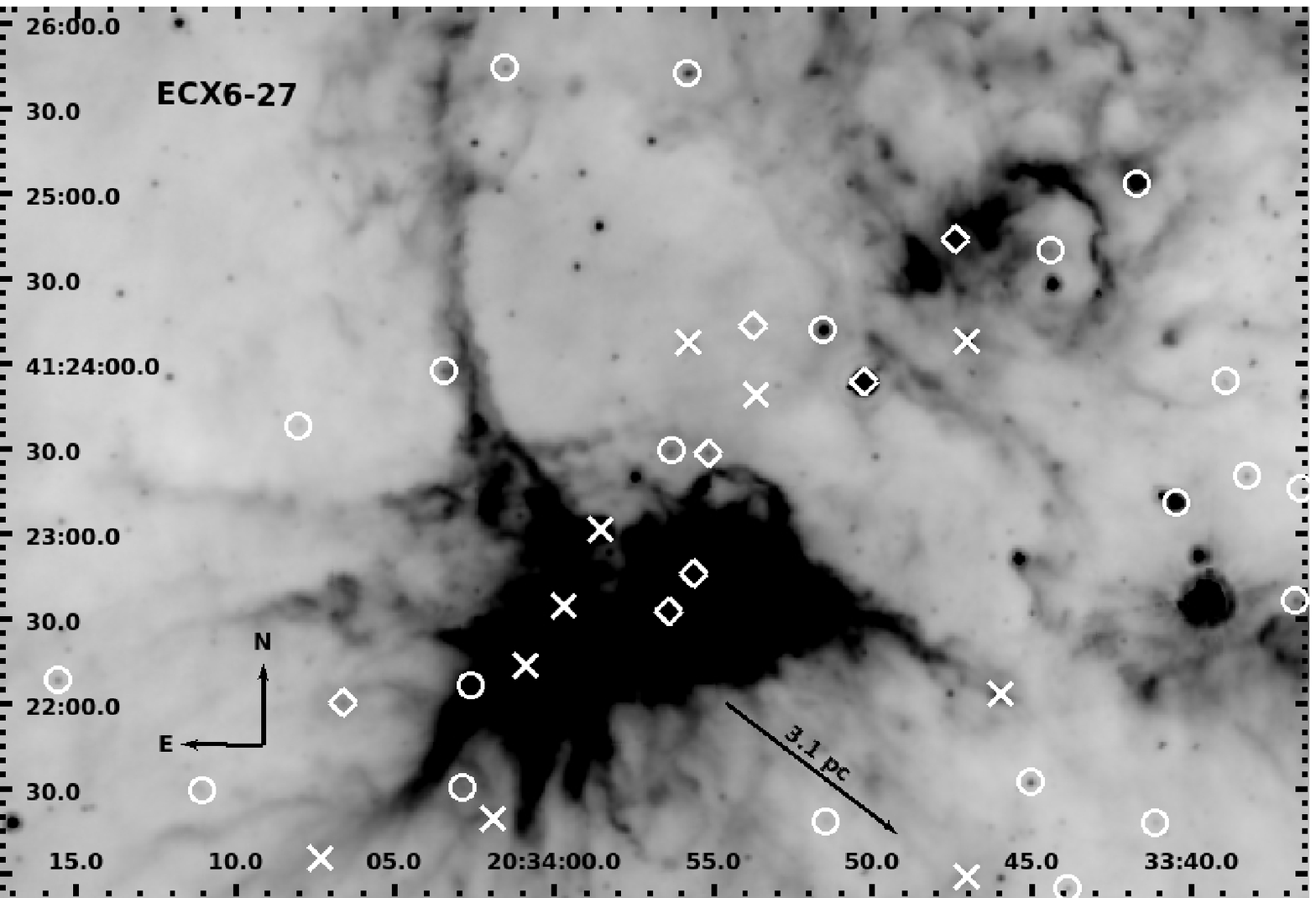}
        \includegraphics[width=5.5cm]{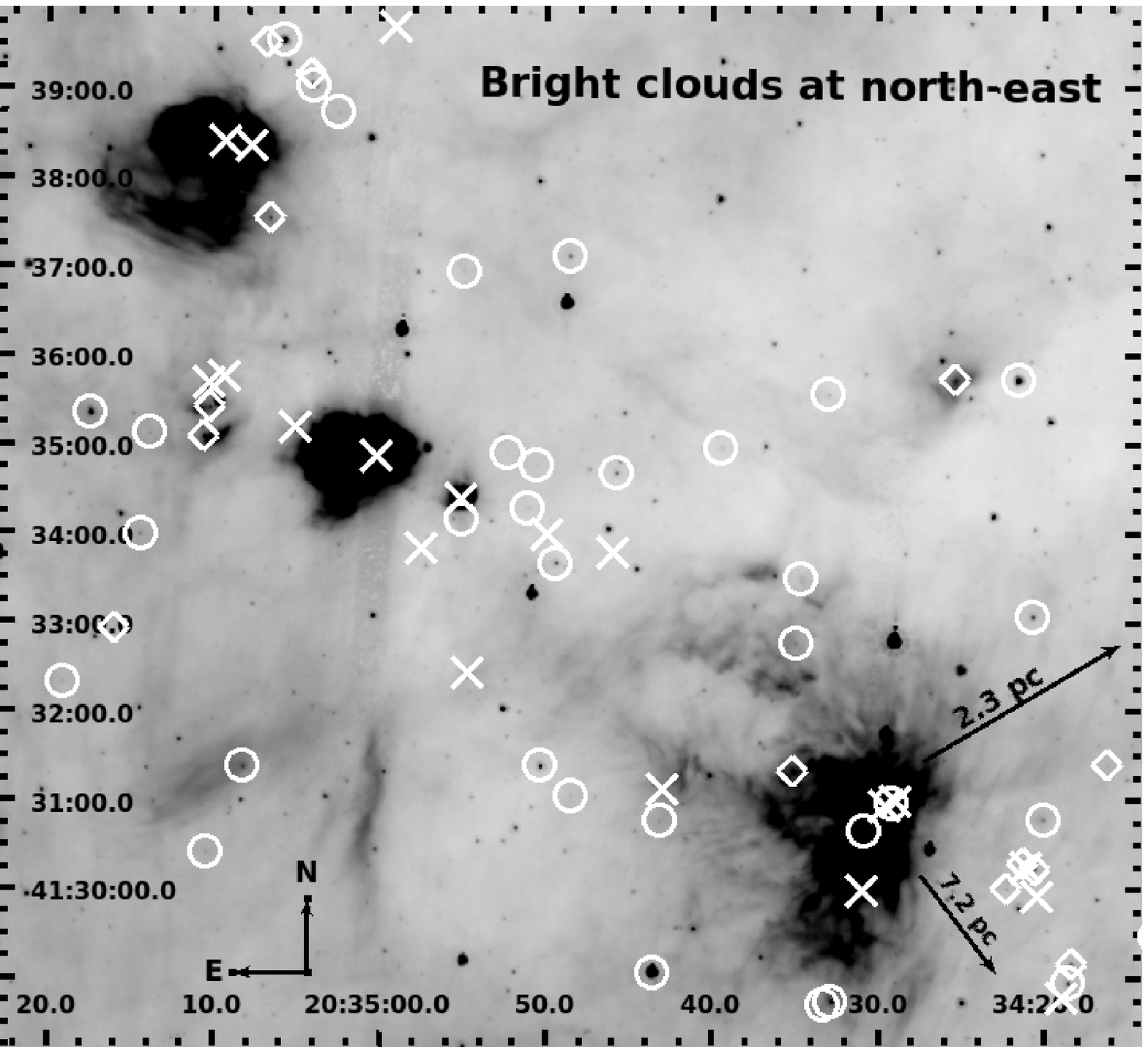}
        \par
        \includegraphics[width=5.5cm]{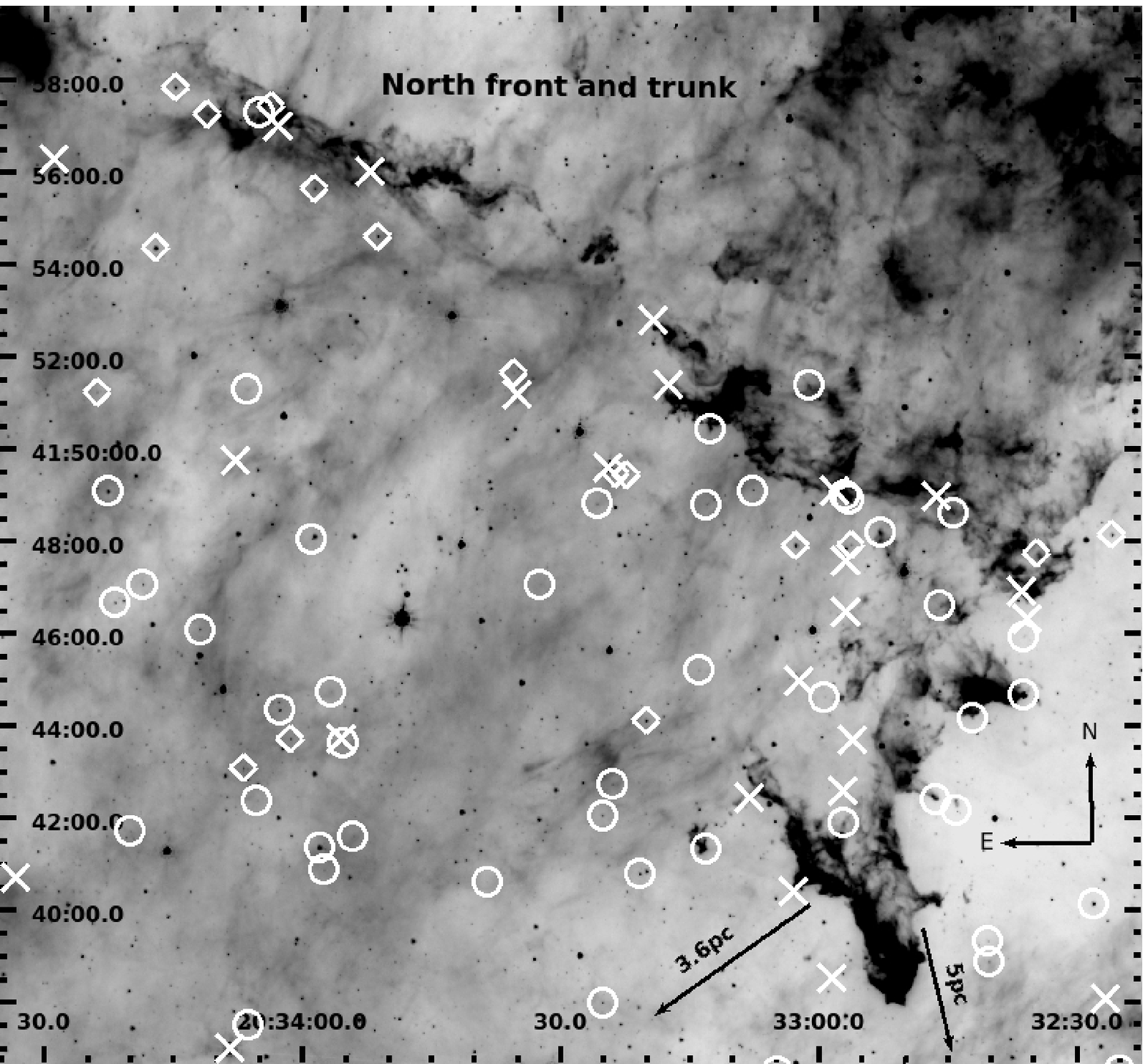}
        \includegraphics[width=5.5cm]{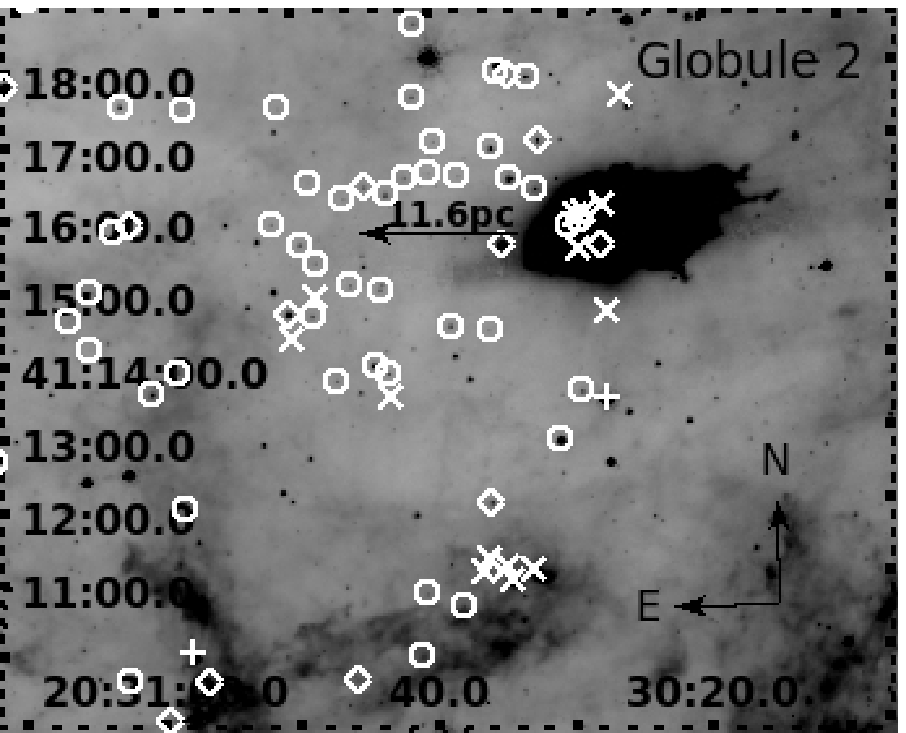}
        \caption{$8.0\mu m$ images of active sites of star formation in the Cygnus~OB2 area. The different symbols mark the positions of stars with different classification: crosses for class~I objects, diamonds for flat-spectrum, and circles for class~II. In some image, O stars are marked with plus symbols. The arrow indicated the direction and the distance of the star-forming region with the central cluster or other groups of O stars. }
        \label{sfr_im}
        \end{figure*}

In the previous sections we identified 1843 stars with disks in Cyg OB2, in various evolutionary stages. The large number of identified embedded objects suggests that star formation is still active in some regions of the cloud. One way to infer the relative age of different young stellar populations is from the class~I/class~II ratio, given the different timescale of these two evolutionary phases. Usually, a ratio larger than 25\% or 30\% indicates a large fraction of embedded population and ongoing star formation \citep{Balog2007,Masiunas2012}.
  Fig. \ref{IvsII_im} shows a gray-scale map of the class~I/class~II ratio in an uniform grid, overplotted with the positions of the O stars of Cyg~OB2 and the emission levels in $8.0\mu m$. The map clearly shows a large difference between the central area plus the cavity, that hosts the large majority of the massive stars, and the outer regions, where the cloud is still dense. In the entire central area the class~I/class~II ratio is lower than 20\%, indicating a low embedded population fraction and that the star formation process has very likely ended. The outer regions are instead rich in embedded objects, with a larger fraction in the east part of the cloud. The class~I/class~II ratio clearly peaks in correspondence of the bright cloud DR18, the dense knots to the north-east, the northern cloud front and the Globule \#1.
Fig. \ref{sfr_im} shows $8.0\mu m$ images of the regions in the Cyg~OB2 area with large fractions of embedded stars. In each panel, the positions of stars with different classification (class~I, flat-spectrum, and class~II YSOs) are marked with different symbols. \par
  The first panel shows the DR18 cloud, which is very bright at $8.0\mu m$ (image size $17^{\prime}.5\times 6^{\prime}.5$). \citet{Schneider2006} observed high gas velocities in the radio emission map of this cloud, and they concluded that it is modeled by the intense ionizing radiation from the center of Cyg~OB2. As shown in Fig. \ref{sfr_im}, the tail of DR18 is not perfectly aligned with the direction of the central cluster (at a projected distance of $8.9\,pc$), but it is likely that the sparse distribution of massive stars in this association contributes in shaping this cloud. For instance, the two closest O stars are in two different directions: at a projected distance of $5.1\,pc$ in the north-west and $5.6\,pc$ in the south-west. Inside this cloud we identified 10 class~I objects, plus $6$ in the proximity, while only 6 class~II objects seems to be related to the cloud. The only flat-spectrum source in this area lies close to the tip of the proplyd-like object number 10 studied by \citet{Wright2012}, and might have recently emerged from it. \par 
	The second panel shows the HII region ECX6-27 \citep{Wendker1991} ($9^{\prime}\times5.5^{\prime}$), at the projected distance from the central cluster of $3.1\,pc$. Following the analysis of the radio emission of this region, its distance is estimated to be larger than that to Cyg~OB2. \citet{PipenbrinkWendker1988} estimated a distance of $\sim 10\,kpc$, which has been more recently reduced to $3.3\,kpc$ by \citet{ComeronTorra2001}. This area hosts 9 class~I objects and the subcluster \#8, which has a fraction of class~I sources of 32\%. It is unlikely that the spatial coincidence between this embedded population and the ECX6-27 is fortuitous. However, we note that the class~II objects which lie in the nebulosity share similar position in the various color-color and color-magnitude diagrams of all the other members of Cyg~OB2 and they are too bright in optical to be at the distance of $3.3\,kpc$, more than twice that of Cyg~OB2. By the way, some of the very embedded objects could be detected even if they lie at such large distance. In conclusion, our data suggest a smaller distance for this nebulosity, but they do not provide any conclusive evidence. \par
The third panel shows a $15^{\prime}.3\times 10^{\prime}.6$ image of three bright clouds in the north-east direction, $7.2\,pc$ distant from the central cluster and $2.3\,pc$ from the north-east group of O stars. This region is rich of class~I YSOs (20, versus 13 flat-spectrum and 37 class~II) and it hosts the subcluster \#20, with a 30\% class~I fraction. This is a region rich in recent star formation activity, experiencing the effects due to both the O stars in the central cluster and those in the north-east. \par
	The fourth panel shows a cloud structure with a north-south trunk-shape and a front, at $5\,pc$ and $3.6\,pc$ from the two closest groups of O stars. Several YSOs (mainly embedded objects) are detected in the head of EGG-like (Evaporating Gaseous Globules, \citealp{Hester1996}) structures. With 17 class~I YSOs, almost all along the trunk and the front, together with 12 flat-spectrum and 12 class~II stars and intense  H$\alpha$ emission in this area (Fig. \ref{rgb_im}), this population may be the best example in Cyg~OB2 of star-formation in an evaporating photodissociation front, triggered by nearby OB stars. \par
The last image shows the ultracompact HII region ECX6-21 \citep{Wendker1991}, classified as ``Globule 2'' by  \citet{Schneider2006}. Indication of ongoing star formation in this cloud has been found by the identification of the source IRAS 20286+4105 with a rising infrared SED \citep{Odenwald1989} and the presence of water masers \citep{Palla1991}. In this area, we detected 6 class~I, 2 flat-spectrum, and 3 class~II objects. Besides, this region hosts the subcluster \#5, which has a 17.8\% fraction of class~I YSOs. All these indications suggest that star-formation is ongoing in this region. Its distance is compatible with the membership to the Cyg~OB2 association \citep{ComeronTorra2001}. The projected distance from the central cluster is large ($11.6\,pc$), but this cloud receives more direct radiation from some sparse O stars in the north-east, east, and south directions with projected distances ranging from $1.3\,pc$ to $5.5\,pc$. Another interesting group of 5 class~I objects lies southward the Globule 2. \par

\section{Summary and Conclusions}

Cygnus~OB2 is the the best target in our Galaxy to study star formation and disk evolution in presence of massive stars, being the closest star forming region to our Sun ($\sim 1.4\,kpc$) that hosts hundreds of massive stars and thousand of young low mass stars. In this paper we select and study the disk-bearing population of Cyg~OB2 in an area of about one square degree centered on the center of the association. We combined deep optical (from OSIRIS@GTC, SDSS, and IPHAS) and infrared data (2MASS/PSC, UKIDSS/GPS, and Spitzer data from the ``The Spitzer Legacy Survey of the Cygnus-X Region'') to obtain a deep and complete understanding of how star formation is affected by massive stars. The combined optical-infrared catalog has optical data covering the $12^m<r<25^m$ range and infrared data in the $8^m<J<20^m$ range with good photometry. \par
  The selection of the disk-bearing population has been performed combining several selection criteria, in order to take full advantage of the large set of multiwavelength data we have. Candidate disks have been selected using the \citet{Gutermuth2009} scheme, the \citet{Harvey2007} scheme, the $J-H$ vs. $H-K$ diagram; six reddening-free color indices $Q$ similar to those defined by \citet{Damiani2006}, and \citet{Guarcello2007,Guarcello2009}, and the [3.6] vs. [3.6]-[24] diagram \citep{Rebull2011}. The combined list of selected stars, amounting to 2703 sources, has been pruned from stars with unreliable excesses, and candidate foreground stars, background giants, and extragalactic sources by using a specific 5-steps selection algorithm. The highly reliable final list of candidate disk-bearing stars contains 1843 stars. \par
The evolutionary status of the selected stars with disks have been determined by adopting the definition from \citet{Wilking2001} based on the spectral index $\alpha$, the IRAC colors \citep{Gutermuth2009}, and the SED analysis \citep{Robitaille2007}, resulting in 155 class~I YSOs (8.4\% of the entire population); 242 flat-spectrum sources (13.1\%), and 1349 class~II objects (72.9\%). We also classified 62 stars with transition disks and 42 whose observed SEDs suggest the presence of pre-transition disks; 39 Blue stars With Excesses, 70 disk-bearing object with H$\alpha$ emission, and 24 highly embedded YSOs. \par
  The structure of the association has been studied with the surface density map \citep{CasertanoHut1985} and the Minimum Spanning Tree \citep{Barrow1985}. These two techniques reveal a complex and clumpy morphology of the association. A central subcluster is surrounded by an annular overdensity of stars with disks, where several candidate subclusters lie. The subclusters lying along this structure show similar stellar content, but increasing extinction from north southward. The surrounding area shows a number of candidate subclusters, peaks in the stellar surface density and nebular structures, with a significant embedded population and large class~I/class~II ratio, suggesting the presence of recent star formation events. The morphology of the association, together with the orientation of these nebular structures with respect to the central massive population of the association, suggest some level of triggered star formation. Future analysis of the sequence of star formation in the association and the dynamics of the stars associated with Cygnus~OB2 will shed some light on the nature of the structures identified in this paper. 

\label{conclusions}


\acknowledgments

We thank the anonymous referee for the useful comments. The author is indebted with Raffaele D'Abrusco and Ignazio Pillitteri for their precious help. This article makes use of data obtained with observations made with the Gran Telescopio CANARIAS (GTC), installed in the Spanish Observatorio del Roque de los Muchachos of the Instituto de Astrof\'{\i}sica de CANARIAS, in the island of La Palma; data obtained as part of the INT Photometric Hα Survey of the Northern Galactic Plane (IPHAS) carried out at the Isaac Newton Telescope (INT; all IPHAS data are processed by the Cambridge Astronomical Survey Unit, at the Institute of Astronomy in Cambridge); SDSS data, founded by the Alfred P. Sloan Foundation, the Participating Institutions, the National Science Foundation, the U.S. Department of Energy, the National Aeronautics and Space Administration, the Japanese Monbukagakusho, the Max Planck Society, and the Higher Education Funding Council for England; data products from the Two Micron All Sky Survey, which is a joint project of the University of Massachusetts and the Infrared Processing and Analysis Center/California Institute of Technology, funded by the National Aeronautics and Space Administration and the National Science Foundation; data based on observations made with the Spitzer Space Telescope, which is operated by the Jet Propulsion Laboratory, California Institute of Technology, under contract with NASA. The paper is also based on data obtained as part of the UKIRT Infrared Deep Sky Survey (UKIDSS, \citealp{LawrenceWAE2007}, we are grateful to Charles Williams for his support of the Apple Mac X-grid cluster in Exeter, on which the UKIRT data were reduced). UKIDSS uses the UKIRT Wide Field Camera (WFCAM; \citealp{Casali2007}). The photometric system is described in \citet{HewettWLH2006}, and the calibration is described in \citet{HodgkinIHW2009}. The pipeline processing and science archive are described in Irwin et al (2009, in prep) and \citet{HamblyCCM2008}. M.G.G. is supported by the Chandra grant GO0-11040X. J.J.D., V.L.K., and T.A. are supported by NASA contract NAS8-39073 to the Chandra X-ray Center (CXC) and thank the Director, Harvey Tananbaum, and the CXC science staff for advice and support. D.G.A. acknowledges support from the Spanish MICINN through grant AYA2008-02038. R.A.G. would like to acknowledge the support of NASA Astrophysics Data Analysis Program (ADAP) grants NNX11AD14G and NNX13AF08G, and Caltech/JPL awards 1373081, 1424329, and 1440160 in support of Spitzer Space Telescope observing programs.

\newpage
\appendix
\section{On-line catalog}
\label{cata_sec}

In this section we describe the information contained in the catalog of the candidate stars with disk (1843 stars). 

\begin{itemize}
\item ID: a sequential ID for the stars.
\item Celestial coordinates: Right Ascension and Declination in J2000. Remember that the astrometric systems of all the merged catalogs are referred to the 2MASS.
\item OSIRIS photometry: If available, magnitudes and errors in $riz$ OSIRIS bands. 
\item IPHAS photometry: If available, magnitudes and errors in $riH_{\alpha}$ IPHAS bands.
\item SDSS photometry: If available, magnitudes and errors in $ugriz$ SDSS bands.
\item 2MASS photometry: If available, magnitudes and errors in $JHK$ 2MASS bands.
\item 2MASS quality flag: four columns showing the {\it ph\_qual, rd\_flg, bl\_flg, cc\_flg} quality flag provided by the 2MASS/PSC catalog.
\item UKIDSS photometry: If available, magnitudes and errors in $JHK$ plus the $J-K$ and $H-K$ colors.
\item IRAC photometry: If available, magnitudes and errors in the IRAC bands.
\item MIPS photometry: If available, magnitude and error in [24].
\item MULTI\_OS\_IP tag: larger than 0 if the source has a multiple match between the OSIRIS and IPHAS counterparts.
\item MULTI\_UK\_2M tag: larger than 0 if the source has a multiple match between the UKIDSS and 2MASS counterparts.
\item MULTI\_NIR\_OPT tag: larger than 0 if the source has a multiple match between the UKIDSS+2MASS and optical counterparts.
\item MULTI\_OI\_SP tag: larger than 0 if the source has a multiple match between the UKIDSS+2MASS+optical and Spitzer counterparts.
\item ec\_gut tag: equal to 1 if the star has been selected with the GMM09 method.
\item ec\_224 tag: equal to 1 if the star has been selected with the $[4.5]$ vs. $[4.5]-[8.0]$ diagram.
\item ec\_11M tag: equal to 1 if the star has been selected with the $[3.6]$ vs. $[3.6]-[24]$ diagram.
\item ec\_M4M tag: equal to 1 if the star has been selected with the $[24]$ vs. $[8.0]-[24]$ diagram.
\item ec\_233M tag: equal to 1 if the star has been selected with the $[4.5]-[5.8]$ vs. $[5.8]-[24]$ diagram.
\item ec\_M24 tag: equal to 1 if the star has been selected with the $[24]$ vs. $[4.5]-[8.0]$ diagram.
\item ec\_QriJ1 tag: equal to 1 if the star has been selected with the $Q_{riJ[3.6]}$ index.
\item ec\_QriJ2 tag: equal to 1 if the star has been selected with the $Q_{riJ[4.5]}$ index.
\item ec\_QriJ3 tag: equal to 1 if the star has been selected with the $Q_{riJ[5.8]}$ index.
\item ec\_QriJ4 tag: equal to 1 if the star has been selected with the $Q_{riJ[8.0]}$ index.
\item ec\_JHHK tag: equal to 1 if the star has been selected with the $J-H$ vs. $H-K$ diagram.
\item ec\_QJHHK tag: equal to 1 if the star has been selected with the $Q_{JHHK}$ index.
\item ec\_QriHK tag: equal to 1 if the star has been selected with the $Q_{riHK}$ index.
\item alpha: spectral index.
\item classification: a string describing the classification of the star. If more classifications are present, they are divided by underscores. The different codes are: ``HE'' (highly embedded), ``Cl1" (class I), ``FS'' (flat-spectrum), ``Cl2'' (class II), ``NC'' (not classifiable), ``BWE'' (BWE star), ``TD'' (transition disk), ``PTD'' (pre-transition disk), ``lowmass'' (low-mass disk with excesses only in [8.0] and [24]), ``high-incl'' (highly inclined disk with excesses only in [8.0] and [24]), ``Ha'' (classified as $H\alpha$ emitter in this paper or by \citet{Vink2008}).
\end{itemize}
 
\newpage
\addcontentsline{toc}{section}{\bf Bibliografia}
\bibliographystyle{aa}
\bibliography{biblio}

\begin{thebibliography}{102}
\expandafter\ifx\csname natexlab\endcsname\relax\def\natexlab#1{#1}\fi

\bibitem[{{Adams}(2010)}]{Adams2010}
{Adams}, F.~C. 2010, \araa, 48, 47

\bibitem[{{Adams} {et~al.}(2006){Adams}, {Proszkow}, {Fatuzzo}, \&
  {Myers}}]{AdamsPFM2006}
{Adams}, F.~C., {Proszkow}, E.~M., {Fatuzzo}, M., \& {Myers}, P.~C. 2006, \apj,
  641, 504

\bibitem[{{Aihara} {et~al.}(2011){Aihara}, {Allende Prieto}, {An}, \&
  {Anderson}}]{Aihara2011}
{Aihara}, H., {Allende Prieto}, C., {An}, D., \& {Anderson}, S.~F. e.~a. 2011,
  \apjs, 193, 29

\bibitem[{{Albacete Colombo} {et~al.}(2007){Albacete Colombo}, {Flaccomio},
  {Micela}, {Sciortino}, \& {Damiani}}]{AlbaceteColombo2007}
{Albacete Colombo}, J.~F., {Flaccomio}, E., {Micela}, G., {Sciortino}, S., \&
  {Damiani}, F. 2007, \aap, 464, 211

\bibitem[{{Alexander} {et~al.}(2006){Alexander}, {Clarke}, \&
  {Pringle}}]{Alexander2006}
{Alexander}, R.~D., {Clarke}, C.~J., \& {Pringle}, J.~E. 2006, \mnras, 369, 216

\bibitem[{{Allen} {et~al.}(2004){Allen}, {Calvet}, {D'Alessio}, {Merin},
  {Hartmann}, {Megeath}, {Gutermuth}, {Muzerolle}, {Pipher}, {Myers}, \&
  {Fazio}}]{Allen2004}
{Allen}, L.~E., {Calvet}, N., {D'Alessio}, P., {et~al.} 2004, \apjs, 154, 363

\bibitem[{{Argiroffi} {et~al.}(2007){Argiroffi}, {Maggio}, \&
  {Peres}}]{Argiroffi2007}
{Argiroffi}, C., {Maggio}, A., \& {Peres}, G. 2007, \aap, 465, L5

\bibitem[{{Balog} {et~al.}(2007){Balog}, {Muzerolle}, {Rieke}, {Su}, {Young},
  \& {Megeath}}]{Balog2007}
{Balog}, Z., {Muzerolle}, J., {Rieke}, G.~H., {et~al.} 2007, \apj, 660, 1532

\bibitem[{{Barrow} {et~al.}(1985){Barrow}, {Bhavsar}, \& {Sonoda}}]{Barrow1985}
{Barrow}, J.~D., {Bhavsar}, S.~P., \& {Sonoda}, D.~H. 1985, \mnras, 216, 17

\bibitem[{{Beerer} {et~al.}(2010){Beerer}, {Koenig}, {Hora}, {Gutermuth},
  {Bontemps}, {Megeath}, {Schneider}, {Motte}, {Carey}, {Simon}, {Keto},
  {Smith}, {Allen}, {Fazio}, {Kraemer}, {Price}, {Mizuno}, {Adams},
  {Hern{\'a}ndez}, \& {Lucas}}]{Beerer2010}
{Beerer}, I.~M., {Koenig}, X.~P., {Hora}, J.~L., {et~al.} 2010, \apj, 720, 679

\bibitem[{{Bica} {et~al.}(2003){Bica}, {Bonatto}, \& {Dutra}}]{Bica2003}
{Bica}, E., {Bonatto}, C., \& {Dutra}, C.~M. 2003, \aap, 405, 991

\bibitem[{{Brott} \& {Hauschildt}(2005)}]{Brott2005}
{Brott}, I. \& {Hauschildt}, P.~H. 2005, in ESA Special Publication, Vol. 576,
  The Three-Dimensional Universe with Gaia, ed. C.~{Turon}, K.~S. {O'Flaherty},
  \& M.~A.~C. {Perryman}, 565

\bibitem[{{Calvet} {et~al.}(2002){Calvet}, {D'Alessio}, {Hartmann}, {Wilner},
  {Walsh}, \& {Sitko}}]{Calvet2002}
{Calvet}, N., {D'Alessio}, P., {Hartmann}, L., {et~al.} 2002, \apj, 568, 1008

\bibitem[{{Casali} {et~al.}(2007){Casali}, {Adamson}, {Alves de Oliveira},
  {Almaini}, {Burch}, {Chuter}, {Elliot}, {Folger}, {Foucaud}, {Hambly},
  {Hastie}, {Henry}, {Hirst}, {Irwin}, {Ives}, {Lawrence}, {Laidlaw}, {Lee},
  {Lewis}, {Lunney}, {McLay}, {Montgomery}, {Pickup}, {Read}, {Rees}, {Robson},
  {Sekiguchi}, {Vick}, {Warren}, \& {Woodward}}]{Casali2007}
{Casali}, M., {Adamson}, A., {Alves de Oliveira}, C., {et~al.} 2007, \aap, 467,
  777

\bibitem[{{Casertano} \& {Hut}(1985)}]{CasertanoHut1985}
{Casertano}, S. \& {Hut}, P. 1985, \apj, 298, 80

\bibitem[{{Cepa} {et~al.}(2000){Cepa}, {Aguiar}, {Escalera},
  {Gonzalez-Serrano}, {Joven-Alvarez}, {Peraza}, {Rasilla}, {Rodriguez-Ramos},
  {Gonzalez}, {Cobos Duenas}, {Sanchez}, {Tejada}, {Bland-Hawthorn},
  {Militello}, \& {Rosa}}]{Cepa2000}
{Cepa}, J., {Aguiar}, M., {Escalera}, V.~G., {et~al.} 2000, in Presented at the
  Society of Photo-Optical Instrumentation Engineers (SPIE) Conference, Vol.
  4008, Society of Photo-Optical Instrumentation Engineers (SPIE) Conference
  Series, ed. {M.~Iye \& A.~F.~Moorwood}, 623--631

\bibitem[{{Comer{\'o}n} {et~al.}(2002){Comer{\'o}n}, {Pasquali}, {Rodighiero},
  {Stanishev}, {De Filippis}, {L{\'o}pez Mart{\'{\i}}}, {G{\'a}lvez Ortiz},
  {Stankov}, \& {Gredel}}]{Comeron2002}
{Comer{\'o}n}, F., {Pasquali}, A., {Rodighiero}, G., {et~al.} 2002, \aap, 389,
  874

\bibitem[{{Comer{\'o}n} \& {Torra}(2001)}]{ComeronTorra2001}
{Comer{\'o}n}, F. \& {Torra}, J. 2001, \aap, 375, 539

\bibitem[{{Currie} \& {Sicilia-Aguilar}(2011)}]{Currie2011}
{Currie}, T. \& {Sicilia-Aguilar}, A. 2011, \apj, 732, 24

\bibitem[{{Cutri} {et~al.}(2003){Cutri}, {Skrutskie}, {van Dyk}, {Beichman},
  {Carpenter}, {Chester}, {Cambresy}, {Evans}, {Fowler}, {Gizis}, {Howard},
  {Huchra}, {Jarrett}, {Kopan}, {Kirkpatrick}, {Light}, {Marsh}, {McCallon},
  {Schneider}, {Stiening}, {Sykes}, {Weinberg}, {Wheaton}, {Wheelock}, \&
  {Zacarias}}]{Cutri2003}
{Cutri}, R.~M., {Skrutskie}, M.~F., {van Dyk}, S., {et~al.} 2003, {2MASS All
  Sky Catalog of point sources.}, ed. {Cutri, R.~M., Skrutskie, M.~F., van Dyk,
  S., Beichman, C.~A., Carpenter, J.~M., Chester, T., Cambresy, L., Evans, T.,
  Fowler, J., Gizis, J., Howard, E., Huchra, J., Jarrett, T., Kopan, E.~L.,
  Kirkpatrick, J.~D., Light, R.~M., Marsh, K.~A., McCallon, H., Schneider, S.,
  Stiening, R., Sykes, M., Weinberg, M., Wheaton, W.~A., Wheelock, S., \&
  Zacarias, N.}

\bibitem[{{Damiani} {et~al.}(2006){Damiani}, {Prisinzano}, {Micela}, \&
  {Sciortino}}]{Damiani2006}
{Damiani}, F., {Prisinzano}, L., {Micela}, G., \& {Sciortino}, S. 2006, \aap,
  459, 477

\bibitem[{{De Marchi} {et~al.}(2012){De Marchi}, {Panagia}, {Guarcello}, \&
  {Bonito}}]{DeMarchiPoster2012}
{De Marchi}, G., {Panagia}, N., {Guarcello}, M.~G., \& {Bonito}, R. 2012, in
  American Astronomical Society Meeting Abstracts, Vol. 219, American
  Astronomical Society Meeting Abstracts \#219, \#337.06

\bibitem[{{De Marchi} {et~al.}(2010){De Marchi}, {Panagia}, \&
  {Romaniello}}]{DeMarchi2010}
{De Marchi}, G., {Panagia}, N., \& {Romaniello}, M. 2010, \apj, 715, 1

\bibitem[{{Donley} {et~al.}(2012){Donley}, {Koekemoer}, {Brusa}, {Capak},
  {Cardamone}, {Civano}, {Ilbert}, {Impey}, {Kartaltepe}, {Miyaji}, {Salvato},
  {Sanders}, {Trump}, \& {Zamorani}}]{Donley2012}
{Donley}, J.~L., {Koekemoer}, A.~M., {Brusa}, M., {et~al.} 2012, \apj, 748, 142

\bibitem[{{Drake} {et~al.}(2009){Drake}, {Wright}, \& {Chandra Cyg Ob2
  Team}}]{Drake2009}
{Drake}, J., {Wright}, N., \& {Chandra Cyg Ob2 Team}. 2009, in Chandra's First
  Decade of Discovery, ed. S.~{Wolk}, A.~{Fruscione}, \& D.~{Swartz}

\bibitem[{{Drew} {et~al.}(2005){Drew}, {Greimel}, {Irwin}, {Aungwerojwit},
  {Barlow}, {Corradi}, {Drake}, {G{\"a}nsicke}, {Groot}, {Hales}, {Hopewell},
  {Irwin}, {Knigge}, {Leisy}, {Lennon}, {Mampaso}, {Masheder}, {Matsuura},
  {Morales-Rueda}, {Morris}, {Parker}, {Phillipps}, {Rodriguez-Gil}, {Roelofs},
  {Skillen}, {Sokoloski}, {Steeghs}, {Unruh}, {Viironen}, {Vink}, {Walton},
  {Witham}, {Wright}, {Zijlstra}, \& {Zurita}}]{Drew2005}
{Drew}, J.~E., {Greimel}, R., {Irwin}, M.~J., {et~al.} 2005, \mnras, 362, 753

\bibitem[{{Drew} {et~al.}(2008){Drew}, {Greimel}, {Irwin}, \&
  {Sale}}]{Drew2008}
{Drew}, J.~E., {Greimel}, R., {Irwin}, M.~J., \& {Sale}, S.~E. 2008, \mnras,
  386, 1761

\bibitem[{{Dye} {et~al.}(2006){Dye}, {Warren}, {Hambly}, {Cross}, {Hodgkin},
  {Irwin}, {Lawrence}, {Adamson}, {Almaini}, {Edge}, {Hirst}, {Jameson},
  {Lucas}, {van Breukelen}, {Bryant}, {Casali}, {Collins}, {Dalton}, {Davies},
  {Davis}, {Emerson}, {Evans}, {Foucaud}, {Gonzales-Solares}, {Hewett},
  {Kendall}, {Kerr}, {Leggett}, {Lodieu}, {Loveday}, {Lewis}, {Mann},
  {McMahon}, {Mortlock}, {Nakajima}, {Pinfield}, {Rawlings}, {Read}, {Riello},
  {Sekiguchi}, {Smith}, {Sutorius}, {Varricatt}, {Walton}, \&
  {Weatherley}}]{DyeWHC2012}
{Dye}, S., {Warren}, S.~J., {Hambly}, N.~C., {et~al.} 2006, \mnras, 372, 1227

\bibitem[{{Elmegreen}(2011)}]{Elmegreen2011}
{Elmegreen}, B.~G. 2011, in EAS Publications Series, Vol.~51, EAS Publications
  Series, ed. C.~{Charbonnel} \& T.~{Montmerle}, 45--58

\bibitem[{{Elmegreen} \& {Lada}(1977)}]{ElmegreenLada1977}
{Elmegreen}, B.~G. \& {Lada}, C.~J. 1977, \apj, 214, 725

\bibitem[{{Espaillat} {et~al.}(2007){Espaillat}, {Calvet}, {D'Alessio},
  {Hern{\'a}ndez}, {Qi}, {Hartmann}, {Furlan}, \& {Watson}}]{Espaillat2007}
{Espaillat}, C., {Calvet}, N., {D'Alessio}, P., {et~al.} 2007, \apjl, 670, L135

\bibitem[{{Flaherty} {et~al.}(2007){Flaherty}, {Pipher}, {Megeath}, {Winston},
  {Gutermuth}, {Muzerolle}, {Allen}, \& {Fazio}}]{Flaherty2007}
{Flaherty}, K.~M., {Pipher}, J.~L., {Megeath}, S.~T., {et~al.} 2007, \apj, 663,
  1069

\bibitem[{{Gennaro} {et~al.}(2011){Gennaro}, {Brandner}, {Stolte}, \&
  {Henning}}]{Gennaro2011}
{Gennaro}, M., {Brandner}, W., {Stolte}, A., \& {Henning}, T. 2011, \mnras,
  412, 2469

\bibitem[{{Girardi} {et~al.}(2002){Girardi}, {Bertelli}, {Bressan}, {Chiosi},
  {Groenewegen}, {Marigo}, {Salasnich}, \& {Weiss}}]{Girardi2002}
{Girardi}, L., {Bertelli}, G., {Bressan}, A., {et~al.} 2002, \aap, 391, 195

\bibitem[{{Girardi} {et~al.}(2005){Girardi}, {Groenewegen}, {Hatziminaoglou},
  \& {da Costa}}]{Girardi05}
{Girardi}, L., {Groenewegen}, M.~A.~T., {Hatziminaoglou}, E., \& {da Costa}, L.
  2005, \aap, 436, 895

\bibitem[{{Groenewegen}(2006)}]{Groenewegen2006}
{Groenewegen}, M.~A.~T. 2006, \aap, 448, 181

\bibitem[{{Guarcello} {et~al.}(2010){Guarcello}, {Damiani}, {Micela}, {Peres},
  {Prisinzano}, \& {Sciortino}}]{GuarcelloDMP2010}
{Guarcello}, M.~G., {Damiani}, F., {Micela}, G., {et~al.} 2010, \aap, 521, A18+

\bibitem[{{Guarcello} {et~al.}(2009){Guarcello}, {Micela}, {Damiani}, {Peres},
  {Prisinzano}, \& {Sciortino}}]{Guarcello2009}
{Guarcello}, M.~G., {Micela}, G., {Damiani}, F., {et~al.} 2009, \aap, 496, 453

\bibitem[{{Guarcello} {et~al.}(2007){Guarcello}, {Prisinzano}, {Micela},
  {Damiani}, {Peres}, \& {Sciortino}}]{Guarcello2007}
{Guarcello}, M.~G., {Prisinzano}, L., {Micela}, G., {et~al.} 2007, \aap, 462,
  245

\bibitem[{{Guarcello} {et~al.}(2012){Guarcello}, {Wright}, {Drake},
  {Garc{\'{\i}}a-Alvarez}, {Drew}, {Aldcroft}, \& {Kashyap}}]{Guarcello2012B}
{Guarcello}, M.~G., {Wright}, N.~J., {Drake}, J.~J., {et~al.} 2012, \apjs, 202,
  19

\bibitem[{{Gutermuth} {et~al.}(2009){Gutermuth}, {Megeath}, {Myers}, {Allen},
  {Pipher}, \& {Fazio}}]{Gutermuth2009}
{Gutermuth}, R.~A., {Megeath}, S.~T., {Myers}, P.~C., {et~al.} 2009, \apjs,
  184, 18

\bibitem[{{Gutermuth} {et~al.}(2008){Gutermuth}, {Myers}, {Megeath}, {Allen},
  {Pipher}, {Muzerolle}, {Porras}, {Winston}, \& {Fazio}}]{Gutermuth2008}
{Gutermuth}, R.~A., {Myers}, P.~C., {Megeath}, S.~T., {et~al.} 2008, \apj, 674,
  336

\bibitem[{{Hambly} {et~al.}(2008){Hambly}, {Collins}, {Cross}, {Mann}, {Read},
  {Sutorius}, {Bond}, {Bryant}, {Emerson}, {Lawrence}, {Rimoldini}, {Stewart},
  {Williams}, {Adamson}, {Hirst}, {Dye}, \& {Warren}}]{HamblyCCM2008}
{Hambly}, N.~C., {Collins}, R.~S., {Cross}, N.~J.~G., {et~al.} 2008, \mnras,
  384, 637

\bibitem[{{Hanson}(2003)}]{Hanson2003}
{Hanson}, M.~M. 2003, \apj, 597, 957

\bibitem[{{Hartmann} \& {Kenyon}(1990)}]{HartmannKenyon1990}
{Hartmann}, L.~W. \& {Kenyon}, S.~J. 1990, \apj, 349, 190

\bibitem[{{Harvey} {et~al.}(2007){Harvey}, {Mer{\'{\i}}n}, {Huard}, {Rebull},
  {Chapman}, {Evans}, \& {Myers}}]{Harvey2007}
{Harvey}, P., {Mer{\'{\i}}n}, B., {Huard}, T.~L., {et~al.} 2007, \apj, 663,
  1149

\bibitem[{{Hester} {et~al.}(1996){Hester}, {Scowen}, {Sankrit}, {Lauer},
  {Ajhar}, {Baum}, {Code}, {Currie}, {Danielson}, {Ewald}, {Faber},
  {Grillmair}, {Groth}, {Holtzman}, {Hunter}, {Kristian}, {Light}, {Lynds},
  {Monet}, {O'Neil}, {Shaya}, {Seidelmann}, \& {Westphal}}]{Hester1996}
{Hester}, J.~J., {Scowen}, P.~A., {Sankrit}, R., {et~al.} 1996, \aj, 111, 2349

\bibitem[{{Hewett} {et~al.}(2006){Hewett}, {Warren}, {Leggett}, \&
  {Hodgkin}}]{HewettWLH2006}
{Hewett}, P.~C., {Warren}, S.~J., {Leggett}, S.~K., \& {Hodgkin}, S.~T. 2006,
  \mnras, 367, 454

\bibitem[{{Hillenbrand}(1997)}]{Hillenbrand1997}
{Hillenbrand}, L.~A. 1997, \aj, 113, 1733

\bibitem[{{Hodgkin} {et~al.}(2009){Hodgkin}, {Irwin}, {Hewett}, \&
  {Warren}}]{HodgkinIHW2009}
{Hodgkin}, S.~T., {Irwin}, M.~J., {Hewett}, P.~C., \& {Warren}, S.~J. 2009,
  \mnras, 394, 675

\bibitem[{{Johnson} \& {Morgan}(1954)}]{JohnsonMorgan1954}
{Johnson}, H.~L. \& {Morgan}, W.~W. 1954, \apj, 119, 344

\bibitem[{{Johnstone} {et~al.}(1998){Johnstone}, {Hollenbach}, \&
  {Bally}}]{Johnstone1998}
{Johnstone}, D., {Hollenbach}, D., \& {Bally}, J. 1998, \apj, 499, 758

\bibitem[{{Kiminki} {et~al.}(2007){Kiminki}, {Kobulnicky}, {Kinemuchi},
  {Irwin}, {Fryer}, {Berrington}, {Uzpen}, {Monson}, {Pierce}, \&
  {Woosley}}]{Kiminki2007}
{Kiminki}, D.~C., {Kobulnicky}, H.~A., {Kinemuchi}, K., {et~al.} 2007, \apj,
  664, 1102

\bibitem[{{King} {et~al.}(2013){King}, {Naylor}, {Broos}, {Getman}, \&
  {Feigelson}}]{KingNBG2013}
{King}, R.~R., {Naylor}, T., {Broos}, P.~S., {Getman}, K.~V., \& {Feigelson},
  E.~D. 2013, \apjs, submitted (MYStIX X-ray matching paper)

\bibitem[{{Kn{\"o}dlseder}(2000)}]{Knodlseder2000}
{Kn{\"o}dlseder}, J. 2000, \aap, 360, 539

\bibitem[{{Kurucz}(1993)}]{Kurucz1993}
{Kurucz}, R. 1993, ATLAS9 Stellar Atmosphere Programs and 2 km/s grid.~Kurucz
  CD-ROM No.~13.~ Cambridge, Mass.: Smithsonian Astrophysical Observatory,
  1993., 13

\bibitem[{{Lada}(1987)}]{Lada1987}
{Lada}, C.~J. 1987, in IAU Symposium, Vol. 115, Star Forming Regions, ed.
  M.~{Peimbert} \& J.~{Jugaku}, 1--17

\bibitem[{{Lada} \& {Lada}(2003)}]{LadaLada2003}
{Lada}, C.~J. \& {Lada}, E.~A. 2003, \araa, 41, 57

\bibitem[{{Lawrence} {et~al.}(2007){Lawrence}, {Warren}, {Almaini}, {Edge},
  {Hambly}, {Jameson}, {Lucas}, {Casali}, {Adamson}, {Dye}, {Emerson},
  {Foucaud}, {Hewett}, {Hirst}, {Hodgkin}, {Irwin}, {Lodieu}, {McMahon},
  {Simpson}, {Smail}, {Mortlock}, \& {Folger}}]{LawrenceWAE2007}
{Lawrence}, A., {Warren}, S.~J., {Almaini}, O., {et~al.} 2007, \mnras, 379,
  1599

\bibitem[{{Lonsdale} {et~al.}(2003){Lonsdale}, {Smith}, {Rowan-Robinson},
  {Surace}, {Shupe}, {Xu}, {Oliver}, {Padgett}, {Fang}, {Conrow},
  {Franceschini}, {Gautier}, {Griffin}, {Hacking}, {Masci}, {Morrison},
  {O'Linger}, {Owen}, {P{\'e}rez-Fournon}, {Pierre}, {Puetter}, {Stacey},
  {Castro}, {Polletta}, {Farrah}, {Jarrett}, {Frayer}, {Siana}, {Babbedge},
  {Dye}, {Fox}, {Gonzalez-Solares}, {Salaman}, {Berta}, {Condon}, {Dole}, \&
  {Serjeant}}]{Lonsdale2003}
{Lonsdale}, C.~J., {Smith}, H.~E., {Rowan-Robinson}, M., {et~al.} 2003, \pasp,
  115, 897

\bibitem[{{Lucas} {et~al.}(2008){Lucas}, {Hoare}, {Longmore}, {Schr{\"o}der},
  {Davis}, {Adamson}, {Bandyopadhyay}, {de Grijs}, {Smith}, {Gosling},
  {Mitchison}, {G{\'a}sp{\'a}r}, {Coe}, {Tamura}, {Parker}, {Irwin}, {Hambly},
  {Bryant}, {Collins}, {Cross}, {Evans}, {Gonzalez-Solares}, {Hodgkin},
  {Lewis}, {Read}, {Riello}, {Sutorius}, {Lawrence}, {Drew}, {Dye}, \&
  {Thompson}}]{Lucas2008}
{Lucas}, P.~W., {Hoare}, M.~G., {Longmore}, A., {et~al.} 2008, \mnras, 391, 136

\bibitem[{{Luhman} {et~al.}(2010){Luhman}, {Allen}, {Espaillat}, {Hartmann}, \&
  {Calvet}}]{Luhman2010}
{Luhman}, K.~L., {Allen}, P.~R., {Espaillat}, C., {Hartmann}, L., \& {Calvet},
  N. 2010, \apjs, 186, 111

\bibitem[{{Masiunas} {et~al.}(2012){Masiunas}, {Gutermuth}, {Pipher},
  {Megeath}, {Myers}, {Allen}, {Kirk}, \& {Fazio}}]{Masiunas2012}
{Masiunas}, L.~C., {Gutermuth}, R.~A., {Pipher}, J.~L., {et~al.} 2012, \apj,
  752, 127

\bibitem[{{Massey} \& {Thompson}(1991)}]{Massey1991}
{Massey}, P. \& {Thompson}, A.~B. 1991, \aj, 101, 1408

\bibitem[{{Mer{\'{\i}}n} {et~al.}(2010){Mer{\'{\i}}n}, {Brown}, {Oliveira},
  {Herczeg}, {van Dishoeck}, {Bottinelli}, {Evans}, {Cieza}, {Spezzi},
  {Alcal{\'a}}, {Harvey}, {Blake}, {Bayo}, {Geers}, {Lahuis}, {Prusti},
  {Augereau}, {Olofsson}, {Walter}, \& {Chiu}}]{Merin2010}
{Mer{\'{\i}}n}, B., {Brown}, J.~M., {Oliveira}, I., {et~al.} 2010, \apj, 718,
  1200

\bibitem[{{Meyer} {et~al.}(1997){Meyer}, {Calvet}, \&
  {Hillenbrand}}]{Meyer1997}
{Meyer}, M.~R., {Calvet}, N., \& {Hillenbrand}, L.~A. 1997, \aj, 114, 288

\bibitem[{{Munari} \& {Carraro}(1996)}]{MunariCarraro1996}
{Munari}, U. \& {Carraro}, G. 1996, \aap, 314, 108

\bibitem[{{Muzerolle} {et~al.}(2010){Muzerolle}, {Allen}, {Megeath},
  {Hern{\'a}ndez}, \& {Gutermuth}}]{Muzerolle2010}
{Muzerolle}, J., {Allen}, L.~E., {Megeath}, S.~T., {Hern{\'a}ndez}, J., \&
  {Gutermuth}, R.~A. 2010, \apj, 708, 1107

\bibitem[{{Nagata} {et~al.}(1995){Nagata}, {Woodward}, {Shure}, \&
  {Kobayashi}}]{Nagata1995}
{Nagata}, T., {Woodward}, C.~E., {Shure}, M., \& {Kobayashi}, N. 1995, \aj,
  109, 1676

\bibitem[{{Naylor}(1998)}]{Naylor1998}
{Naylor}, T. 1998, \mnras, 296, 339

\bibitem[{{Negueruela} {et~al.}(2010){Negueruela}, {Clark}, \&
  {Ritchie}}]{Negueruela2010}
{Negueruela}, I., {Clark}, J.~S., \& {Ritchie}, B.~W. 2010, \aap, 516, A78

\bibitem[{{Negueruela} {et~al.}(2008){Negueruela}, {Marco}, {Herrero}, \&
  {Clark}}]{Negueruela2008}
{Negueruela}, I., {Marco}, A., {Herrero}, A., \& {Clark}, J.~S. 2008, \aap,
  487, 575

\bibitem[{{O'dell} \& {Wen}(1994)}]{ODell1994}
{O'dell}, C.~R. \& {Wen}, Z. 1994, \apj, 436, 194

\bibitem[{{Odenwald} \& {Schwartz}(1989)}]{Odenwald1989}
{Odenwald}, S.~F. \& {Schwartz}, P.~R. 1989, \apjl, 345, L47

\bibitem[{{O'Donnell}(1994)}]{Donnell94}
{O'Donnell}, J.~E. 1994, \apj, 422, 158

\bibitem[{{Palla} {et~al.}(1991){Palla}, {Brand}, {Comoretto}, {Felli}, \&
  {Cesaroni}}]{Palla1991}
{Palla}, F., {Brand}, J., {Comoretto}, G., {Felli}, M., \& {Cesaroni}, R. 1991,
  \aap, 246, 249

\bibitem[{{Palla} {et~al.}(2005){Palla}, {Randich}, {Flaccomio}, \&
  {Pallavicini}}]{Palla2005}
{Palla}, F., {Randich}, S., {Flaccomio}, E., \& {Pallavicini}, R. 2005, \apjl,
  626, L49

\bibitem[{{Peth} {et~al.}(2011){Peth}, {Ross}, \& {Schneider}}]{PethRS2011}
{Peth}, M.~A., {Ross}, N.~P., \& {Schneider}, D.~P. 2011, \aj, 141, 105

\bibitem[{{Pickles}(1998)}]{Pickles1998}
{Pickles}, A.~J. 1998, \pasp, 110, 863

\bibitem[{{Pipenbrink} \& {Wendker}(1988)}]{PipenbrinkWendker1988}
{Pipenbrink}, A. \& {Wendker}, H.~J. 1988, \aap, 191, 313

\bibitem[{{Rebull} {et~al.}(2011){Rebull}, {Guieu}, {Stauffer}, {Hillenbrand},
  {Noriega-Crespo}, {Stapelfeldt}, {Carey}, {Carpenter}, {Cole}, {Padgett},
  {Strom}, \& {Wolff}}]{Rebull2011}
{Rebull}, L.~M., {Guieu}, S., {Stauffer}, J.~R., {et~al.} 2011, \apjs, 193, 25

\bibitem[{{Reddish} {et~al.}(1967){Reddish}, {Lawrence}, \&
  {Pratt}}]{Reddish67}
{Reddish}, V.~C., {Lawrence}, L.~C., \& {Pratt}, N.~M. 1967, Publications of
  the Royal Observatory of Edinburgh, 5, 112

\bibitem[{{Richards} {et~al.}(2004){Richards}, {Nichol}, {Gray}, {Brunner},
  {Lupton}, {Vanden Berk}, {Chong}, {Weinstein}, {Schneider}, {Anderson},
  {Munn}, {Harris}, {Strauss}, {Fan}, {Gunn}, {Ivezi{\'c}}, {York},
  {Brinkmann}, \& {Moore}}]{Richards2004}
{Richards}, G.~T., {Nichol}, R.~C., {Gray}, A.~G., {et~al.} 2004, \apjs, 155,
  257

\bibitem[{{Rieke} \& {Lebofsky}(1985)}]{RiekeLebofsky1985}
{Rieke}, G.~H. \& {Lebofsky}, M.~J. 1985, \apj, 288, 618

\bibitem[{{Robberto} {et~al.}(2004){Robberto}, {Song}, {Mora Carrillo},
  {Beckwith}, {Makidon}, \& {Panagia}}]{Robberto2004}
{Robberto}, M., {Song}, J., {Mora Carrillo}, G., {et~al.} 2004, in Astronomical
  Society of the Pacific Conference Series, Vol. 322, The Formation and
  Evolution of Massive Young Star Clusters, ed. H.~J.~G.~L.~M. {Lamers}, L.~J.
  {Smith}, \& A.~{Nota}, 383

\bibitem[{{Robitaille} {et~al.}(2007){Robitaille}, {Whitney}, {Indebetouw}, \&
  {Wood}}]{Robitaille2007}
{Robitaille}, T.~P., {Whitney}, B.~A., {Indebetouw}, R., \& {Wood}, K. 2007,
  \apjs, 169, 328

\bibitem[{{Rygl} {et~al.}(2012){Rygl}, {Brunthaler}, {Sanna}, {Menten}, {Reid},
  {van Langevelde}, {Honma}, {Torstensson}, \& {Fujisawa}}]{Rygl2012}
{Rygl}, K.~L.~J., {Brunthaler}, A., {Sanna}, A., {et~al.} 2012, \aap, 539, A79

\bibitem[{{Sale} {et~al.}(2009){Sale}, {Drew}, {Unruh}, {Irwin}, {Knigge},
  {Phillipps}, {Zijlstra}, {G{\"a}nsicke}, {Greimel}, {Groot}, {Mampaso},
  {Morris}, {Napiwotzki}, {Steeghs}, \& {Walton}}]{Sale09}
{Sale}, S.~E., {Drew}, J.~E., {Unruh}, Y.~C., {et~al.} 2009, \mnras, 392, 497

\bibitem[{{Schneider} {et~al.}(2006){Schneider}, {Bontemps}, {Simon}, {Jakob},
  {Motte}, {Miller}, {Kramer}, \& {Stutzki}}]{Schneider2006}
{Schneider}, N., {Bontemps}, S., {Simon}, R., {et~al.} 2006, \aap, 458, 855

\bibitem[{{Siess} {et~al.}(2000){Siess}, {Dufour}, \& {Forestini}}]{Siess2000}
{Siess}, L., {Dufour}, E., \& {Forestini}, M. 2000, \aap, 358, 593

\bibitem[{{Smith}(2002)}]{Smith2002}
{Smith}, N. 2002, \mnras, 337, 1252

\bibitem[{{Stern} {et~al.}(2005){Stern}, {Eisenhardt}, {Gorjian}, {Kochanek},
  {Caldwell}, {Eisenstein}, {Brodwin}, {Brown}, {Cool}, {Dey}, {Green},
  {Jannuzi}, {Murray}, {Pahre}, \& {Willner}}]{Stern2005}
{Stern}, D., {Eisenhardt}, P., {Gorjian}, V., {et~al.} 2005, \apj, 631, 163

\bibitem[{{Vink} {et~al.}(2008){Vink}, {Drew}, {Steeghs}, {Wright}, {Martin},
  {G{\"a}nsicke}, {Greimel}, \& {Drake}}]{Vink2008}
{Vink}, J.~S., {Drew}, J.~E., {Steeghs}, D., {et~al.} 2008, \mnras, 387, 308

\bibitem[{{Walborn}(1973)}]{Walborn1973}
{Walborn}, N.~R. 1973, \apjl, 180, L35+

\bibitem[{{Walborn} {et~al.}(2002){Walborn}, {Howarth}, {Lennon}, {Massey},
  {Oey}, {Moffat}, {Skalkowski}, {Morrell}, {Drissen}, \&
  {Parker}}]{WalbornHLM2002}
{Walborn}, N.~R., {Howarth}, I.~D., {Lennon}, D.~J., {et~al.} 2002, \aj, 123,
  2754

\bibitem[{{Wendker} {et~al.}(1991){Wendker}, {Higgs}, \&
  {Landecker}}]{Wendker1991}
{Wendker}, H.~J., {Higgs}, L.~A., \& {Landecker}, T.~L. 1991, \aap, 241, 551

\bibitem[{{Wilking} {et~al.}(2001){Wilking}, {Bontemps}, {Schuler}, {Greene},
  \& {Andr{\'e}}}]{Wilking2001}
{Wilking}, B.~A., {Bontemps}, S., {Schuler}, R.~E., {Greene}, T.~P., \&
  {Andr{\'e}}, P. 2001, \apj, 551, 357

\bibitem[{{Winston} {et~al.}(2012){Winston}, {Wolk}, {Bourke}, {Megeath},
  {Gutermuth}, \& {Spitzbart}}]{Winston2012}
{Winston}, E., {Wolk}, S.~J., {Bourke}, T.~L., {et~al.} 2012, \apj, 744, 126

\bibitem[{{Wright} \& {Drake}(2009)}]{Wright2009}
{Wright}, N.~J. \& {Drake}, J.~J. 2009, \apjs, 184, 84

\bibitem[{{Wright} {et~al.}(2012){Wright}, {Drake}, {Drew}, {Guarcello},
  {Gutermuth}, {Hora}, \& {Kraemer}}]{Wright2012}
{Wright}, N.~J., {Drake}, J.~J., {Drew}, J.~E., {et~al.} 2012, \apjl, 746, L21

\bibitem[{{Wright} {et~al.}(2010){Wright}, {Drake}, {Drew}, \&
  {Vink}}]{Wright2010}
{Wright}, N.~J., {Drake}, J.~J., {Drew}, J.~E., \& {Vink}, J.~S. 2010, \apj,
  713, 871

\bibitem[{{Wright} {et~al.}(2008){Wright}, {Greimel}, {Barlow}, {Drew},
  {Cioni}, {Zijlstra}, {Corradi}, {Gonz{\'a}lez-Solares}, {Groot}, {Irwin},
  {Irwin}, {Mampaso}, {Morris}, {Steeghs}, {Unruh}, \& {Walton}}]{Wright2008}
{Wright}, N.~J., {Greimel}, R., {Barlow}, M.~J., {et~al.} 2008, \mnras, 390,
  929

\end{thebibliography}

\newpage

        \begin{table}[]
        \centering
        \caption {Source matching results}
        \vspace{0.5cm}
        \begin{tabular}{cccc}
        \hline
        \hline
        Catalog A & Catalog B & $r_{match}$ & N. matches \\
       \hline
	2MASS		& UKIDSS 	& $0.4^{\prime \prime}$ & 34446 \\
	OSIRIS		& IPHAS 	& $0.5^{\prime \prime}$ & 11207 \\
	OSIRIS+IPHAS	& SDSS	 	& $0.3^{\prime \prime}$ & 19314 \\
	NIR		& OPTICAL 	& $0.3^{\prime \prime}$ & 73208 \\
	NIR+OPTICAL	& Spitzer	& $0.4^{\prime \prime}$ & 114842\\
        \hline
        \hline
        \multicolumn{4}{l}{} 
        \end{tabular}
        \label{match_tb}
        \end{table}

        \begin{table}[]
        \centering
        \caption {Counterparts distributions}
        \vspace{0.5cm}
        \begin{tabular}{cccc}
        \hline
        \hline
        N. sources & Optical & $JHK$ & IRAC/MIPS \\
       \hline
	47187	& X & X & X \\
	64396	&   & X & X \\
	1813	& X &   & X \\
	26651	& X & X &   \\
	9200	& X &   &   \\
	144282	&   & X &   \\
	25985	&   &   & X \\
        \hline
        \hline
        \multicolumn{4}{l}{} 
        \end{tabular}
        \label{cntrp_tb}
        \end{table}

        \begin{table}[]
        \centering
        \caption {Selection of the disk bearing stars with ``relaxed'' limits}
        \vspace{0.5cm}
        \begin{tabular}{ccc}
        \hline
        \hline
        Method & New selections & Unique new selections \\
       \hline
	GMM09 scheme	& 215 & 79 \\
	$[4.5]$ vs. $[4.5]-[8.0]$	& 270 & 51 \\
	$J-H$ vs. $H-K$			& 151 & 103\\
	$Q_{JHHK}$ and $Q_{riHK}$	& 157 & 39 \\
	$Q_{riJ[sp]}$			& 61  & 39 \\
	MIPS diagrams			& 24  & 7  \\
        \hline
        \hline
        \multicolumn{3}{l}{} 
        \end{tabular}
        \label{relax_tb}
        \end{table}

        \begin{table}[]
        \centering
        \caption {Number of sources selected with the adopted criteria}
        \vspace{0.5cm}
        \begin{tabular}{cccc}
        \hline
        \hline
        Criterion & Initial$^*$ & Final$^{**}$ & Unique$^{***}$ \\
       \hline
	GMM09 scheme	& 1461& 1405& 478\\
	$[4.5]$ vs. $[4.5]-[8.0]$	& 655 & 646 & 38 \\
	$[3.6]$ vs. $[3.6]-[24]$	& 393 & 377 & 12 \\
	$[24]$ vs. $[8.0-24]$		& 230 & 212 & 11 \\
	$[4.5]-[5.8]$ vs. $[5.8]-[24]$	& 366 & 352 & 1  \\
	$[24]$ vs. $[4.5]-[8.0]$	& 245 & 223 & 2  \\
	$Q_{riJ[3.6]}$			& 291 & 235 & 0  \\
	$Q_{riJ[4.5]}$			& 385 & 330 & 5  \\
	$Q_{riJ[5.8]}$			& 485 & 426 & 2  \\
	$Q_{riJ[8.0]}$			& 680 & 638 & 49 \\
	$J-H$ vs. $H-K$			& 382 & 138 & 8  \\
	$Q_{JHHK}$			& 244 & 134 & 1  \\
	$Q_{riHK}$			& 128 & 83  & 3  \\
	\hline
        \hline
        \multicolumn{4}{l}{$^*$ Stars selected by the given criterion.}  \\
        \multicolumn{4}{l}{$^{**}$ Stars that have been accepted by the merging algorithm.} \\
        \multicolumn{4}{l}{$^{***}$ Stars in the final list selected {\bf only} by the given criterion.} \\
        \end{tabular}
        \label{criteria_tb}
        \end{table}

       \begin{table}[]
        \centering
        \caption {Characteristics of the subclusters}
        \vspace{0.5cm}
        \begin{tabular}{cccccccc}
        \hline
        \hline
        ID & stars& class~I& flat-spect.& class~II & TD& H$\alpha$-emit.& cl.~I frac.\\
        \hline
	1&  213& 11& 20& 176&  3& 4& 5.3\%\\
	2&  187&  8& 30& 144&  4& 8& 4.4\%\\
	3&   93&  5& 12&  74&  4& 1& 5.5\%\\
	4&   47&  2&  6&  38&  1& 2& 4.3\%\\
	5&   46&  8&  6&  31&  1& 1& 17.8\%\\
	6&   42&  2&  6&  34&  2& 4& 4.8\%\\
	7&   25&  3&  0&  21&  0& 1& 12.5\%\\
	8&   24&  8&  7&   8&  2& 3& 32\%\\
	9&   19&  0&  1&  18&  1& 0& 0\%\\
	10&  16&  0&  1&  15&  0& 1& 0\%   \\ 
	11&  15&  4&  2&   9&  0& 0& 26.7\%\\
	12&  15&  1&  2&  12&  3& 0& 6.7\%\\
	13&  14&  0&  2&  12&  1& 2& 0\%\\
	14&  13&  1&  1&  11&  2& 1& 7.7\%\\
	15&  13&  1&  2&  10&  1& 0& 7.7\%\\
	16&  12&  0&  2&  10&  0& 0& 0\%   \\
	17&  12&  0&  2&  10&  0& 0& 0\%\\
	18&  12&  0&  1&  11&  0& 0& 0\%\\
	19&  11&  1&  0&  10&  0& 9& 9.1\%   \\
	20&  11&  3&  5&   2&  0& 0& 30\%\\
        \hline
        \hline
        \multicolumn{4}{l}{} 
        \end{tabular}
        \label{clust_tbl}
        \end{table}

\newpage
\clearpage

\clearpage

\clearpage

\end{document}